\documentclass[twocolumn,pra,showpacs,superscriptaddress,amssymb,amsmath,amsmath]{revtex4-1}
\usepackage[T1]{fontenc}
\usepackage{graphicx}
\usepackage{epstopdf}
\usepackage{bm}
\usepackage{hyperref}
\usepackage{comment}
\usepackage{color}
\usepackage{physics}
\usepackage{amsmath}
\usepackage{tikz}
\usepackage{enumitem}
\usepackage{siunitx}
\usepackage{times}

\hypersetup{%
   pdfpagemode=None, 
   pdfstartpage=1,
   pdfmenubar=true,
   pdftoolbar=true,
   colorlinks = true,
   linkcolor=blue,
   citecolor=blue,
   urlcolor=blue,
   bookmarksopen=false
 }

\newcommand{\be}{\begin{equation}}
\newcommand{\ee}{\end{equation}}

\newcommand{\ham}{\mathcal{\hat{H}}}
\DeclareMathOperator{\erfc}{erfc}

\begin{document}

\title{Two ultracold highly magnetic atoms in a one-dimensional harmonic trap}

\author{Micha{\l} Suchorowski}
\email{m.suchorowski@uw.edu.pl}
\affiliation{Faculty of Physics,  University of Warsaw, Pasteura 5, 02-093 Warsaw, Poland}
\author{Anna Dawid}
\affiliation{Faculty of Physics,  University of Warsaw, Pasteura 5, 02-093 Warsaw, Poland}
\affiliation{ICFO - Institut de Ci\`encies Fot\`oniques, The Barcelona Institute of Science and Technology, 08860 Castelldefels, Spain}
\author{Micha{\l} Tomza}
\email{michal.tomza@fuw.edu.pl}
\affiliation{Faculty of Physics,  University of Warsaw, Pasteura 5, 02-093 Warsaw, Poland}

\date{\today}

\begin{abstract}

We theoretically investigate the properties of two interacting ultracold highly magnetic atoms trapped in a one-dimensional harmonic potential. The atoms interact via an anisotropic long-range dipole-dipole interaction, which in one dimension effectively can be modeled by the contact interaction. We investigate the interplay of the external magnetic field, the spin-spin interaction, and the trapping potential and how they affect the magnetization of the system. We show the role of indistinguishability and symmetries in the dynamics by studying the time evolution of the observables that could be measured experimentally. The presented model may depict the on-site interaction of the extended Hubbard models, therefore giving a better understanding of the fundamental building block of the respective many-body quantum simulators.

\end{abstract}

\pacs{}

\maketitle

\section{INTRODUCTION}

\hspace{1cm}

Modern technology allows us not only to produce ultracold synthetic quantum matter but also to control it on an unprecedented level. Few-body tunable systems~\cite{Serwane_2011}, atom-by-atom assembled cold atomic arrays of various dimensionality~\cite{Endres_2016, Barredo_2016}, single-atom imaging of bosons and fermions~\cite{Bakr_2009,Sherson_2010, Cheuk15PRL, Haller15NatPhys, Boll_2016}, and forming ultracold molecules from atoms~\cite{Regal_2003, Hodby_2005, Danzl08Science, Ni08, deMarco2019} are examples of many achievements that bring us closer and closer to fully controllable quantum systems. Such systems with tunable interactions and geometry are especially useful for quantum simulations of solid-state models such as Hubbard~\cite{Tarruell_2018, Gall_2021}, Ising~\cite{Labuhn_2016}, Heisenberg~\cite{Jepsen_2020}, or even extended Hubbard~\cite{Rossini_2012, Hofmann_2012, Dhar_2016, Dhar_2018, Biedron_2018,Suthar_2020} models. Quantum simulators~\cite{Lewenstein07, Bloch_2008, Blatt_2012, Bloch_2012, Schneider_2012, Dutta_2015, Baier_2016, Zhang_2017, Bernien_2017} have been used to investigate a broad range of physical phenomena such as the many-body localization~\cite{Choi_2016}, mobile spin impurity ~\cite{Fukuhara_2013a}, magnons~\cite{Fukuhara_2013b}, Mott transition~\cite{Greiner_2002a}, Fermi–Hubbard antiferromagnets~\cite{Mazurenko_2017}, or doping in the Hubbard model~\cite{Chiu_2019}. Recent simulators have already exploited 256 atoms~\cite{Ebadi_2021}, which exceeds the possibilities of most numerical approaches.

Next to many-body ultracold simulators, few-body systems are thoroughly investigated both experimentally and theoretically~\cite{Mistakidis2022}. Among them, one-dimensional (1D) setups~\cite{Sowinski_2019} have attracted significant attention due to the increased role of interactions and interesting novel phenomena resulting from quantum fluctuations~\cite{Cazalilla_2011}.
In particular, the analytical solutions for low-spin atoms interacting via contact~\cite{Busch_1998} or finite-range soft-core~\cite{Koscik_2018} potential have been developed. The two-body physics has been investigated using many systems~\cite{shea_2009, Sowinski_2010, Oldziejewski_2016, Gorecki_2016, Budewig_2019}, including ultracold molecules~\cite{Gorecki_2017, Dawid_2018, Dawid_2020,SroczyskaNJP22}. Recently, few spin-1/2 fermions in a 1D trap~\cite{Sowinski_2013,Rojo-Francas_2020} and one and two spin-1/2 fermions interacting via the spin-exchange with the impurity~\cite{Peng_2020} have also been studied. Few-body SU($N$) systems~\cite{Laird_2017} and larger systems of several particles confined in 1D geometries have been investigated as well~\cite{Deuretzbacher_2014, Sowinski_2013, Volosniev_2014, Grining_2015}. 
In parallel, experimental efforts have allowed for the deterministic preparation of a few ultracold fermions in a trap~\cite{Serwane_2011} that has led to groundbreaking studies of fermionization~\cite{Zurn_2012}, formation of a Fermi sea~\cite{Wenz_2013}, two fermions in a double well~\cite{Murmann_2015a}, and more~\cite{Zurn_2013, Murmann_2015b}.
Moreover, developments of the optical tweezers technology~\cite{Ashkin_1970, Ashkin_1986, Matthews_2009} have opened the possibility to trap and manipulate single particles also in quasi-1D harmoniclike potentials~\cite{Liao_2008, Matheson_2021}, which has allowed, among other possibilities for studying two bosons in a double well~\cite{Kaufman_2014, Kaufman_2016}.

Quantum simulators can be constructed using various particles including Rydberg atoms~\cite{Browaeys_2020, Barbier_2021}, ultracold ions~\cite{Blatt_2012}, and molecules~\cite{Ortner_2009, Blackmore_2019}. A new species has joined the toolbox with the realization of the Bose-Einstein condensate in chromium, dysprosium, erbium, and europium gases~\cite{Griesmaier_2005,Lu_2011,Aikawa_2012, MiyazawaARXIV2022}. Ultracold lanthanides have opened new possibilities coming from their large magnetic moments and more complex electronic structure~\cite{Norcia_2021, Chomaz_2022}. The anisotropic and long-range nature of magnetic dipolar interactions results in a range of interesting properties such as the existence of quantum chaos~\cite{Frisch_2014}, the emergence of roton modes~\cite{Chomaz_2018}, the Rosenweig instability~\cite{Kadau_2016, Schmitt_2016}, the abnormal stability of a Bose-Einstein condensate~\cite{Koch_2008}, or quantum magnetism~\cite{dePazPRL13,Lepoutre_2018, Lepoutre_2019,Patscheider_2020,GabardosPRL20}. Moreover, lanthanides like dysprosium and erbium have nonzero orbital angular momenta that opened studies on large orbital anisotropy of their short-range interactions~\cite{Li_2018, Tiesinga_2021}. The native spin structure and long lifetimes compensate for weaker dipolar interaction strength and make highly magnetic atoms a promising platform for quantum simulations. They have been already used to simulate the extended Bose-Hubbard model~\cite{Baier_2016}. While the many-body regime of such magnetic simulators has been extensively studied theoretically~\cite{dePazPRL13,Lepoutre_2018,Lepoutre_2019,Zhu_2019,Fersterer_2019,Patscheider_2020,GabardosPRL20}, the few-body picture is less known.

\begin{figure*}[t]
\begin{center}
\includegraphics[width=2\columnwidth]{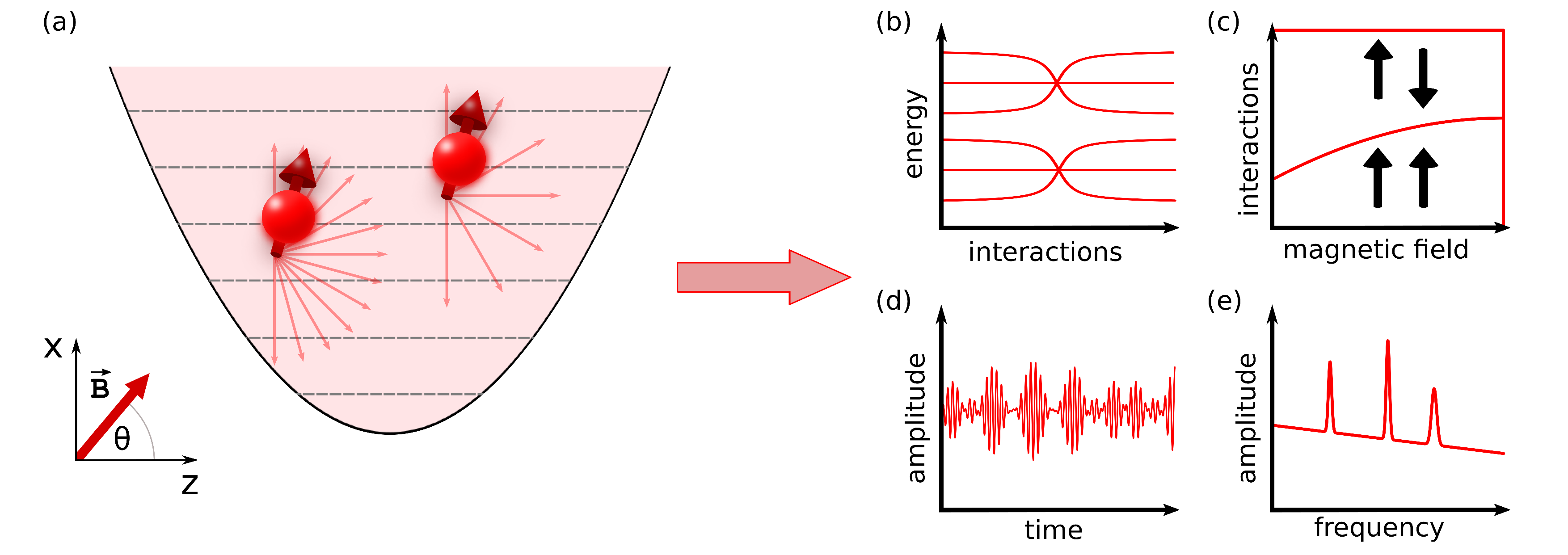}
\end{center}
\caption{Schematic representation of the investigated system and its features. (a) Two interacting ultracold highly magnetic atoms (e.g., chromium) in a 1D harmonic trap under the influence of the external magnetic field can be described by (b) an energy spectrum or (c) a magnetization diagram depending on the interaction and field strengths. (d) Time evolution of the system's observable and (e) its Fourier transform can reveal information on the coupling between states of the system and effects of symmetries.}
\label{fig:intro}
\end{figure*}

To address this limit, we study the building block of the quantum simulator of the Hubbard models based on ultracold highly magnetic atoms, that is, two particles interacting on site. In this paper, we describe two indistinguishable and two distinguishable highly magnetic atoms in a 1D harmonic trap interacting via the contact spin-spin interaction. We investigate the interplay between the spin-spin interaction, the external magnetic field, and the trapping potential. Energy spectra and eigenstates are calculated by means of the exact diagonalization. We show that, for indistinguishable atoms, the system has a high symmetry that can be reduced by an appropriately applied external magnetic field. The ground state can manifest antiferromagnetic or ferromagnetic configurations. Moreover, quantum statistics has an important impact on the dynamics of the system, where indistinguishable atoms have much simpler dynamics than distinguishable ones.

The plan of the paper is as follows. Section \ref{sec:model} describes the theoretical model and its experimental feasibility. Section \ref{sec:results} presents and discusses the static properties of the ground state and the dynamics of the system. Section \ref{sec:summary} summarizes the paper and explains further perspectives of research in this area. The computer code to recreate the presented results is available on GitLab~\cite{OurRepo}.

\section{Theoretical model and methods}
\label{sec:model}

We start by considering two interacting indistinguishable ultracold highly magnetic atoms with the same spin $s$ and mass $m$. We also study a more general case of two interacting distinguishable atoms with different spins $s_i$ and masses $m_i$. Their motion is restricted to 1D along the $z$ axis due to strong perpendicular trapping potential in the $x$ and $y$ directions. Additionally, atoms are confined in a 1D harmonic trap in the $z$-direction with the trapping frequency $\omega$. Their magnetic dipole-dipole interaction can be effectively approximated by the contact spin-spin interaction.

\subsection{Hamiltonian}

The Hamiltonian describing our system is
\begin{equation}\label{eq:tot_ham}
    \ham = \ham_{\text{trap}} + \ham_{\text{dd}} + \ham_{\text{Zeeman}},
\end{equation}
where $\ham_{\text{trap}}$ describes the motion of atoms in the 1D harmonic trap, $\ham_{\text{dd}}$ is the magnetic dipole-dipole interaction between atoms, and $\ham_{\text{Zeeman}}$ describes the interaction of atoms with an external magnetic field.

The Hamiltonian describing two structureless atoms in the 1D harmonic trap is
\begin{equation}
   \ham_{\text{trap}} = \sum_{i=1}^2 \qty(\frac{\hat{p}_i^2}{2m_i}  + \frac{1}{2} m_i \omega^2 z_i^2 ) ,
   \label{eq:trap}
\end{equation}
where $\omega$ is the trapping frequency, $z_i$ is the position, $\hat{p}_i$ is the momentum operator, and $m_i$ is a mass of the $i$th atom.

The Hamiltonian describing the magnetic dipole-dipole interaction in three dimensions (3D) is
\begin{equation}
    \ham_{\text{dd}}^{\text{3D}} =  \frac{\mu_0}{4\pi} \qty( \frac{\hat{\vb{d}}_1 \cdot \hat{\vb{d}}_2}{r^3}  - \frac{3 (\hat{\vb{d}}_1 \cdot \vb{r})(\hat{\vb{d}}_2 \cdot \vb{r})}{r^5} ),
    \label{eq:dd3D}
\end{equation}
where $\hat{\vb{d}}_i$ is the magnetic dipole moment operator of the $i$th atom, $\vb{r}$ is a vector connecting two atoms, and $\mu_0$ is the magnetic constant. In the 1D limit, the dipole-dipole interaction can be effectively described by the contact interaction~\cite{Gadway_2016, Deuretzbacher_2010, Sinha_2007}. In our research, we consider the dipole-dipole interaction of the form
\begin{equation}
    \ham_{\text{dd}}^{1D} =  \frac{2J}{\hbar^2} \delta(z) \hat{\vb{s}}_1 \cdot \hat{\vb{s}}_2,
    \label{eq:dd}
\end{equation}
where $z=\abs{z_1-z_2}$ is the interatomic distance, $\hat{\vb{s}}_i$ is the spin operator of the $i$th atom and $J$ is the spin-spin interaction strength. The full form of the dipole-dipole interaction from Eq.~\eqref{eq:dd3D} reduced to 1D has the additional term
\begin{equation}
    \ham_{\text{dd,}\delta}^{1D} = -\frac{6J}{\hbar^2} \delta(z) s_{1z} s_{2z}\,,
    \label{eq:additional_term}
\end{equation}
which is diagonal in the spin basis. Therefore, it effectively acts like the contact interaction of form $g \delta(z)$, where $g$ is the interaction strength. Unless it is stated otherwise, we omit this term to focus on the Heisenberg-like interaction.
We show how $J$ relates to the system constants and prove the contact form of the interatomic interaction in Appendix \ref{sec:dd_in_1D}.

The Hamiltonian describing the interaction of the atoms with an external magnetic field is
\begin{equation}
    \ham_{\text{Zeeman}} =   \mu_\text{B} \sum_{i=1}^2 g_i \hat{\vb{s}}_i \cdot \vb{B},
    \label{eq:zeeman}
\end{equation}
where $\vb{B}$ is an external magnetic field vector and $g_i$ is the magnetic $g$-factor. Further, the magnetic field will be described by its magnitude $B$ and the angle $\theta$, which is an angle between the $z$ axis of the trap and the magnetic field vector visible at Fig.~\ref{fig:intro}. We neglect the potential nonzero value of nuclear spins.

We use the harmonic oscillator units. Energies are then expressed in units of $\hbar \omega$, lengths in units of the harmonic oscillator characteristic length $a_{\mathrm{ho}} =
\sqrt{\hbar / (\mu \omega)}$ (where $\mu$ is a reduced mass of the atoms), magnetic field magnitude $B$ in units of $\mu_\text{B}/(\hbar \omega)$, and interaction strength $J$ in units of $\hbar \omega a_{\mathrm{ho}}$.

\subsection{Basis set and state symmetry}

The form of the Hamiltonian in Eq.~\eqref{eq:tot_ham} allows us to separate the center-of-mass and relative motions of atoms. The center-of-mass movement of two atoms in the harmonic trap is described by a well-known symmetric solution of the harmonic oscillator problem. We solve the remaining relative motion problem by decomposing the relative motion wave function $\ket{\Psi_k}$ for a $k$th state in the product basis constructed as
\begin{equation}
    \ket{\Psi_k} = \sum_{n,S,M_z} a_{n,S,M_z}^k \ket{n} \ket{S, M_z, s_1, s_2},
\end{equation}
where $\ket{n}$ are eigenstates of the 1D harmonic oscillator and $\ket{S, M_z, s_1, s_2}$ are eigenstates of the total spin angular momentum operator, $\hat{\vb{S}}=\hat{\vb{s}}_1+\hat{\vb{s}}_2$, where $M_z$ is its projection onto the $z$ axis. Moreover,
\begin{equation}
  \begin{aligned}
    \ket{S,M_z, s_1, s_2} &  \\
    = &\sum_{m_{s_1}, m_{s_2}} C^{s_1,m_{s_1}, s_2, m_{s_2}}_{S,M_z} \ket{s_1, m_{s_1}} \ket{s_2, m_{s_2}},
  \end{aligned}
\end{equation}
where $C^{s_1,m_{s_1}, s_2, m_{s_2}}_{S,M_z}$ are the Clebsch-Gordan coefficients.
Coefficients $a_{n,S,M_z}^k$ are calculated by means of the exact diagonalization. In some cases, the symmetry of the Hamiltonian leads to conservation of $S$ or $M_z$ that is used to reduce the size of the basis set and therefore of the diagonalized matrix. In some cases, when $M_z$ is not conserved, $M_x$ might be used as a good quantum number instead.

To ensure a proper symmetry of the total wave function, the relative motion wave function has to be symmetric (antisymmetric) with respect to particle exchange for bosons (fermions). 
The symmetry of $\ket{S, M_z, s_1, s_2}$ depends on the total spin of the system $S$. If the parity of the total spin quantum number $S$ is the same as the parity of $s_1+s_2$, then $\ket{S, M_z,s_1, s_2}$ is symmetric; otherwise it is antisymmetric. The harmonic part  $\ket{n}$ is symmetric (antisymmetric) for even (odd) $n$.
Therefore, for both fermions and bosons, the odd $S$ is paired with odd $n$ and the even $S$ is paired with even $n$.

The matrix elements of the Hamiltonian of Eq.~\eqref{eq:tot_ham} in the discussed basis are shown in Appendix \ref{sec:matrix_el}.

\subsection{Dynamics}

We consider a unitary time evolution of the system that is prepared in the initial product state 
\begin{equation}
    \ket{\psi_0} = \ket{n} \ket{s_1,m_{s_1}} \ket{s_2,m_{s_2}}.
\end{equation}
The time evolution of the system is then 
\begin{equation}
  \begin{aligned}
   \ket{\psi(t)} &= \exp\left(-i\ham t / \hbar\right) \ket{\psi_0} \\
      &= \sum_p \braket{\Psi_p}{\psi_0} \exp \Bigl( -i E_p t / \hbar \Bigr)\ket{\Psi_p},
  \end{aligned}
\end{equation}
where $\ham$ is the Hamiltonian describing the system after the quench and $\ket{\Psi_p}$ are its eigenstates with the corresponding energies $E_p$.
The time evolution of the observable $\hat{O}$ is then 
\begin{equation}
  \begin{aligned}
   \ev{\hat{O}}{\psi(t)} = \sum_{p,q} \braket{\Psi_p}{\psi_0} \braket{\Psi_q}{\psi_0}  \\ \times \exp\Bigl(-i(E_p-E_q)t / \hbar \Bigr) \mel{\Psi_p}{\hat{O}}{\Psi_q}.
  \end{aligned}
\label{eq:time_ev}
\end{equation}
The dynamics is calculated until time $t = 10000 \times \frac{2\pi}{\omega}$ with a time step of $0.01 \times \frac{2\pi}{\omega}$. We perform the discrete Fourier transforms of the observables' evolution using the \textsc{scipy} package~\cite{scipy}.

\subsection{Convergence with the basis set size}

Calculations are done in the basis set composed of $N_n=30$ harmonic oscillator eigenfunctions. However, for two indistinguishable atoms due to fermionic or bosonic symmetry of states, we take every second harmonic state, which means that we take $n_\text{max}=60$. For two distinguishable atoms, $n_\text{max}=30$. In the investigated system, $N_n=30$ functions guarantee a sufficiently good convergence for the lowest-energy states. We note, however, that even not fully converged results may be enough to describe and understand the physical properties of the few-body systems~\cite{Sowinski_2013, Sowinski_2019, Grining_2015, GriningPRA2015}. For time evolution, the finite basis set is causing problems that we have already seen in calculations for molecules~\cite{Dawid_2020}. Namely, the frequencies of the oscillations depend on the energy differences between all states, including the highest-energy states that are divergent.  However, from the lower states we learn the asymptotic behavior of all the harmonic states, and this allows us to use the condition that, if the energy of the state is higher than $\hbar \omega(n_\text{max}+\frac{1}{2})$, it is omitted in the time evolution calculations.

\subsection{Experimental feasibility}

In the ultracold experiment, to achieve an effective 1D optical trap, we can use a 3D cigar-shaped harmonic trap,
\begin{equation}
    V_{\text{trap}} = \frac{\mu}{2} \qty[\omega z^2 + \omega_\perp (x^2+y^2)],
\end{equation}
where $\omega \ll	\omega_\perp$.

In our model, the spin-spin interaction strength is expressed as
\begin{equation}
    J= \frac{\mu_0 d^2 \omega_\perp \mu}{8 \pi  \hbar S(S+1)} ,
\end{equation}
the derivation of which can be found in Appendix \ref{sec:dd_in_1D}. We consider the regime where the spin-spin interaction strength is a magnitude lower than the harmonic oscillator energy. This condition is equivalent to $J$ being comparable to $0.1$ in the harmonic oscillator units
\begin{equation}
    \frac{J}{\hbar \omega a_{\mathrm{ho}}}  =  \frac{\mu_0 d^2 \omega_\perp \mu^{3/2}}{8 \pi  \hbar^{5/2} S(S+1) \omega^{1/2}}
    \propto 0.1.
\end{equation}

In this work, we investigate the spin-3 atoms, an equivalent of bosonic chromium atoms with  experimental values of $d=6 \, \mu_\text{B}$ and $\mu=51.94 \, \si{u}$~\cite{Gadway_2016}. Note that chromium has a zero orbital electronic angular momentum, and its bosonic isotopes do not have nuclear spins.

Therefore, to have $J$ of the order of $0.1 \hbar \omega a_{\mathrm{ho}}$, for the perpendicular trap frequency $\omega_\perp=1 \, \si{MHz}$, the trapping frequency $\omega$ should be around $13 \, \si{kHz}$, and for $\omega_\perp=10 \, \si{MHz}$, the frequency should be around $\omega=1.3 \, \si{MHz}$. Those traps would have anisotropy parameters $1/\lambda$ equal to 8.8 and 2.8, respectively, where $ \lambda =\sqrt{\frac{\omega}{\omega_\perp}} $. Typical trapping frequencies vary from kHz to MHz in range~\cite{BoldaPRA2002} and the highest anisotropy of the trap achieved in experiment, that we know of, is $1/\lambda \approx 50$~\cite{Kinoshita_2004}. Analogously, the magnitude of the magnetic field strength $B$ would vary from 1.5 to 150 $\si{mG}$. Using lower trap frequencies would allow one to obtain higher anisotropy of the trap. However, simultaneously, it would make the system highly sensitive to a magnetic field.

\section{Results}
\label{sec:results}

We start by analyzing the symmetries present in the system and the energy spectra to investigate the effect of the spin-spin interaction and the external magnetic field on the system. Then, we investigate how the interplay of the interatomic interactions and the external field affects the magnetization of the ground state. Finally, we probe the dynamics of the system, show the importance of the symmetries of the system, and study the observables that may be accessible in the experiment.

\begin{figure}[tb!]
\begin{center}
\includegraphics[trim={0cm 0cm 0cm 2cm},clip,width=0.48\textwidth]{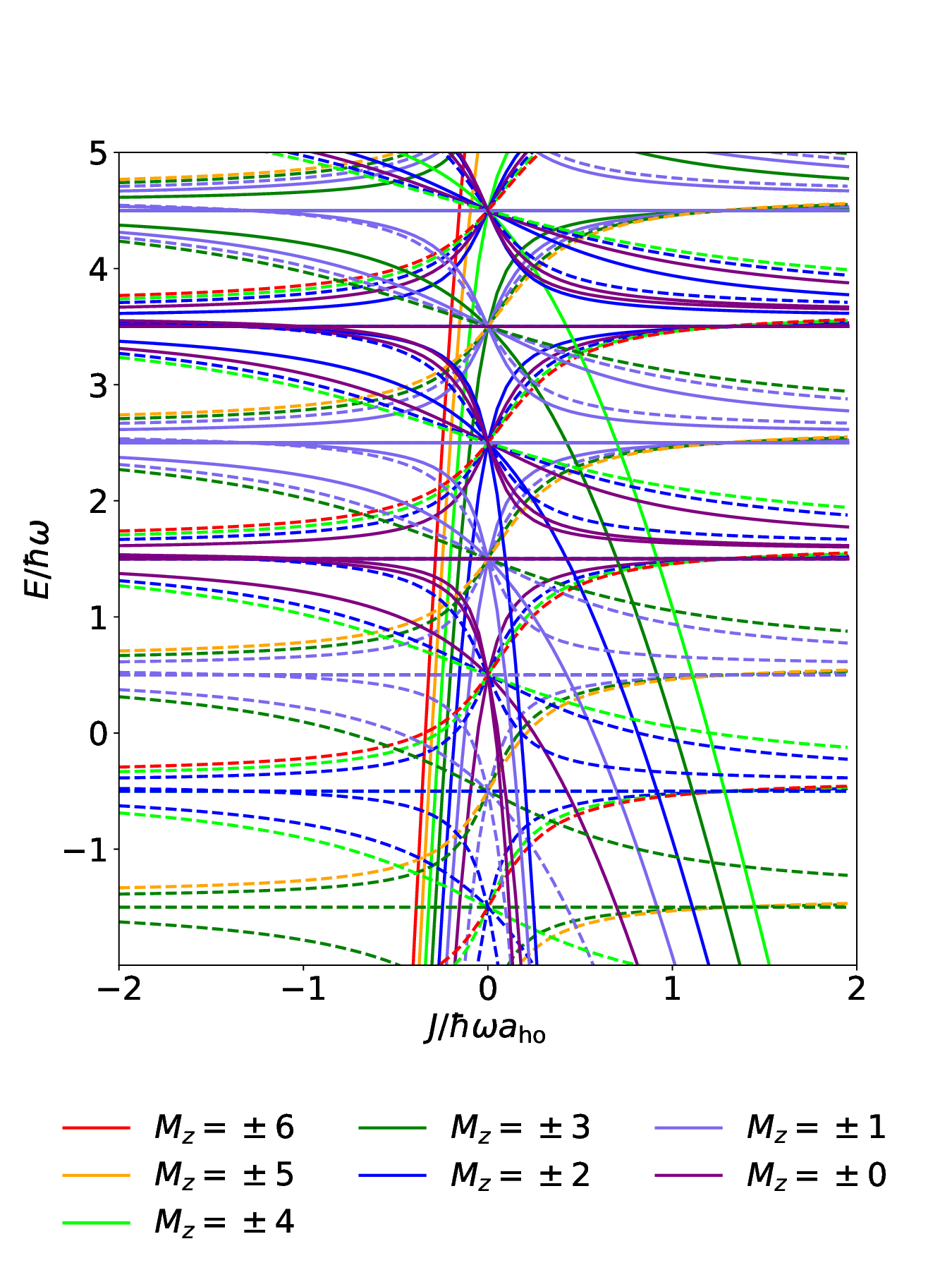}
\end{center}
\caption{Energy spectrum of two indistinguishable bosonic spin-3 atoms in a 1D harmonic trap as a function of the spin-spin interaction $J$ for $\mu_\text{B} B=0.1 \, \hbar \omega$ and $\theta=0$ (field parallel to the trap axis). Colors (grayscale) denote the total spin projection $M_z$ of the corresponding states. Solid lines shows the positive values of $M_z$ and dashed lines show the negative values.}
\label{fig:wholespectrum}
\end{figure}

\begin{figure*}[tb!]
\centering
  \centering
  \includegraphics[trim={0cm 0cm 0cm 0cm},clip,width=0.98\linewidth]{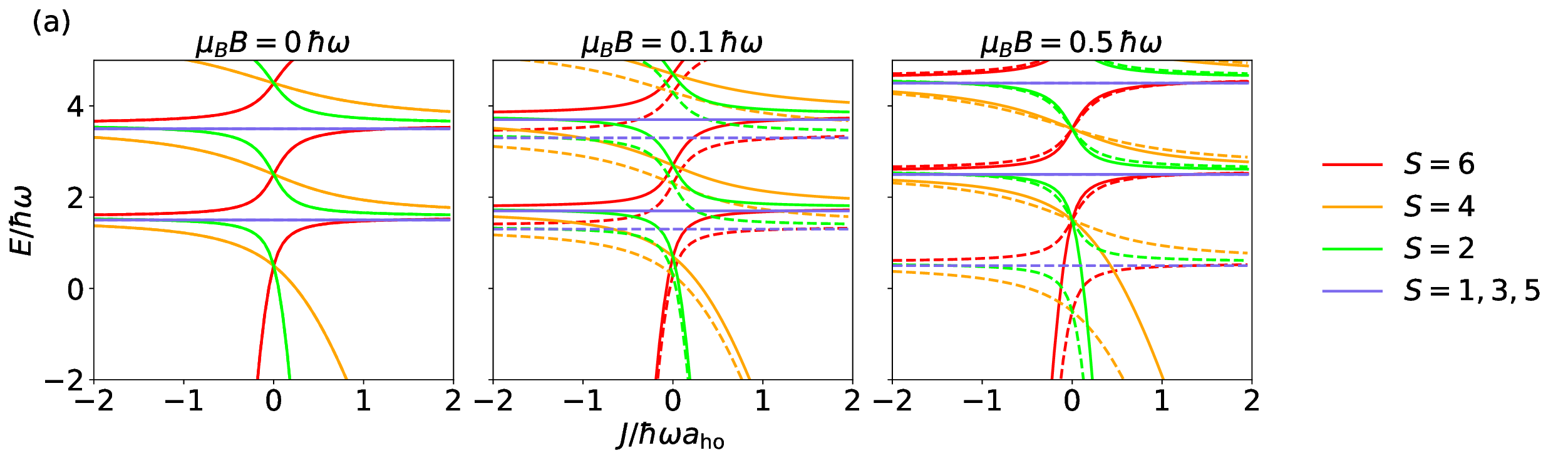}
\\
  \includegraphics[trim={0cm 0cm 0cm 0cm},clip,width=0.98\linewidth]{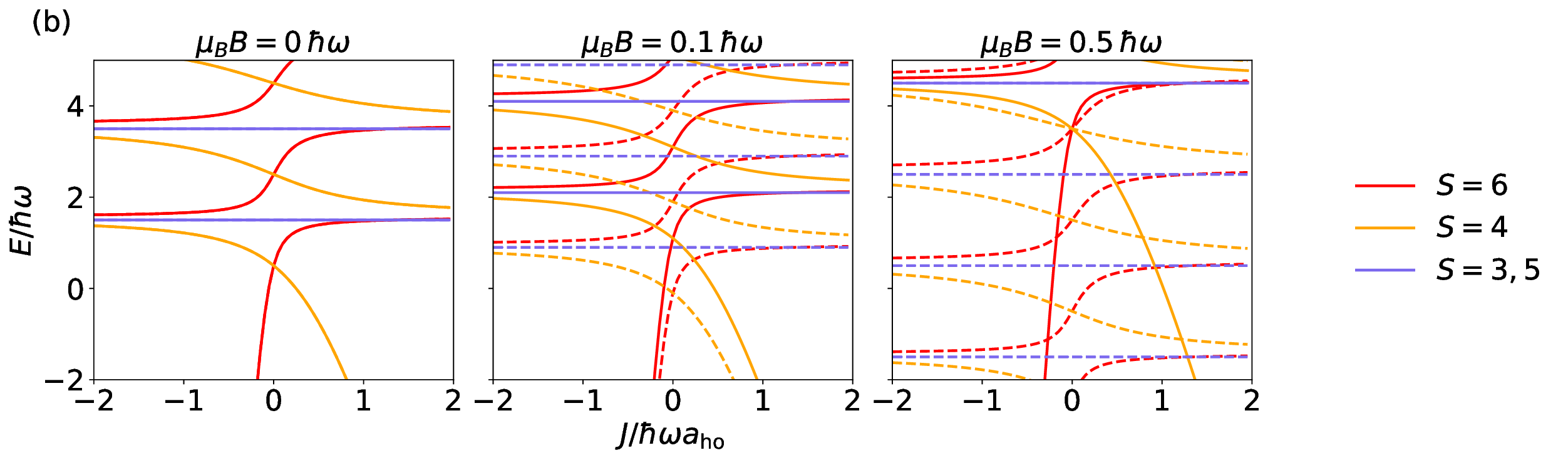}
\caption{Energy spectra of subspaces (a) $M_z=1$ and $M_z=-1$ and (b) $M_z=3$ and $M_z=-3$ of two indistinguishable bosonic spin-3 atoms in a 1D harmonic trap as a function of the spin-spin interaction $J$ for $\theta=0$ and three values of the magnetic field $\mu_\text{B} B=0,0.1,$ and $0.5  \, \hbar \omega$. Solid lines shows the positive values of $M_z$ and dashed lines show the negative values.}
\label{fig:spectrum_seperated_M}
\end{figure*}

\begin{figure*}[tb!]
\centering
  \includegraphics[trim={0cm 0cm 0cm 0cm},clip,width=0.98\linewidth]{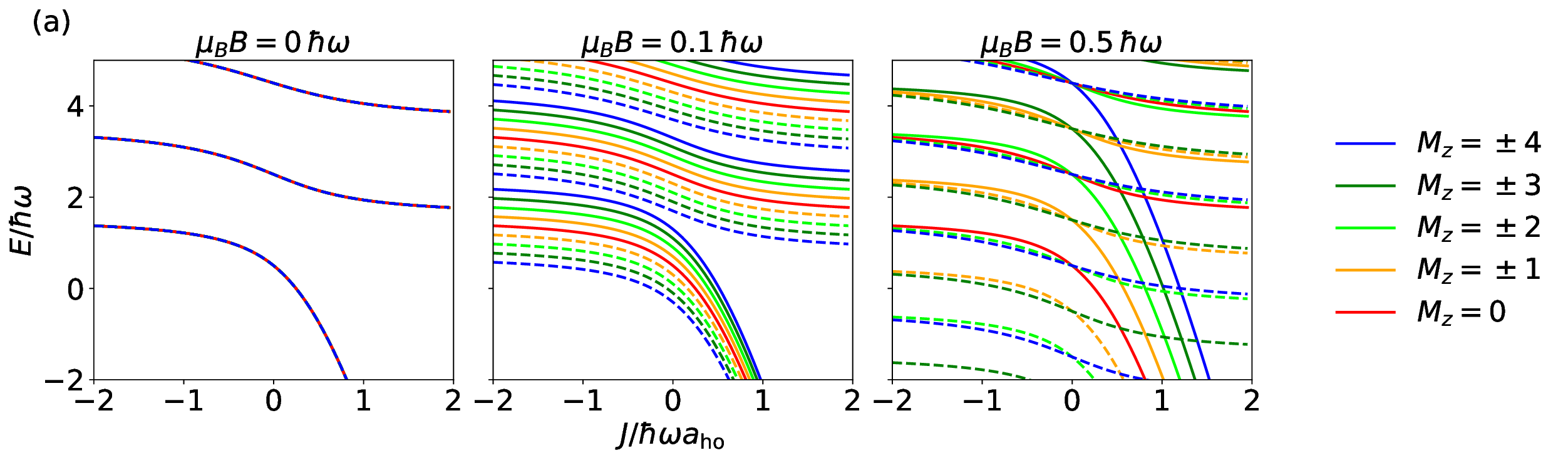}
\\
\includegraphics[trim={0cm 0cm 0cm 0cm},clip,width=0.98\linewidth]{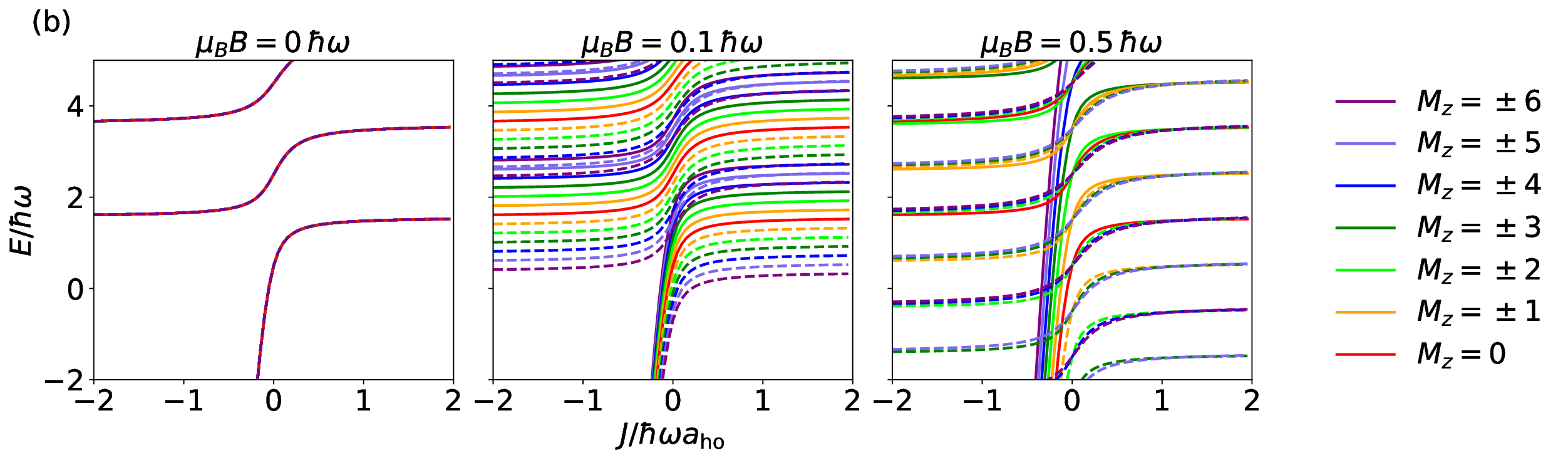}
\caption{Energy spectra of subspaces (a) $S=4$ and (b) $S=6$ of two indistinguishable bosonic spin-3 atoms in a 1D harmonic trap as a function of the spin-spin interaction $J$ for $\theta=0$ and three values of the magnetic field $\mu_\text{B} B=0,0.1,$ and $0.5  \, \hbar \omega$. Solid lines show the positive values of $M_z$ and dashed lines show the negative values.}
\label{fig:spectrum_seperated_S}
\end{figure*}

\subsection{Symmetries}
The symmetries of the system lead to the conservation of certain quantities and related quantum numbers. As introduced in Sec. \ref{sec:model}, we examine the total spin $S$ and its projection onto the quantization axis $M_z$. We see that, for a magnetic field parallel to the trap axis [$\theta=0$ in Fig.~\ref{fig:intro}(a)], $M_z$ is always conserved, independently of the distinguishability of atoms. On the other hand, $S$ is conserved as long as both atoms have the same magnetic susceptibility, which in our model is given by the $g$-factor. In general, the spin-spin interaction conserves both $S$ and $M_z$, but it mixes different harmonic states. If the magnetic field is not parallel to the trap axis ($\theta \neq 0$), the states with different $M_z$ begin to mix. The symmetries and the resulting conservation of the quantum numbers turn out to determine the properties of the system, especially when it comes to its dynamics.

\subsection{Energy spectrum}

For the static properties of the system, we consider two indistinguishable bosonic spin-3 atoms, equivalent to chromium atoms with a $g$-factor of $g=2$. In Fig.~\ref{fig:wholespectrum}, we show the spectrum with all possible states as a function of the spin-spin interaction strength $J$ in the magnetic field $\mu_\text{B} B=0.1 \, \hbar \omega$ and $\theta=0$. The spectrum is rich; however, we do not observe any anticrossings between states with different spin quantum numbers as the considered interaction does not couple them. The spectrum can be further analyzed after the separation into subspaces with specific quantum numbers. The examples of the separated spectra are shown in Figs.~\ref{fig:spectrum_seperated_M} and \ref{fig:spectrum_seperated_S}. In all panels, we plot the energy spectra as a function of $J$ for three values of the magnetic field $\mu_\text{B} B=0$, $0.1$, and $0.5 \, \hbar \omega$. In Fig.~\ref{fig:spectrum_seperated_M}, we present states with selected total spin projections to compare how states with different total spins $S$ behave due to the spin-spin interaction. Without a magnetic field (left column in each panel), we see that the energy levels are being repelled by the energy levels of adjacent harmonic states. Moreover, we observe that the states with odd $S$ are insensitive to the spin-spin interaction. This is due to the bosonic symmetry imposed on the system. States with odd $S$ have antisymmetric harmonic components and therefore the contact interaction does not affect those states. In Fig.~\ref{fig:spectrum_seperated_S}, we present the spectra of states with specific total spins $S$. Without a magnetic field, the states are strongly degenerate because the energy does not depend on the projection $M_z$. Therefore, we observe only one energy level for each harmonic state. After switching on the magnetic field, the degeneration is lifted and we observe the energy level splitting due to the Zeeman effect. 

The energy spectrum in the perpendicular magnetic field ($\theta=\pi$) behaves almost identically, with the difference that $M_x$ becomes a good quantum number and can be used instead of $M_z$ to segregate the states.

\subsection{Ground-state magnetization}

\begin{figure}[tb!]
\centering
\includegraphics[trim={0cm 0cm 0cm 0cm},clip,width=0.48\textwidth]{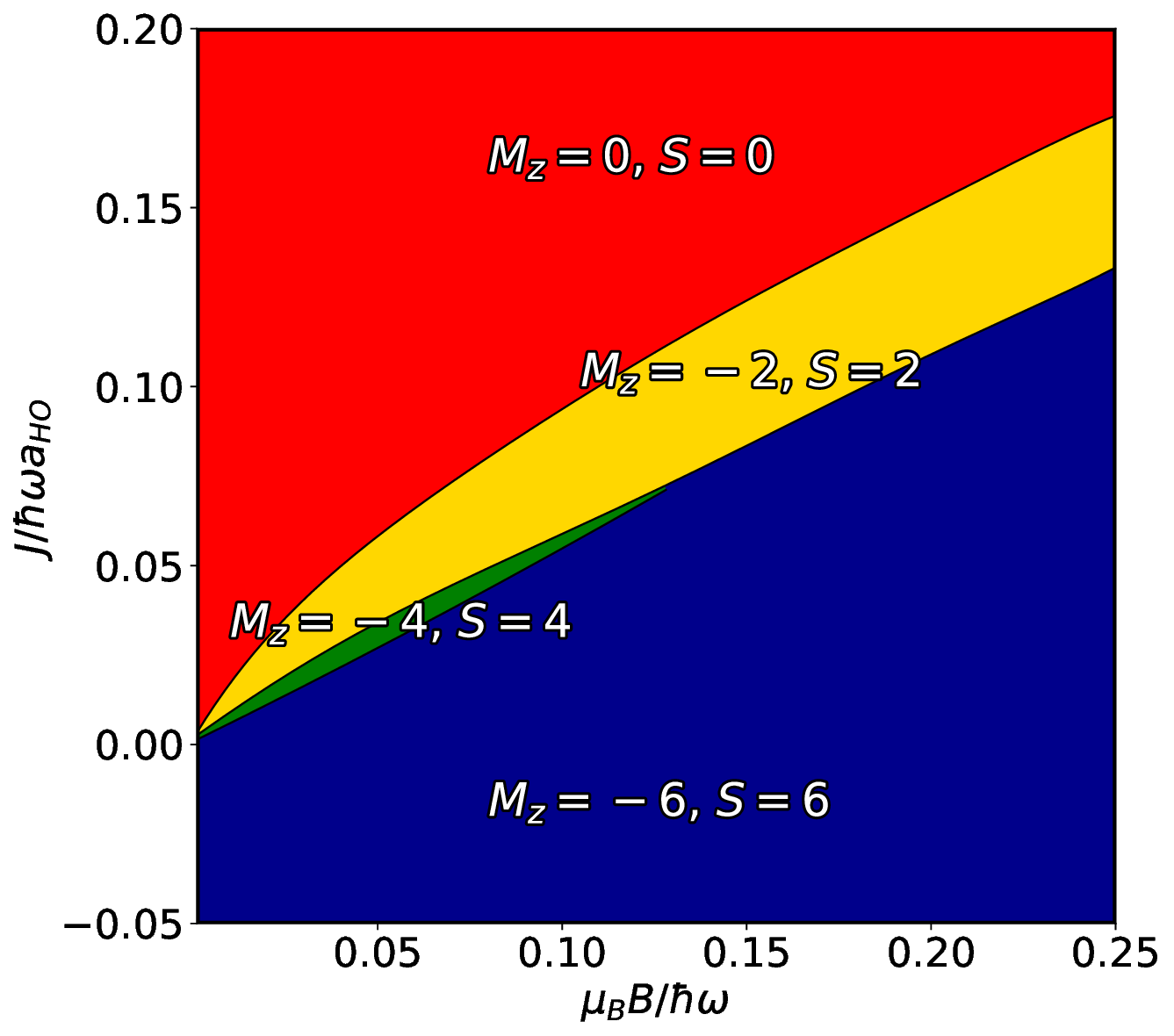}
\caption{Ground-state magnetization of two indistinguishable bosonic spin-3 atoms in a 1D harmonic trap as a function of the spin-spin interaction $J$ and the external magnetic field $B$ parallel to the trap axis ($\theta=0$).}
\label{fig:phase}
\end{figure}

Figure~\ref{fig:phase} shows the ground-state magnetization, understood as the value of the projection of the total spin $M_z$, which we use as a probe into the magnetic properties of the two-body system. Additionally, we mark the total spin $S$ of the ground state. We look at the interplay of the spin-spin interaction and the effect of the magnetic field parallel to the trap axis ($\theta=0$). For positive and strong $J$, the lowest-energy state is $M_z=0$ and $S=0$, which might be considered the equivalent of the antiferromagnetic configuration. For negative $J$ at any magnetic field as well as for low, positive $J$ and strong magnetic field, the ground state is equivalent to the ferromagnetic configuration with the highest possible spin  $S=6$ and an absolute value of its projection $M_z=-6$. Between those two configurations, there are transition areas. We observe only the even values of $S$ due to the bosonic symmetry, the states with odd $S$ are not affected by the spin-spin interaction, which makes them unlikely to be the ground state. However, we cannot be sure about the existence of the $M_z=-4$ area in the complete basis limit. The size of this area shows a strong dependence on the basis set size and it gets smaller when increasing the basis set size.

\begin{figure*}[tb!]
\centering
  \includegraphics[trim={0cm 0cm 0cm 0cm},clip,width=.48\textwidth]{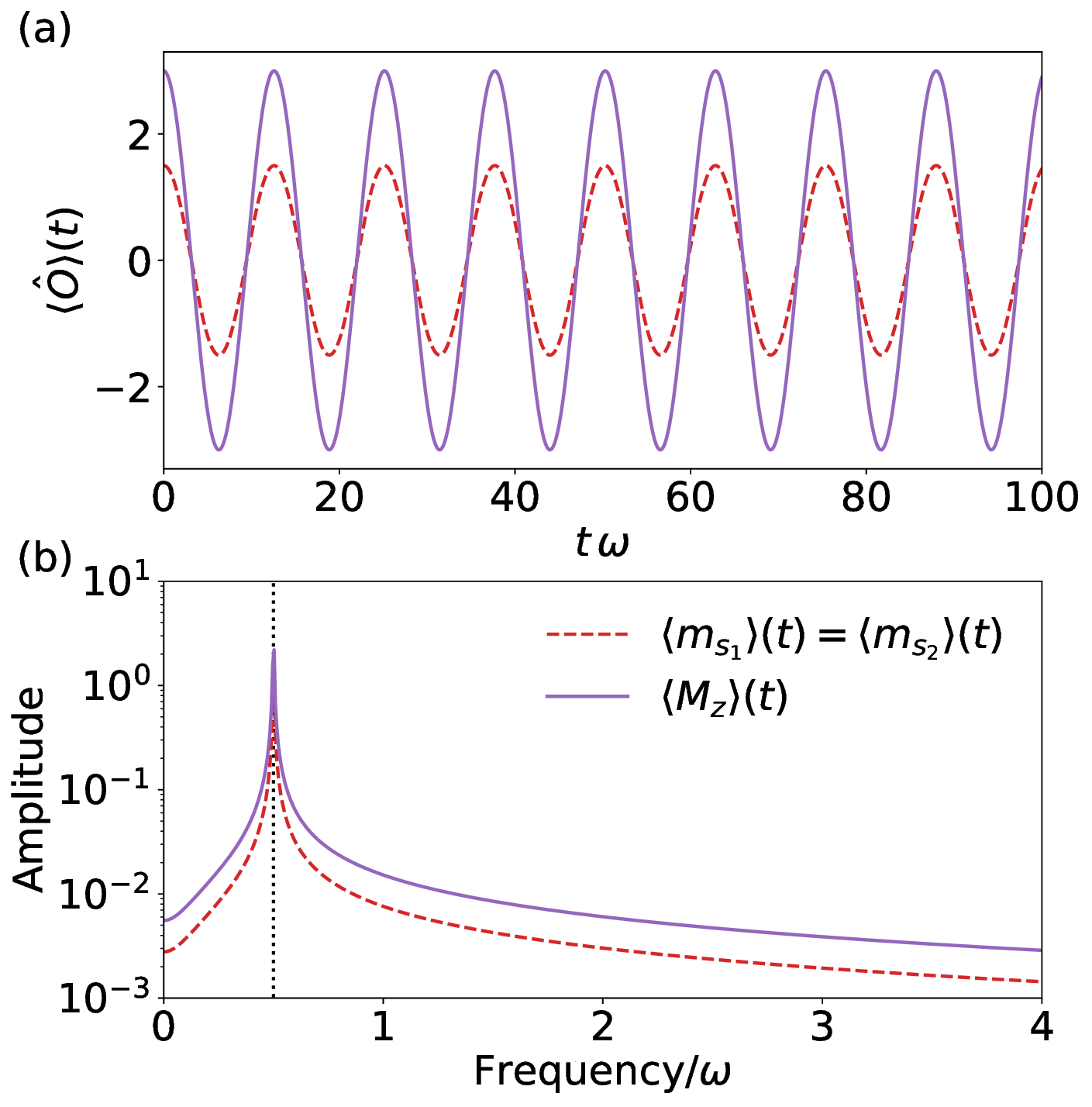}
\includegraphics[trim={0cm 0cm 0cm 0cm},clip,width=.48\textwidth]{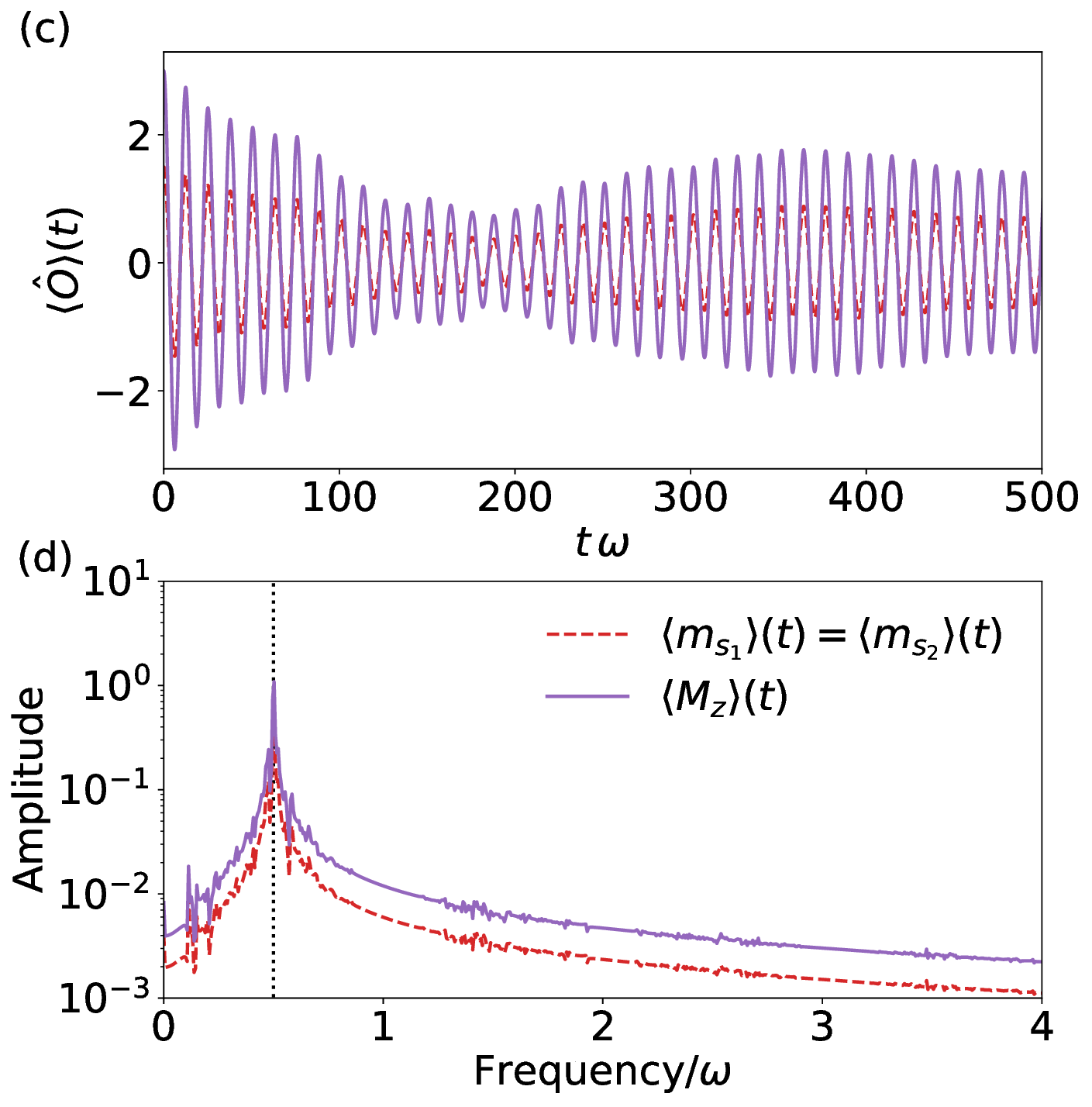}

\caption{The time evolution of the expectation value of the projection of the total $\expval{M_z}$ and individual $\expval{m_{s_i}}$ spins of the product state $\ket{s_1,m_{s_1}}\ket{s_2,m_{s_2}}=\ket{3,1}\ket{3,2}$ of two indistinguishable bosonic spin-3 atoms in a 1D harmonic trap for the spin-spin interaction strength $J=1 \, \hbar \omega a_\text{ho}$, the magnetic field angle $\theta=\pi$, and $\mu_\text{B} B=0.5 \, \hbar \omega$ (a) without and (c) with effective contact interaction $g\delta(z)$.  The black dotted line shows the frequency $g \mu_\text{B} B$. Panels (b) and (d) show the Fourier transform of the signals without and with effective contact interaction, respectively.}
\label{fig:time_ev}
\end{figure*}

\begin{figure}[tb!]
\centering
\includegraphics[width=0.5\textwidth]{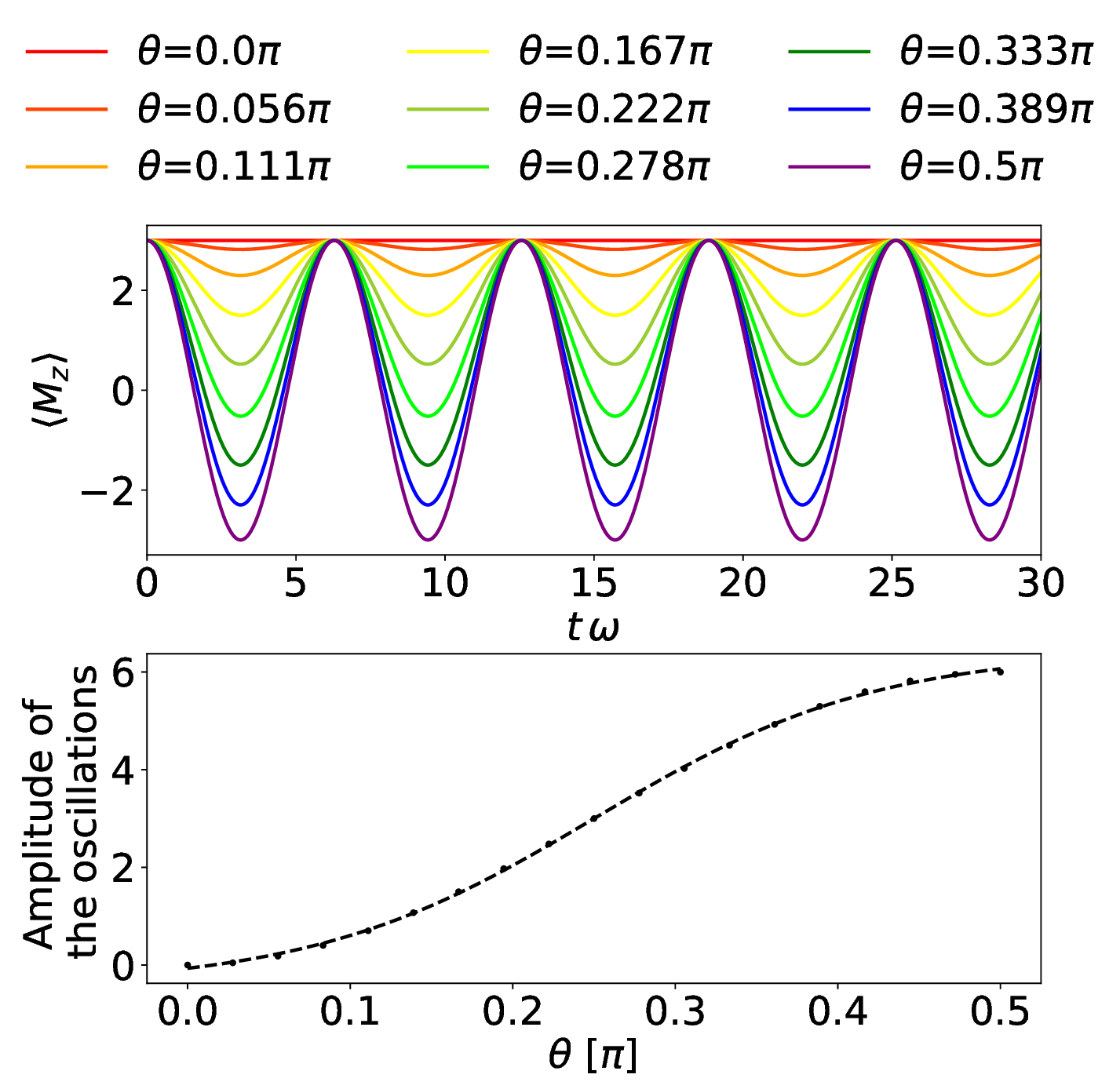}
\caption{The dependence of the amplitude of the oscillation of the total spin projection $\expval{M_z}$ of two spin-3 atoms in a 1D harmonic trap on the angle $\theta$ of the magnetic field vector with constant magnitude. The dashed black line in the lower panel shows the fitted hyperbolic tangent function $a\tanh(b\theta+c)+d$. The fitted parameters are $a=3.43$, $b=1.83$, $c=-1.44$, and $d=3.00$.}
\label{fig:amplitude}
\end{figure}

\begin{figure}[tb!]
\begin{center}
\includegraphics[width=0.48\textwidth]{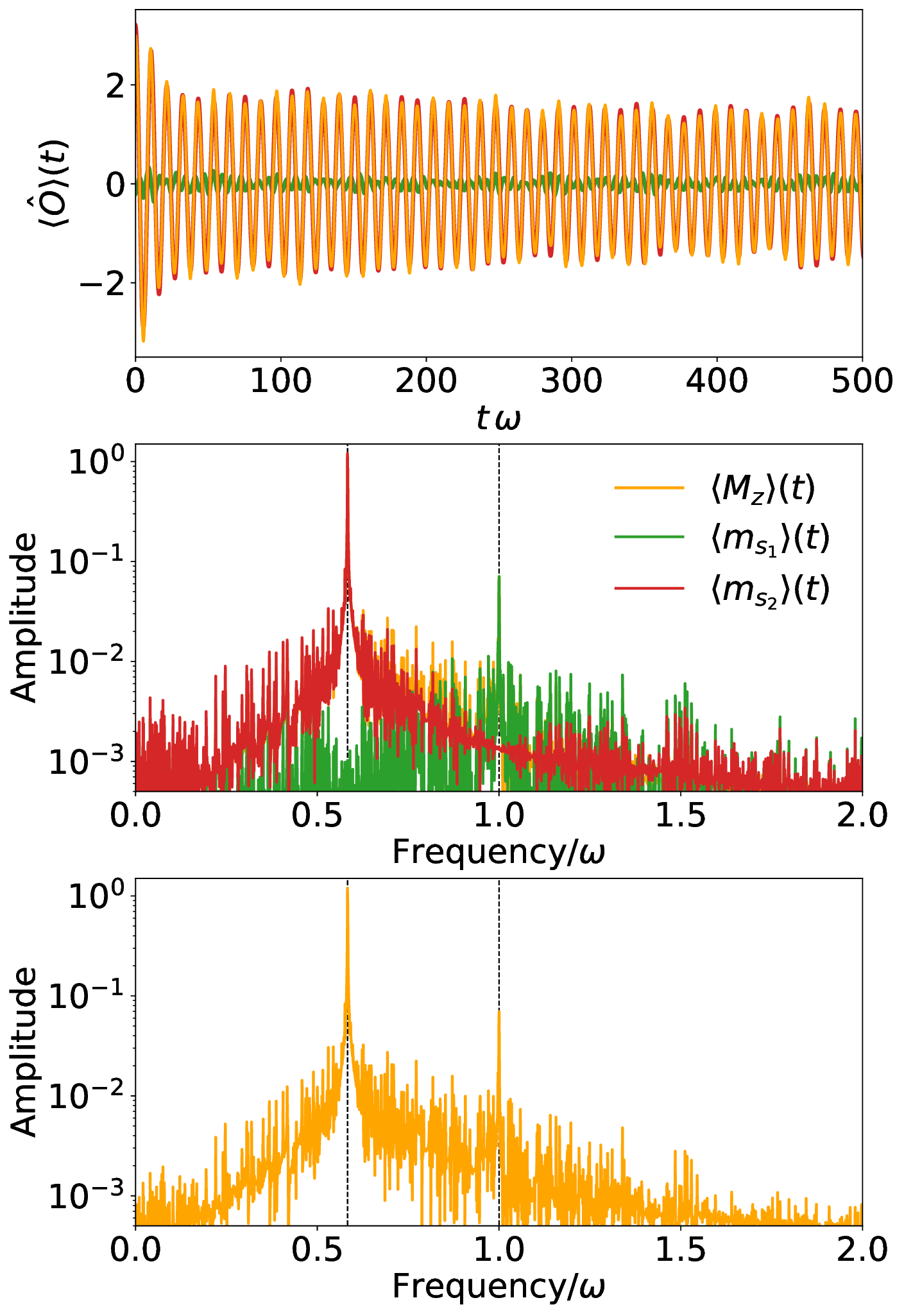}
\end{center}
\caption{The time evolution of the expectation value of the projections of the total $\expval{M_z}$ and individual $ \expval{m_{s_i}}$ spins of the system initialized as a product state $\ket{s_1,m_{s_1}}\ket{s_2,m_{s_2}}=\ket{3,1}\ket{6,2}$ of spin-3 and spin-6 distinguishable bosonic atoms in a 1D harmonic trap for the spin-spin interaction strength $J=1\, \hbar \omega a_\text{ho}$ and the magnetic field $\mu_\text{B} B=0.5 \, \hbar \omega$ and $\theta=\pi$. The black dotted line shows the frequencies $g_i \mu_\text{B} B$. The central and lower panels show the Fourier transform of the signal.}
\label{fig:diff_atoms}
\end{figure}

For the perpendicular magnetic field ($\theta=\pi$), the states with the same $\abs{M_z}$ but an opposite sign have the same contribution to the ground state, effectively resulting in $M_z$ being always equal to 0. In this case, $M_x$ becomes the good quantum number and the plot of $M_x$ is identical to the plot of $M_z$ in the parallel magnetic field.

Thus, we can control the system magnetization using an external magnetic field as well as the frequency of the trapping potential that can be used to effectively change the strength of the spin-spin interaction.

\subsection{Time evolution}

In this section, we analyze the time evolution of the system initialized as a product state $\ket{s_1,m_{s_1}}\ket{s_2,m_{s_2}}$ in the constant magnetic field with the fixed strength of the spin-spin interaction. This initial state corresponds to the initialization of the system, which is much faster than typical dynamics timescales. We choose to examine the expectation values of projection of the total spin $\ev{M_z}$ and the individual atom spins $\ev{m_{s_1}}, \, \ev{m_{s_2}}$. 

In the parallel magnetic field ($\theta=0$), both $S$ and $M_z$ are conserved and the mixing of harmonic states is not enough to induce any nontrivial dynamics of the magnetization. However, in the perpendicular magnetic field, $M_z$ states are mixed and we can observe some dynamics.

The dynamics of two chromium atoms appears to be surprisingly simple, and both investigated observables oscillate with a single frequency. The signal and its Fourier transform are shown in Figs.~\ref{fig:time_ev}(a) and ~\ref{fig:time_ev}(b), respectively. The localization of the single peak in the Fourier transform points out that the frequency of oscillation has exactly the value of $g\mu_\text{B} B$, which is related to the energy difference between two neighboring states in the energy level ladder induced by the Zeeman effect. This frequency is marked by the black dotted line in the plots. Those states and their energy differences are best visible in the central subpanels in both Figs.~\ref{fig:spectrum_seperated_S}(a) and~\ref{fig:spectrum_seperated_S}(b). The coupling with further neighboring states is canceled due to the different parity of states.

The values of $\expval{m_{s_1}}$ and $\expval{m_{s_2}}$ are identical at all times due to the indistinguishability of atoms. Moreover, the expectation value of the projection of an individual spin $\expval{m_{s_i}}$ is always equal to half of the value of the expectation value of the projection of the total spin which is required to preserve the indistinguishability. We additionally prove this in Appendix~\ref{sec:m1_proof}.

Moreover, we notice the dependence of the amplitude of the oscillations on the orientation of the magnetic field vector, given by the angle $\theta$, and we plot it in Fig.~\ref{fig:amplitude}. In extreme cases of the parallel and perpendicular magnetic fields, we observe zero amplitude, i.e., no oscillation at all, and maximal amplitude corresponding to a complete flip of the total spin from $\expval{M_z}$ to $-\expval{M_z}$, respectively. In the intermediate cases, we observe that with an increasing angle $\theta$ the amplitude of the oscillation raises to the maximal value. The dependence can be well described by the hyperbolic tangent function in the form $a\tanh(b\theta+c)+d$ with the fitted parameters $a=3.43$, $b=1.83$, $c=-1.44$, and $d=3.00$.

Furthermore, it appears that the omitted so far effective contact interaction $g\delta(z)$ [$\ham_{\text{dd,}\delta}^{1D}$ from Eq. \eqref{eq:additional_term}] can result in a more complex dynamics of the system. Although the term is not breaking any spin symmetry, it causes stronger mixing of the harmonic states and as a result more couplings between the states are present in the system. In Fig.~\ref{fig:time_ev}, we show the signal in panel (c) and its Fourier transform in panel (d) for the time evolution with this additional term.

As mentioned previously, as long as both atoms have the same $g$-factor, the total spin $S$ is conserved. In the more complicated case, where we consider two different distinguishable atoms with different $g$-factors, the time evolution gets more complex as can be seen in Fig.~\ref{fig:diff_atoms}. We consider the equivalent of chromium and erbium atoms with spins $s_1=3$ and $s_2=6$, respectively. The erbium atom has the total angular momentum $j=6$; however, we omit the orbital short-range anisotropy to study long-range magnetic interactions, effectively treating it as the spin-6 atom. The known formula for the Landé $g$-factor~\cite{hamilton_2010} gives the value $g_2=1.167$ (where for chromium $g_1=2$). In the Fourier transform of the signal (see the lower panels in Fig.~\ref{fig:diff_atoms}), we observe a lot of frequencies with a relatively small amplitude, which were not present in the case of two indistinguishable spin-3 atoms. However, there are two peaks with large amplitudes, which are located at frequencies $f_1=g_1 \mu_\text{B} B=1 \, \omega$ and $f_2=g_2 \mu_\text{B} B=0.583\, \omega$, that correspond to values $g_i \mu_\text{B} B$. Moreover, looking at the Fourier spectra in the central panel in Fig.~\ref{fig:diff_atoms}, we see that the first peak comes from the oscillation of $\expval{m_{s_1}}$ and the second one from $\expval{m_{s_2}}$. It shows that the values of these frequencies have the same origin as in the two spin-3 atoms and the coupling of the nearest Zeeman states is dominating the time evolution. Again, those frequencies corresponding to Zeeman shifts are marked by the black dotted lines.

However, the peaks with lower amplitude show that additional weaker couplings exist in this system but still we do not observe the ladderlike Fourier spectrum typical for coupling between harmonic states. It can be explained by the strong mixing between states causing the energy differences to be inconsistent with the harmonic trap energy spacing. 

Although we find out that the energy spectra, in this case, are relatively complex, we see no sign of the chaotic behavior in the unfolded spectrum~\cite{haake_qc}.


\section{Summary and conclusions}
\label{sec:summary}

Within this work, we have studied the properties of two interacting ultracold highly magnetic atoms in a 1D harmonic trap. We have investigated the interplay of the contact spin-spin interaction and the external magnetic field. We have shown how different symmetries affect the system and how to use the magnetic field to break them. We have shown that an external magnetic field and a trapping potential frequency can be used to control the magnetization of the system. We have shown the time evolution of the observables that could be measured in the experiment. 

Our findings can be summarized as follows.

\begin{itemize}
\item For two spin-3 indistinguishable atoms, the energy spectrum shows the repulsion between energy levels of different harmonic states. We observe states that are insensitive to the spin-spin interaction due to the bosonic symmetry of the system.

\item The magnetization of the system can be controlled by the interplay of the magnetic field and the harmonic trap frequency. The magnetization mimics the many-body ferromagnetic and antiferromagnetic configurations.

\item The time evolution of the expectation value of the total spin projection for two indistinguishable spin-3 atoms is unexpectedly simple. The value of the observable oscillates with a single frequency connected with the coupling of the nearest Zeeman states. However, the extra contact interaction causes additional mixing in the system. The amplitude of the oscillation depends strongly on the orientation of the magnetic field vector.

\item The distinguishability plays vital role in the time evolution of the system. For distinguishable atoms, more states are coupled than for the indistinguishable ones. However, the dominating frequencies of the oscillations have the same origin as for the indistinguishable atoms. We do not observe a typical ladderlike Fourier spectrum for coupling between harmonic states due to strong mixing between states. 

\item For the indistinguishable and distinguishable atoms described by the considered model, we did not observe any signature of quantum chaos.

\end{itemize}

The presented results concerning the interplay of the spin-spin interaction and the external magnetic field and resulting properties of the trapped two-body atomic system give insight into the controlled magnetization of the system as well as may guide the potential experimental measurements. Additionally, our results stress the importance of the symmetries of the system.

The results show the behavior of the building block of the extended Bose-Hubbard model simulator built from highly magnetic atoms. Its analysis had been a missing element of the theoretical description of such a simulator. These results may be extended by studying two fermionic atoms, which initially have shown similar results to bosons. However, further investigation could show more details and potential differences in their properties. Another extension that might be considered is a system of several traps to analyze the long-range character of the dipolar interaction which we now omit. Another direction would be to study the full form of the dipole-dipole interaction, in both 1D or 3D harmonic traps, including a more detailed analysis of the contact interactions. We do not expect a significant change in the properties of the two-atom system extended to quasi-1D or including the long-range dipolar interaction. However, in 3D, the anisotropic nature of the dipole-dipole interaction and its coupling to the relative orbital angular momentum should play a more critical role. Similarly, we suspect that the long-range nature of the interatomic interaction would be more important in systems of many trapped atoms in both 1D and 3D. Finally, we could include the orbital short-range interatomic interaction anisotropy present between lanthanides such as erbium and dysprosium.

\begin{acknowledgments}
We thank Piotr Gniewek for useful remarks. We acknowledge the financial support from the Foundation for Polish Science within the First Team programme co-financed by the EU Regional Development Fund and the PL-Grid Infrastructure. A.D. acknowledges ﬁnancial support from the Foundation for Polish Science (FNP) and the National Science Centre, Poland, within Etiuda Grant No. 2020/36/T/ST2/00588, as well as from Agencia Estatal de Investigaci\'on (the R\&D Project No. CEX2019-000910-S, funded by MCIN/ AEI/10.13039/501100011033, Plan National FIDEUA PID2019-106901GB-I00, FPI), Fundaci\'o Privada Cellex, Fundaci\'o Mir-Puig, and from Generalitat de Catalunya (AGAUR Grant No. 2017 SGR 1341, CERCA program).
\end{acknowledgments}

\onecolumngrid

\appendix

\section{Dipole-dipole interaction in one dimension} \label{sec:dd_in_1D}


The dipole-dipole interaction in the 1D limit can be effectively modeled by the Dirac delta potential ~\cite{Deuretzbacher_2010, Sinha_2007}. For completeness, we present here the argument.

\begin{figure*}[tb!]
\centering
\includegraphics[width=0.5\textwidth]{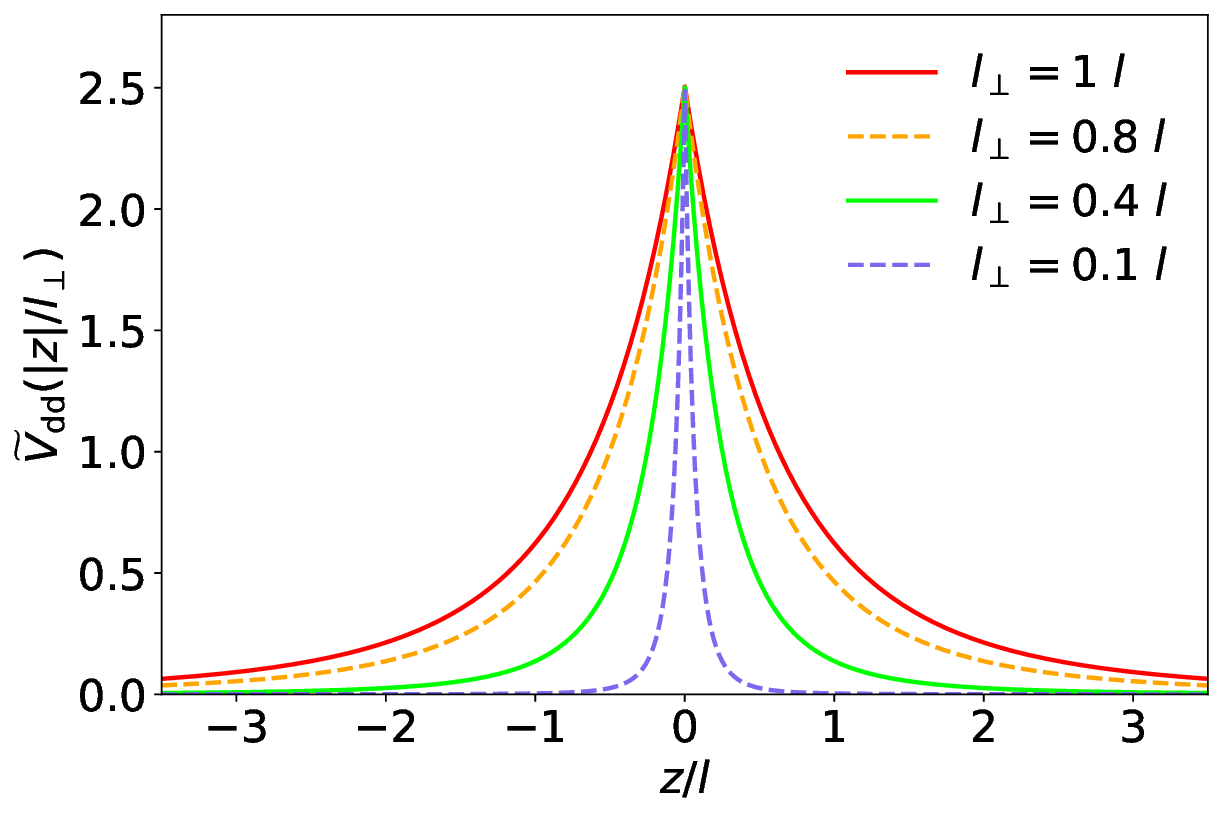}
\caption{Dimensionless interaction potential $ \widetilde{ V}_{\text{dd}}(u)$.}
\label{fig:Vdd}
\end{figure*}

The dipole-dipole interaction potential between two pointlike dipoles is given as
\begin{equation}
    V_{\text{dd}}(\vb{r}) =  C_{\text{dd}} \qty( \frac{\hat{\vb{d}}_1 \cdot \hat{\vb{d}}_2}{r^3}  - \frac{3 (\hat{\vb{d}}_1 \cdot \vb{r})(\hat{\vb{d}}_2 \cdot \vb{r})}{r^5} ) ,
\end{equation}
where $\hat{\vb{d}}_i$ is a dipole moment operator and $C_{\text{dd}}$ is a constant depending on the type of the dipoles.
Dipoles are trapped in the 3D cigar-shaped harmonic trap
\begin{equation}
    V_{\text{trap}} = \frac{\mu}{2} \qty[\omega z^2 + \omega_\perp (x^2+y^2)],
\end{equation}
resulting in a quasi-1D trap for $ \omega \ll	\omega_\perp = \omega_x = \omega_y$. Then, the wave function can be separated into an independent axial (dependent on $z$) and transverse (dependent on $x$ and $y$) part. In that case, the excitation energy in the transverse direction is too large and might be omitted. After integration over the transverse directions we get 
\begin{equation}
    V_{\text{dd}}(z) = U_{\text{dd}} \qty(-2u + \sqrt{2 \pi} \qty(1+u^2) \exp\qty(\frac{u^2}{2}) \erfc \qty( \frac{u}{\sqrt{2}} ) ) = U_{\text{dd}} \widetilde{ V}_{\text{dd}}(u),
\end{equation}
where $u=\frac{\abs{z}}{l_\perp}$ and $l_\perp=\sqrt{\frac{\hbar}{\mu \omega_\perp}} $, and 
\begin{equation}
    U_{\text{dd}} = \frac{C_{\text{dd}}}{4 l_\perp^3} \qty( \hat{\vb{d}}_1 \cdot \hat{\vb{d}}_2 - \frac{3 (\hat{\vb{d}}_1 \cdot \vb{r})(\hat{\vb{d}}_2 \cdot \vb{r})}{r^2} ).
    \label{Udd}
\end{equation}
Now, we consider the long-range and short-range behavior of the interaction.
For $l_\perp \rightarrow 0$ and long-range interactions $z \gg l_\perp$, we can use an asymptotic expansion of the complementary error function for large argument,
\begin{equation}
    \erfc (t) = \frac{\exp(-t^2)}{t \sqrt{\pi}} \sum_{n=0}^{\infty} (-1)^n \frac{(2n-1)!!}{(2t^2)^n},
\end{equation}
to approximate 
\begin{equation}
    \widetilde{ V}_{\text{dd}}(u) \approx -2u+\sqrt{2\pi} (1+u^2)\exp\qty(\frac{u^2}{2}) \frac{\sqrt{2} \exp\qty(-\frac{u^2}{2})}{u \sqrt{\pi}} \qty(1-\frac{1}{u^2}+\frac{3}{u^4}) =  \frac{4}{u^3} + \frac{6}{u^5} \xrightarrow[u \to \infty]{} \frac{4}{u^3}.
\end{equation}
Next, we want to consider a short-ranged interaction (small $z$), again for $l_\perp \rightarrow 0$.
We can observe the behavior of $ \widetilde{ V}_{\text{dd}}(u)$ in Fig.~\ref{fig:Vdd}. 
We can see that the peak at $z=0$ is finite and for smaller $l_\perp$ the peak becomes narrower. It can be shown that
\begin{equation}
    \int_{\infty}^\infty \dd z \frac{1}{4 l_\perp} \widetilde{V}_{\text{dd}}\qty(\frac{\abs{z}}{l_\perp}) = 1,
\end{equation}
and then in the small $l_\perp$ limit, the integrand behaves as the Dirac delta distribution 
\begin{equation}
    \frac{1}{4 l_\perp} \widetilde{V}_{\text{dd}}\qty(\frac{\abs{z}}{l_\perp}) \xrightarrow[l_\perp \to 0]{} \delta(z).
\end{equation}
Thus for short distances, we get
\begin{equation}
    V_{\text{dd}}(z) = 4 \delta(z)  l_\perp U_{\text{dd}} ,
\end{equation}
and for a long distance, we get
\begin{equation}
    V_{\text{dd}}(z) = \frac{4}{\abs{z}^3} l_\perp^3 U_{\text{dd}}.
\end{equation}
Now we can introduce the parameter
\begin{equation}
    \lambda = \frac{l_\perp}{l} = \sqrt{\frac{\omega}{\omega_\perp}},
\end{equation}
which describes the trap anisotropy. With the 1D limit ($\lambda \to 0)$ short-range interaction in the Dirac delta form dominates over long-range interactions. As the result, the interaction has the form
\begin{equation}
    V_{\text{dd}}(z) = 4 \lambda U_{\text{dd}} l \delta(z).
\end{equation}

For the spin magnetic moments,  $    \hat{\vb{d}} = -  \frac{g \mu_\text{B}}{\hbar} \hat{\vb{s}} $, where $g$ is the $g$-factor and $C_{\text{dd}} = \frac{\mu_0}{4\pi}$. The length of the magnetic dipole moment is $\abs{\vb{d}} = d = \mu_\text{B} g  \sqrt{s(s+1)}$, where $s$ is the spin quantum number. Then the magnetic dipole moment can be expressed using $d$ and $s$:
\begin{equation}
    \hat{\vb{d}} = - \frac{d}{\hbar \sqrt{s(s+1)}} \hat{\vb{s}}.
\end{equation}
Substituting $\hat{\vb{d}}$ in Eq. \eqref{Udd} with the conditions that $d_1=d_2=d$ and $s_1=s_2=s$, we get
\begin{equation}
    U_{\text{dd}}=\frac{\mu_0 d^2}{16 \pi \hbar^2 s(s+1) l_\perp^3}  \qty( \hat{\vb{s}}_1 \cdot \hat{\vb{s}}_2 - \frac{3 (\hat{\vb{s}}_1 \cdot \vb{r})(\hat{\vb{s}}_2 \cdot \vb{r})}{r^2} ),
\end{equation}
and in the one-dimensional situation
\begin{equation}
    \frac{ (\hat{\vb{s}}_1 \cdot \vb{r})(\hat{\vb{s}}_2 \cdot \vb{r})}{r^2}=\frac{ (s_{1z} z)(s_{2z} z)}{z^2} = s_{1z} s_{2z}.
\end{equation}
Finally
\begin{equation}
    V_{\text{dd}}(z) =  \lambda \frac{\mu_0 \mu^2 l}{4 \pi  \hbar^2 s(s+1) l_\perp^3 }   \delta(z)  \qty( \hat{\vb{s}}_1 \cdot \hat{\vb{s}}_2 - 3 s_{1z} s_{2z} ),
\end{equation}
and after substitution for $\lambda$, $l$, and $l_\perp$,
\begin{equation}
    V_{\text{dd}}(z) =   \frac{\mu_0 d^2 \omega_\perp \mu}{4 \pi  \hbar^3 s(s+1)}    \delta(z)  \qty( \hat{\vb{s}}_1 \cdot \hat{\vb{s}}_2 - 3 s_{1z} s_{2z} ).
\end{equation}

Term $3 s_{1z} s_{2z}$ is diagonal in the spin basis and effectively acts like the contact interaction mixing of the harmonic trap states. In this work, we focus on Heisenberg-like interaction in the form
\begin{equation}
    \ham_{\text{dd}}^{\text{1D}}= \frac{\mu_0 d^2 \omega_\perp \mu}{4 \pi  \hbar^3 s(s+1)}    \delta(z)  \hat{\vb{s}}_1 \cdot \hat{\vb{s}}_2 = \frac{2J}{\hbar^2}\delta(z_1-z_2) \hat{\vb{s}}_1 \cdot \hat{\vb{s}}_2,
\end{equation}
therefore the dipole-dipole interaction strength has the form
\begin{equation}
    J= \frac{\mu_0 d^2 \omega_\perp \mu}{8 \pi  \hbar s(s+1)}.
\end{equation}

\section{Matrix elements} \label{sec:matrix_el}
Here, we provide the matrix elements of the terms of the relative motion part of the Hamiltonian given by Eq. \eqref{eq:tot_ham} defined in Eqs. \eqref{eq:trap}, \eqref{eq:dd}, and \eqref{eq:zeeman} in the computation basis of $\ket{n,S,M_z,s_1,s_2} \equiv \ket{i}$:
\begin{equation}
    \mel{i}{\ham_{\text{rel}}}{i^\prime} = 
    \mel{i}{\ham_{\text{trap}}^{\text{rel}}}{i^\prime} + \mel{i}{\ham_{\text{dd}}}{i^\prime} + \mel{i}{\ham_{\text{Zeeman}}^{\text{1D}}}{i^\prime},
\end{equation}
where
\begin{equation}
    \mel{i}{\ham_{\text{trap}}^{\text{rel}}}{i^\prime}  = \hbar \omega \qty(n+\frac{1}{2} )  \delta_{n, n^\prime} \delta_{s_1, s_1^\prime} \delta_{s_2, s_2^\prime} \delta_{S, S^\prime} \delta_{M_z, M_z^\prime},
\end{equation}

\begin{align}
     & \mel{i}{\ham_{\text{dd}}^{\text{1D}}}{i^\prime} = \frac{2}{\sqrt{2}} \frac{J}{\hbar^2} \varphi_n(0) \varphi_{n^\prime}(0) \sum_{\substack{m_{s_1}, m_{s_2} \\ m_{s_1}^\prime, m_{s_2}^\prime}} \Bigg[ C_{SM_z}^{s_1 m_{s_1} s_2 m_{s_2}} C_{S^\prime M_z^\prime}^{s_1^\prime m_{s_1}^\prime s_2^\prime m_{s_2}^\prime} \Big[  m_{s_1}^\prime m_{s_2}^\prime \delta_{s_1,s_1^\prime} \delta_{s_2,s_2^\prime} \delta_{m_{s_1},m_{s_1}^\prime} \delta_{m_{s_2},m_{s_2}^\prime}  \nonumber \\ 
    &+ \frac{1}{2} \sqrt{s_2^\prime(s_2^\prime+1) - m_{s_2}\prime(m_{s_2}^\prime+1) } \sqrt{s_1^\prime (s_1^\prime+1) - m_{s_1}^\prime (m_{s_1}^\prime - 1)} \delta_{s_1,s_1^\prime} \delta_{s_2,s_2^\prime} \delta_{m_{s_1},m_{s_1}^\prime-1} \delta_{m_{s_2},m_{s_2}^\prime+1}  \\
    &+ \frac{1}{2} \sqrt{s_2^\prime (s_2^\prime+1) - m_{s_2}^\prime (m_{s_2}^\prime-1) } \sqrt{s_1^\prime (s_1^\prime +1) - m_{s_1}^\prime (m_{s_1}^\prime+1)} \delta_{s_1,s_1^\prime} \delta_{s_2,s_2^\prime} \delta_{m_{s_1},m_{s_1}^\prime+1} \delta_{m_{s_2},m_{s_2}^\prime-1}  \Big] \Bigg] \nonumber,
\end{align}

and assuming that $\vb{B} = (B_x,0,B_z)$

\begin{align}
    &\mel{i}{\ham_{\text{Zeeman}}}{i^\prime} = \mu_\text{B} \delta_{n, n^\prime} \sum_{\substack{m_{s_1}, m_{s_2} \\ m_{s_1}^\prime, m_{s_2}^\prime}} \Bigg[ C_{SM_z}^{s_1 m_{s_1} s_2 m_{s_2}} C_{S^\prime M_z^\prime}^{s_1^\prime m_{s_1}^\prime s_2^\prime m_{s_2}^\prime} \frac{1}{2} B_x \big[ g_1 \sqrt{s_1(s_1+1)-m_{s_1}^\prime (m_{s_1}^\prime+1)} \delta_{m_{s_1},m_{s_1}^\prime+1} \delta_{m_{s_2},m_{s_2}^\prime} + \nonumber \\    
&+	 g_1 \sqrt{s_1(s_1+1)-m_{s_1}^\prime (m_{s_1}^\prime-1)} \delta_{m_{s_1},m_{s_1}^\prime-1} \delta_{m_{s_2},m_{s_2}^\prime} +  g_2 \sqrt{s_2(s_2+1)-m_{s_2}^\prime (m_{s_2}^\prime+1)} \delta_{m_{s_1},m_{s_1}^\prime} \delta_{m_{s_2},m_{s_2}^\prime+1} +  \\ 
&+	 g_2 \sqrt{s_2(s_2+1)-m_{s_2}^\prime (m_{s_2}^\prime-1)} \delta_{m_{s_1},m_{s_1}^\prime+1} \delta_{m_{s_2},m_{s_2}^\prime-1} \big] +  B_z (g_1 m_{s_1}^\prime + g_2 m_{s_2}^\prime) \delta_{s_1,s_1^\prime} \delta_{s_2,s_2^\prime} \delta_{m_{s_1},m_{s_1}^\prime} \delta_{m_{s_2},m_{s_2}^\prime} \Bigg]. \nonumber
\end{align}


\section{Analytical proof that $\ev{m_{s_1}}(t)=\ev{m_{s_2}}(t)=\frac{1}{2}\ev{M_z}(t)$} 
\label{sec:m1_proof}

Below we show the exact analytical proof that $\ev{m_{s_1}}(t)=\ev{m_{s_2}}(t)=\frac{1}{2}\ev{M_z}(t)$ for two indistinguishable bosonic spin-3 atoms in a 1D harmonic trap. We consider the case where the quantum number $S$ is conserved and $M_z$ is not. We assuming that eigenstates of the system are expressed as
\begin{equation}
    \ket{\Psi_p^S} = \sum_{nM_z} a_{nM_z} \ket{n}\ket{S,M_z}.
\end{equation}
The time evolution of $\ev{m_{s_1}}$ for the system in an initial state $\ket{\psi_0}$, which is not an eigenstate of the system,  is calculated as
\begin{equation}
    \ev{\hat{s}_{1z}}{\psi(t)} = \sum_{S,S^\prime} \sum_{p,q} \braket{\Psi_p^S}{\psi_0} \braket{\Psi_q^{S^\prime}}{\psi_0} \exp(-i(E_{p}^S-E_q^{S^\prime})t) \mel{\Psi_p^{S}}{\hat{s}_{1z}}{\Psi_q^{S^\prime}},
\label{S1z_ev}
\end{equation}
where $\ket{\psi(t)}=\hat{U}(t) \ket{\psi_0}$ and $\hat{U}(t)$ is the unitary time evolution operator.
First, we look at the expectation value:
\begin{equation}
    \mel{\Psi_p^{S}}{\hat{s}_{1z}}{\Psi_q^{S^\prime}} = \sum_{n n^\prime} \sum_{M_z M_z^\prime} \qty(a_{nM_z})^* b_{n^\prime M_z^\prime} \braket{n}{n^\prime} \mel{SM_z}{\hat{s}_{1z}}{S^\prime M_z^\prime} 
\end{equation}
\begin{equation}
    = \sum_{n} \sum_{M_z M_z^\prime} \sum_{m_{s_1},m_{s_2},m^\prime_1,m_{s_2}^\prime} \qty(a_{nM_z})^* b_{n M_z^\prime} C_{SM_z}^{m_{s_1} m_{s_2}} C_{S^\prime M_z^\prime}^{m^\prime_1 m_{s_2}^\prime}  \mel{m_{s_1} m_{s_2}}{\hat{s}_{1z}}{m^\prime_1 m_{s_2}^\prime} 
\end{equation}
\begin{equation}
   =\sum_{n}  \sum_{M_z M_z^\prime} \qty(a_{n M_z})^* b_{n M_z^\prime} \sum_{m_{s_1},m_{s_2}}  C_{S M_z}^{m_{s_1} m_{s_2}} C_{S^\prime M_z^\prime}^{m_{s_1} m_{s_2}} m_{s_1} 
\end{equation}
\begin{equation}
   =\sum_{n}  \sum_{M_z M_z^\prime} \qty(a_{n M_z})^* b_{n M_z^\prime} \frac{1}{2} \sum_{m_{s_1},m_{s_2}} \qty( C_{S M_z}^{m_{s_1} m_{s_2}} C_{S^\prime M_z^\prime}^{m_{s_1} m_{s_2}} m_{s_1} + C_{S M_z}^{m_{s_2} m_{s_1}} C_{S^\prime M_z^\prime}^{m_{s_2} m_{s_1}} m_{s_2} ).
\end{equation}
From the symmetry of the Clebsch-Gordan coefficients, we know that
\begin{equation}
    C_{S M_z}^{m_{s_1} m_{s_2}} = \qty(-1)^{S-(s_1+s_2)} C_{SM_z}^{m_{s_2} m_{s_1}},
\end{equation}
and therefore,
\begin{equation}
    C_{SM_z}^{m_{s_1} m_{s_2}} C_{S^\prime M_z^\prime}^{m_{s_1} m_{s_2}} = \qty(-1)^{S+S^\prime-2(s_1+s_2)} C_{SM_z}^{m_{s_2} m_{s_1}} C_{S^\prime M_z^\prime}^{m_{s_2} m_{s_1}},
\end{equation}
however, we know that due to the symmetry of the states, $a_{n M_z}$ are nonzero only if $n$ and $S$  have the same parity because only those states are part of the basis (then $S$ and $S^\prime$ must have the same parity, too). For non-zero elements of the sum, $S+S^\prime$ is always even. Then
\begin{equation}
   \mel{\Psi_p^{S}}{\hat{s}_{1z}}{\Psi_q^{S^\prime}} =\sum_{n} {}^\prime  \sum_{M_z M_z^\prime} {}^\prime \qty(a_{n M_z})^* b_{n M_z^\prime} \frac{1}{2} \sum_{m_{s_1},m_{s_2}} \qty( C_{S M_z}^{m_{s_1} m_{s_2}} C_{S^\prime M_z^\prime}^{m_{s_1} m_{s_2}} m_{s_1} + C_{S M_z}^{m_{s_1} m_{s_2}} C_{S^\prime M_z^\prime}^{m_{s_1} m_{s_2}} m_{s_2} )  
\end{equation}
 \begin{equation}
   =\sum_{n} {}^\prime  \sum_{M_z M_z^\prime} {}^\prime \qty(a_{n S})^* b_{n S^\prime} \frac{1}{2} \sum_{m_{s_1},m_{s_2}} C_{S M_z}^{m_{s_1} m_{s_2}} C_{S^\prime M_z^\prime}^{m_{s_1} m_{s_2}} \qty(  m_{s_1} + m_{s_2} ), 
\end{equation}
where $\sum {}^\prime$ means the summation with the preserved parity of $n$ and $S,S^\prime$. We know that $C_{S M_z}^{m_{s_1} m_{s_2}}$ are equal to 0 if $m_{s_1}+m_{s_2} \neq M_z$, and then  
 \begin{equation}
   \mel{\Psi_p^{S}}{\hat{s}_{1z}}{\Psi_q^{S^\prime}} =\sum_{n} {}^\prime  \sum_{M_z} {}^\prime \qty(a_{nS})^* b_{n S^\prime} \frac{1}{2} \sum_{m_{s_1},m_{s_2}} C_{S M_z}^{m_{s_1} m_{s_2}} C_{S^\prime M_z}^{m_{s_1} m_{s_2}} M_z. 
\end{equation}
Moreover,
\begin{equation}
\braket{S M_z}{S^\prime M_z} = \sum_{m_{s_1}, m_{s_2}, m_{s_1}^\prime, m_{s_2}^\prime} C_{S M_z}^{m_{s_1} m_{s_2}} C_{S^\prime M_z}^{m_{s_1}^\prime m_{s_2}^\prime} \braket{m_{s_1} m_{s_2}}{m_{s_1}^\prime m_{s_2}^\prime} = \sum_{m_{s_1},m_{s_2}} C_{S M_z}^{m_{s_1} m_{s_2}} C_{S^\prime M_z}^{m_{s_1} m_{s_2}};
\end{equation}
therefore 
 \begin{equation}
   \mel{\Psi_p^{S}}{\hat{s}_{1z}}{\Psi_q^{S^\prime}} =\sum_{n} {}^\prime  \sum_{M_z} {}^\prime \qty(a_{n M_z})^* b_{n M_z} \frac{1}{2} M_z \delta_{S,S^\prime}  
\end{equation}
 \begin{equation}
   = \frac{1}{2} M_z \braket{\Psi_p^S}{\Psi_q^{S^\prime}} = \frac{1}{2} \mel{\Psi_p^S}{M_z}{\Psi_q^{S^\prime}} = \frac{1}{2} \mel{\Psi_p^S}{\hat{S}_z}{\Psi_q^{S^\prime}} ,
\end{equation}
and therefore 
\begin{equation}
    \ev{\hat{s}_{1z}}{\psi(t)}  = \frac{1}{2} \ev{\hat{S}_{z}}{\psi(t)} = \ev{\hat{s}_{2z}}{\psi(t)} .
\end{equation}

\twocolumngrid
\bibliography{magnetic_atoms_bib}

\begin{thebibliography}{106}%
\makeatletter
\providecommand \@ifxundefined [1]{%
 \@ifx{#1\undefined}
}%
\providecommand \@ifnum [1]{%
 \ifnum #1\expandafter \@firstoftwo
 \else \expandafter \@secondoftwo
 \fi
}%
\providecommand \@ifx [1]{%
 \ifx #1\expandafter \@firstoftwo
 \else \expandafter \@secondoftwo
 \fi
}%
\providecommand \natexlab [1]{#1}%
\providecommand \enquote  [1]{``#1''}%
\providecommand \bibnamefont  [1]{#1}%
\providecommand \bibfnamefont [1]{#1}%
\providecommand \citenamefont [1]{#1}%
\providecommand \href@noop [0]{\@secondoftwo}%
\providecommand \href [0]{\begingroup \@sanitize@url \@href}%
\providecommand \@href[1]{\@@startlink{#1}\@@href}%
\providecommand \@@href[1]{\endgroup#1\@@endlink}%
\providecommand \@sanitize@url [0]{\catcode `\\12\catcode `\$12\catcode
  `\&12\catcode `\#12\catcode `\^12\catcode `\_12\catcode `\%12\relax}%
\providecommand \@@startlink[1]{}%
\providecommand \@@endlink[0]{}%
\providecommand \url  [0]{\begingroup\@sanitize@url \@url }%
\providecommand \@url [1]{\endgroup\@href {#1}{\urlprefix }}%
\providecommand \urlprefix  [0]{URL }%
\providecommand \Eprint [0]{\href }%
\providecommand \doibase [0]{http://dx.doi.org/}%
\providecommand \selectlanguage [0]{\@gobble}%
\providecommand \bibinfo  [0]{\@secondoftwo}%
\providecommand \bibfield  [0]{\@secondoftwo}%
\providecommand \translation [1]{[#1]}%
\providecommand \BibitemOpen [0]{}%
\providecommand \bibitemStop [0]{}%
\providecommand \bibitemNoStop [0]{.\EOS\space}%
\providecommand \EOS [0]{\spacefactor3000\relax}%
\providecommand \BibitemShut  [1]{\csname bibitem#1\endcsname}%
\let\auto@bib@innerbib\@empty
\bibitem [{\citenamefont {Serwane}\ \emph {et~al.}(2011)\citenamefont
  {Serwane}, \citenamefont {Z{\"u}rn}, \citenamefont {Lompe}, \citenamefont
  {Ottenstein}, \citenamefont {Wenz},\ and\ \citenamefont
  {Jochim}}]{Serwane_2011}%
  \BibitemOpen
  \bibfield  {author} {\bibinfo {author} {\bibfnamefont {F.}~\bibnamefont
  {Serwane}}, \bibinfo {author} {\bibfnamefont {G.}~\bibnamefont {Z{\"u}rn}},
  \bibinfo {author} {\bibfnamefont {T.}~\bibnamefont {Lompe}}, \bibinfo
  {author} {\bibfnamefont {T.~B.}\ \bibnamefont {Ottenstein}}, \bibinfo
  {author} {\bibfnamefont {A.~N.}\ \bibnamefont {Wenz}}, \ and\ \bibinfo
  {author} {\bibfnamefont {S.}~\bibnamefont {Jochim}},\ }\href {\doibase
  10.1126/science.1201351} {\bibfield  {journal} {\bibinfo  {journal}
  {Science}\ }\textbf {\bibinfo {volume} {332}},\ \bibinfo {pages} {336}
  (\bibinfo {year} {2011})}\BibitemShut {NoStop}%
\bibitem [{\citenamefont {Endres}\ \emph {et~al.}(2016)\citenamefont {Endres},
  \citenamefont {Bernien}, \citenamefont {Keesling}, \citenamefont {Levine},
  \citenamefont {Anschuetz}, \citenamefont {Krajenbrink}, \citenamefont
  {Senko}, \citenamefont {Vuletic}, \citenamefont {Greiner},\ and\
  \citenamefont {Lukin}}]{Endres_2016}%
  \BibitemOpen
  \bibfield  {author} {\bibinfo {author} {\bibfnamefont {M.}~\bibnamefont
  {Endres}}, \bibinfo {author} {\bibfnamefont {H.}~\bibnamefont {Bernien}},
  \bibinfo {author} {\bibfnamefont {A.}~\bibnamefont {Keesling}}, \bibinfo
  {author} {\bibfnamefont {H.}~\bibnamefont {Levine}}, \bibinfo {author}
  {\bibfnamefont {E.~R.}\ \bibnamefont {Anschuetz}}, \bibinfo {author}
  {\bibfnamefont {A.}~\bibnamefont {Krajenbrink}}, \bibinfo {author}
  {\bibfnamefont {C.}~\bibnamefont {Senko}}, \bibinfo {author} {\bibfnamefont
  {V.}~\bibnamefont {Vuletic}}, \bibinfo {author} {\bibfnamefont
  {M.}~\bibnamefont {Greiner}}, \ and\ \bibinfo {author} {\bibfnamefont
  {M.~D.}\ \bibnamefont {Lukin}},\ }\href {\doibase 10.1126/science.aah3752}
  {\bibfield  {journal} {\bibinfo  {journal} {Science}\ }\textbf {\bibinfo
  {volume} {354}},\ \bibinfo {pages} {1024} (\bibinfo {year}
  {2016})}\BibitemShut {NoStop}%
\bibitem [{\citenamefont {Barredo}\ \emph {et~al.}(2016)\citenamefont
  {Barredo}, \citenamefont {de~L{\'e}s{\'e}leuc}, \citenamefont {Lienhard},
  \citenamefont {Lahaye},\ and\ \citenamefont {Browaeys}}]{Barredo_2016}%
  \BibitemOpen
  \bibfield  {author} {\bibinfo {author} {\bibfnamefont {D.}~\bibnamefont
  {Barredo}}, \bibinfo {author} {\bibfnamefont {S.}~\bibnamefont
  {de~L{\'e}s{\'e}leuc}}, \bibinfo {author} {\bibfnamefont {V.}~\bibnamefont
  {Lienhard}}, \bibinfo {author} {\bibfnamefont {T.}~\bibnamefont {Lahaye}}, \
  and\ \bibinfo {author} {\bibfnamefont {A.}~\bibnamefont {Browaeys}},\ }\href
  {\doibase 10.1126/science.aah3778} {\bibfield  {journal} {\bibinfo  {journal}
  {Science}\ }\textbf {\bibinfo {volume} {354}},\ \bibinfo {pages} {1021}
  (\bibinfo {year} {2016})}\BibitemShut {NoStop}%
\bibitem [{\citenamefont {Bakr}\ \emph {et~al.}(2009)\citenamefont {Bakr},
  \citenamefont {Gillen}, \citenamefont {Peng}, \citenamefont {F{\"o}lling},\
  and\ \citenamefont {Greiner}}]{Bakr_2009}%
  \BibitemOpen
  \bibfield  {author} {\bibinfo {author} {\bibfnamefont {W.~S.}\ \bibnamefont
  {Bakr}}, \bibinfo {author} {\bibfnamefont {J.~I.}\ \bibnamefont {Gillen}},
  \bibinfo {author} {\bibfnamefont {A.}~\bibnamefont {Peng}}, \bibinfo {author}
  {\bibfnamefont {S.}~\bibnamefont {F{\"o}lling}}, \ and\ \bibinfo {author}
  {\bibfnamefont {M.}~\bibnamefont {Greiner}},\ }\href {\doibase
  10.1038/nature08482} {\bibfield  {journal} {\bibinfo  {journal} {Nature}\
  }\textbf {\bibinfo {volume} {462}},\ \bibinfo {pages} {74} (\bibinfo {year}
  {2009})}\BibitemShut {NoStop}%
\bibitem [{\citenamefont {Sherson}\ \emph {et~al.}(2010)\citenamefont
  {Sherson}, \citenamefont {Weitenberg}, \citenamefont {Endres}, \citenamefont
  {Cheneau}, \citenamefont {Bloch},\ and\ \citenamefont {Kuhr}}]{Sherson_2010}%
  \BibitemOpen
  \bibfield  {author} {\bibinfo {author} {\bibfnamefont {J.~F.}\ \bibnamefont
  {Sherson}}, \bibinfo {author} {\bibfnamefont {C.}~\bibnamefont {Weitenberg}},
  \bibinfo {author} {\bibfnamefont {M.}~\bibnamefont {Endres}}, \bibinfo
  {author} {\bibfnamefont {M.}~\bibnamefont {Cheneau}}, \bibinfo {author}
  {\bibfnamefont {I.}~\bibnamefont {Bloch}}, \ and\ \bibinfo {author}
  {\bibfnamefont {S.}~\bibnamefont {Kuhr}},\ }\href {\doibase
  10.1038/nature09378} {\bibfield  {journal} {\bibinfo  {journal} {Nature}\
  }\textbf {\bibinfo {volume} {467}},\ \bibinfo {pages} {68} (\bibinfo {year}
  {2010})}\BibitemShut {NoStop}%
\bibitem [{\citenamefont {Cheuk}\ \emph {et~al.}(2015)\citenamefont {Cheuk},
  \citenamefont {Nichols}, \citenamefont {Okan}, \citenamefont {Gersdorf},
  \citenamefont {Ramasesh}, \citenamefont {Bakr}, \citenamefont {Lompe},\ and\
  \citenamefont {Zwierlein}}]{Cheuk15PRL}%
  \BibitemOpen
  \bibfield  {author} {\bibinfo {author} {\bibfnamefont {L.~W.}\ \bibnamefont
  {Cheuk}}, \bibinfo {author} {\bibfnamefont {M.~A.}\ \bibnamefont {Nichols}},
  \bibinfo {author} {\bibfnamefont {M.}~\bibnamefont {Okan}}, \bibinfo {author}
  {\bibfnamefont {T.}~\bibnamefont {Gersdorf}}, \bibinfo {author}
  {\bibfnamefont {V.~V.}\ \bibnamefont {Ramasesh}}, \bibinfo {author}
  {\bibfnamefont {W.~S.}\ \bibnamefont {Bakr}}, \bibinfo {author}
  {\bibfnamefont {T.}~\bibnamefont {Lompe}}, \ and\ \bibinfo {author}
  {\bibfnamefont {M.~W.}\ \bibnamefont {Zwierlein}},\ }\href {\doibase
  10.1103/PhysRevLett.114.193001} {\bibfield  {journal} {\bibinfo  {journal}
  {Phys. Rev. Lett.}\ }\textbf {\bibinfo {volume} {114}},\ \bibinfo {pages}
  {193001} (\bibinfo {year} {2015})}\BibitemShut {NoStop}%
\bibitem [{\citenamefont {Haller}\ \emph {et~al.}(2015)\citenamefont {Haller},
  \citenamefont {Hudson}, \citenamefont {Kelly}, \citenamefont {Cotta},
  \citenamefont {Peaudecerf}, \citenamefont {Bruce},\ and\ \citenamefont
  {Kuhr}}]{Haller15NatPhys}%
  \BibitemOpen
  \bibfield  {author} {\bibinfo {author} {\bibfnamefont {E.}~\bibnamefont
  {Haller}}, \bibinfo {author} {\bibfnamefont {J.}~\bibnamefont {Hudson}},
  \bibinfo {author} {\bibfnamefont {A.}~\bibnamefont {Kelly}}, \bibinfo
  {author} {\bibfnamefont {D.~A.}\ \bibnamefont {Cotta}}, \bibinfo {author}
  {\bibfnamefont {B.}~\bibnamefont {Peaudecerf}}, \bibinfo {author}
  {\bibfnamefont {G.~D.}\ \bibnamefont {Bruce}}, \ and\ \bibinfo {author}
  {\bibfnamefont {S.}~\bibnamefont {Kuhr}},\ }\href {\doibase
  10.1038/nphys3403} {\bibfield  {journal} {\bibinfo  {journal} {Nat. Phys.}\
  }\textbf {\bibinfo {volume} {11}},\ \bibinfo {pages} {738–742} (\bibinfo
  {year} {2015})}\BibitemShut {NoStop}%
\bibitem [{\citenamefont {Boll}\ \emph {et~al.}(2016)\citenamefont {Boll},
  \citenamefont {Hilker}, \citenamefont {Salomon}, \citenamefont {Omran},
  \citenamefont {Nespolo}, \citenamefont {Pollet}, \citenamefont {Bloch},\ and\
  \citenamefont {Gross}}]{Boll_2016}%
  \BibitemOpen
  \bibfield  {author} {\bibinfo {author} {\bibfnamefont {M.}~\bibnamefont
  {Boll}}, \bibinfo {author} {\bibfnamefont {T.~A.}\ \bibnamefont {Hilker}},
  \bibinfo {author} {\bibfnamefont {G.}~\bibnamefont {Salomon}}, \bibinfo
  {author} {\bibfnamefont {A.}~\bibnamefont {Omran}}, \bibinfo {author}
  {\bibfnamefont {J.}~\bibnamefont {Nespolo}}, \bibinfo {author} {\bibfnamefont
  {L.}~\bibnamefont {Pollet}}, \bibinfo {author} {\bibfnamefont
  {I.}~\bibnamefont {Bloch}}, \ and\ \bibinfo {author} {\bibfnamefont
  {C.}~\bibnamefont {Gross}},\ }\href {\doibase 10.1126/science.aag1635}
  {\bibfield  {journal} {\bibinfo  {journal} {Science}\ }\textbf {\bibinfo
  {volume} {353}},\ \bibinfo {pages} {1257} (\bibinfo {year}
  {2016})}\BibitemShut {NoStop}%
\bibitem [{\citenamefont {Regal}\ \emph {et~al.}(2003)\citenamefont {Regal},
  \citenamefont {Ticknor}, \citenamefont {Bohn},\ and\ \citenamefont
  {Jin}}]{Regal_2003}%
  \BibitemOpen
  \bibfield  {author} {\bibinfo {author} {\bibfnamefont {C.~A.}\ \bibnamefont
  {Regal}}, \bibinfo {author} {\bibfnamefont {C.}~\bibnamefont {Ticknor}},
  \bibinfo {author} {\bibfnamefont {J.~L.}\ \bibnamefont {Bohn}}, \ and\
  \bibinfo {author} {\bibfnamefont {D.~S.}\ \bibnamefont {Jin}},\ }\href
  {\doibase 10.1038/nature01738} {\bibfield  {journal} {\bibinfo  {journal}
  {Nature}\ }\textbf {\bibinfo {volume} {424}},\ \bibinfo {pages} {47}
  (\bibinfo {year} {2003})}\BibitemShut {NoStop}%
\bibitem [{\citenamefont {Hodby}\ \emph {et~al.}(2005)\citenamefont {Hodby},
  \citenamefont {Thompson}, \citenamefont {Regal}, \citenamefont {Greiner},
  \citenamefont {Wilson}, \citenamefont {Jin}, \citenamefont {Cornell},\ and\
  \citenamefont {Wieman}}]{Hodby_2005}%
  \BibitemOpen
  \bibfield  {author} {\bibinfo {author} {\bibfnamefont {E.}~\bibnamefont
  {Hodby}}, \bibinfo {author} {\bibfnamefont {S.~T.}\ \bibnamefont {Thompson}},
  \bibinfo {author} {\bibfnamefont {C.~A.}\ \bibnamefont {Regal}}, \bibinfo
  {author} {\bibfnamefont {M.}~\bibnamefont {Greiner}}, \bibinfo {author}
  {\bibfnamefont {A.~C.}\ \bibnamefont {Wilson}}, \bibinfo {author}
  {\bibfnamefont {D.~S.}\ \bibnamefont {Jin}}, \bibinfo {author} {\bibfnamefont
  {E.~A.}\ \bibnamefont {Cornell}}, \ and\ \bibinfo {author} {\bibfnamefont
  {C.~E.}\ \bibnamefont {Wieman}},\ }\href {\doibase
  10.1103/PhysRevLett.94.120402} {\bibfield  {journal} {\bibinfo  {journal}
  {Phys. Rev. Lett.}\ }\textbf {\bibinfo {volume} {94}},\ \bibinfo {pages}
  {120402} (\bibinfo {year} {2005})}\BibitemShut {NoStop}%
\bibitem [{\citenamefont {Danzl}\ \emph {et~al.}(2008)\citenamefont {Danzl},
  \citenamefont {Haller}, \citenamefont {Gustavsson}, \citenamefont {Mark},
  \citenamefont {Hart}, \citenamefont {Bouloufa}, \citenamefont {Dulieu},
  \citenamefont {Ritsch},\ and\ \citenamefont {N\"agerl}}]{Danzl08Science}%
  \BibitemOpen
  \bibfield  {author} {\bibinfo {author} {\bibfnamefont {J.~G.}\ \bibnamefont
  {Danzl}}, \bibinfo {author} {\bibfnamefont {E.}~\bibnamefont {Haller}},
  \bibinfo {author} {\bibfnamefont {M.}~\bibnamefont {Gustavsson}}, \bibinfo
  {author} {\bibfnamefont {M.~J.}\ \bibnamefont {Mark}}, \bibinfo {author}
  {\bibfnamefont {R.}~\bibnamefont {Hart}}, \bibinfo {author} {\bibfnamefont
  {N.}~\bibnamefont {Bouloufa}}, \bibinfo {author} {\bibfnamefont
  {O.}~\bibnamefont {Dulieu}}, \bibinfo {author} {\bibfnamefont
  {H.}~\bibnamefont {Ritsch}}, \ and\ \bibinfo {author} {\bibfnamefont {H.-C.}\
  \bibnamefont {N\"agerl}},\ }\href {\doibase 10.1126/science.1159909}
  {\bibfield  {journal} {\bibinfo  {journal} {Science}\ }\textbf {\bibinfo
  {volume} {321}},\ \bibinfo {pages} {1062} (\bibinfo {year}
  {2008})}\BibitemShut {NoStop}%
\bibitem [{\citenamefont {Ni}\ \emph {et~al.}(2008)\citenamefont {Ni},
  \citenamefont {Ospelkaus}, \citenamefont {de~Miranda}, \citenamefont {Pe'er},
  \citenamefont {Neyenhuis}, \citenamefont {Zirbel}, \citenamefont
  {Kotochigova}, \citenamefont {Julienne}, \citenamefont {Jin},\ and\
  \citenamefont {Ye}}]{Ni08}%
  \BibitemOpen
  \bibfield  {author} {\bibinfo {author} {\bibfnamefont {K.-K.}\ \bibnamefont
  {Ni}}, \bibinfo {author} {\bibfnamefont {S.}~\bibnamefont {Ospelkaus}},
  \bibinfo {author} {\bibfnamefont {M.~H.~G.}\ \bibnamefont {de~Miranda}},
  \bibinfo {author} {\bibfnamefont {A.}~\bibnamefont {Pe'er}}, \bibinfo
  {author} {\bibfnamefont {B.}~\bibnamefont {Neyenhuis}}, \bibinfo {author}
  {\bibfnamefont {J.~J.}\ \bibnamefont {Zirbel}}, \bibinfo {author}
  {\bibfnamefont {S.}~\bibnamefont {Kotochigova}}, \bibinfo {author}
  {\bibfnamefont {P.~S.}\ \bibnamefont {Julienne}}, \bibinfo {author}
  {\bibfnamefont {D.~S.}\ \bibnamefont {Jin}}, \ and\ \bibinfo {author}
  {\bibfnamefont {J.}~\bibnamefont {Ye}},\ }\href {\doibase
  10.1126/science.1163861} {\bibfield  {journal} {\bibinfo  {journal}
  {{Science}}\ }\textbf {\bibinfo {volume} {{322}}},\ \bibinfo {pages} {{231}}
  (\bibinfo {year} {{2008}})}\BibitemShut {NoStop}%
\bibitem [{\citenamefont {De~Marco}\ \emph {et~al.}(2019)\citenamefont
  {De~Marco}, \citenamefont {Valtolina}, \citenamefont {Matsuda}, \citenamefont
  {Tobias}, \citenamefont {Covey},\ and\ \citenamefont {Ye}}]{deMarco2019}%
  \BibitemOpen
  \bibfield  {author} {\bibinfo {author} {\bibfnamefont {L.}~\bibnamefont
  {De~Marco}}, \bibinfo {author} {\bibfnamefont {G.}~\bibnamefont {Valtolina}},
  \bibinfo {author} {\bibfnamefont {K.}~\bibnamefont {Matsuda}}, \bibinfo
  {author} {\bibfnamefont {W.~G.}\ \bibnamefont {Tobias}}, \bibinfo {author}
  {\bibfnamefont {J.~P.}\ \bibnamefont {Covey}}, \ and\ \bibinfo {author}
  {\bibfnamefont {J.}~\bibnamefont {Ye}},\ }\href {\doibase
  10.1126/science.aau7230} {\bibfield  {journal} {\bibinfo  {journal}
  {Science}\ }\textbf {\bibinfo {volume} {363}},\ \bibinfo {pages} {853}
  (\bibinfo {year} {2019})}\BibitemShut {NoStop}%
\bibitem [{\citenamefont {Tarruell}\ and\ \citenamefont
  {Sanchez-Palencia}(2018)}]{Tarruell_2018}%
  \BibitemOpen
  \bibfield  {author} {\bibinfo {author} {\bibfnamefont {L.}~\bibnamefont
  {Tarruell}}\ and\ \bibinfo {author} {\bibfnamefont {L.}~\bibnamefont
  {Sanchez-Palencia}},\ }\href {\doibase
  https://doi.org/10.1016/j.crhy.2018.10.013} {\bibfield  {journal} {\bibinfo
  {journal} {C. R. Phys.}\ }\textbf {\bibinfo {volume} {19}},\ \bibinfo {pages}
  {365} (\bibinfo {year} {2018})}\BibitemShut {NoStop}%
\bibitem [{\citenamefont {Gall}\ \emph {et~al.}(2021)\citenamefont {Gall},
  \citenamefont {Wurz}, \citenamefont {Samland}, \citenamefont {Chan},\ and\
  \citenamefont {K{\"o}hl}}]{Gall_2021}%
  \BibitemOpen
  \bibfield  {author} {\bibinfo {author} {\bibfnamefont {M.}~\bibnamefont
  {Gall}}, \bibinfo {author} {\bibfnamefont {N.}~\bibnamefont {Wurz}}, \bibinfo
  {author} {\bibfnamefont {J.}~\bibnamefont {Samland}}, \bibinfo {author}
  {\bibfnamefont {C.~F.}\ \bibnamefont {Chan}}, \ and\ \bibinfo {author}
  {\bibfnamefont {M.}~\bibnamefont {K{\"o}hl}},\ }\href {\doibase
  10.1038/s41586-020-03058-x} {\bibfield  {journal} {\bibinfo  {journal}
  {Nature}\ }\textbf {\bibinfo {volume} {589}},\ \bibinfo {pages} {40}
  (\bibinfo {year} {2021})}\BibitemShut {NoStop}%
\bibitem [{\citenamefont {Labuhn}\ \emph {et~al.}(2016)\citenamefont {Labuhn},
  \citenamefont {Barredo}, \citenamefont {Ravets}, \citenamefont
  {de~L{\'{e}}s{\'{e}}leuc}, \citenamefont {Macr{\`{\i}}}, \citenamefont
  {Lahaye},\ and\ \citenamefont {Browaeys}}]{Labuhn_2016}%
  \BibitemOpen
  \bibfield  {author} {\bibinfo {author} {\bibfnamefont {H.}~\bibnamefont
  {Labuhn}}, \bibinfo {author} {\bibfnamefont {D.}~\bibnamefont {Barredo}},
  \bibinfo {author} {\bibfnamefont {S.}~\bibnamefont {Ravets}}, \bibinfo
  {author} {\bibfnamefont {S.}~\bibnamefont {de~L{\'{e}}s{\'{e}}leuc}},
  \bibinfo {author} {\bibfnamefont {T.}~\bibnamefont {Macr{\`{\i}}}}, \bibinfo
  {author} {\bibfnamefont {T.}~\bibnamefont {Lahaye}}, \ and\ \bibinfo {author}
  {\bibfnamefont {A.}~\bibnamefont {Browaeys}},\ }\href {\doibase
  10.1038/nature18274} {\bibfield  {journal} {\bibinfo  {journal} {Nature}\
  }\textbf {\bibinfo {volume} {534}},\ \bibinfo {pages} {667} (\bibinfo {year}
  {2016})}\BibitemShut {NoStop}%
\bibitem [{\citenamefont {Jepsen}\ \emph {et~al.}(2020)\citenamefont {Jepsen},
  \citenamefont {Amato-Grill}, \citenamefont {Dimitrova}, \citenamefont {Ho},
  \citenamefont {Demler},\ and\ \citenamefont {Ketterle}}]{Jepsen_2020}%
  \BibitemOpen
  \bibfield  {author} {\bibinfo {author} {\bibfnamefont {P.~N.}\ \bibnamefont
  {Jepsen}}, \bibinfo {author} {\bibfnamefont {J.}~\bibnamefont {Amato-Grill}},
  \bibinfo {author} {\bibfnamefont {I.}~\bibnamefont {Dimitrova}}, \bibinfo
  {author} {\bibfnamefont {W.~W.}\ \bibnamefont {Ho}}, \bibinfo {author}
  {\bibfnamefont {E.}~\bibnamefont {Demler}}, \ and\ \bibinfo {author}
  {\bibfnamefont {W.}~\bibnamefont {Ketterle}},\ }\href {\doibase
  10.1038/s41586-020-3033-y} {\bibfield  {journal} {\bibinfo  {journal}
  {Nature}\ }\textbf {\bibinfo {volume} {588}},\ \bibinfo {pages} {403}
  (\bibinfo {year} {2020})}\BibitemShut {NoStop}%
\bibitem [{\citenamefont {Rossini}\ and\ \citenamefont
  {Fazio}(2012)}]{Rossini_2012}%
  \BibitemOpen
  \bibfield  {author} {\bibinfo {author} {\bibfnamefont {D.}~\bibnamefont
  {Rossini}}\ and\ \bibinfo {author} {\bibfnamefont {R.}~\bibnamefont
  {Fazio}},\ }\href {\doibase 10.1088/1367-2630/14/6/065012} {\bibfield
  {journal} {\bibinfo  {journal} {New J. Phys.}\ }\textbf {\bibinfo {volume}
  {14}},\ \bibinfo {pages} {065012} (\bibinfo {year} {2012})}\BibitemShut
  {NoStop}%
\bibitem [{\citenamefont {Hofmann}\ and\ \citenamefont
  {Potthoff}(2012)}]{Hofmann_2012}%
  \BibitemOpen
  \bibfield  {author} {\bibinfo {author} {\bibfnamefont {F.}~\bibnamefont
  {Hofmann}}\ and\ \bibinfo {author} {\bibfnamefont {M.}~\bibnamefont
  {Potthoff}},\ }\href {\doibase 10.1103/PhysRevB.85.205127} {\bibfield
  {journal} {\bibinfo  {journal} {Phys. Rev. B}\ }\textbf {\bibinfo {volume}
  {85}},\ \bibinfo {pages} {205127} (\bibinfo {year} {2012})}\BibitemShut
  {NoStop}%
\bibitem [{\citenamefont {Dhar}\ \emph {et~al.}(2016)\citenamefont {Dhar},
  \citenamefont {Kinnunen},\ and\ \citenamefont {T\"orm\"a}}]{Dhar_2016}%
  \BibitemOpen
  \bibfield  {author} {\bibinfo {author} {\bibfnamefont {A.}~\bibnamefont
  {Dhar}}, \bibinfo {author} {\bibfnamefont {J.~J.}\ \bibnamefont {Kinnunen}},
  \ and\ \bibinfo {author} {\bibfnamefont {P.}~\bibnamefont {T\"orm\"a}},\
  }\href {\doibase 10.1103/PhysRevB.94.075116} {\bibfield  {journal} {\bibinfo
  {journal} {Phys. Rev. B}\ }\textbf {\bibinfo {volume} {94}},\ \bibinfo
  {pages} {075116} (\bibinfo {year} {2016})}\BibitemShut {NoStop}%
\bibitem [{\citenamefont {Dhar}\ \emph {et~al.}(2018)\citenamefont {Dhar},
  \citenamefont {T\"orm\"a},\ and\ \citenamefont {Kinnunen}}]{Dhar_2018}%
  \BibitemOpen
  \bibfield  {author} {\bibinfo {author} {\bibfnamefont {A.}~\bibnamefont
  {Dhar}}, \bibinfo {author} {\bibfnamefont {P.}~\bibnamefont {T\"orm\"a}}, \
  and\ \bibinfo {author} {\bibfnamefont {J.~J.}\ \bibnamefont {Kinnunen}},\
  }\href {\doibase 10.1103/PhysRevA.97.043624} {\bibfield  {journal} {\bibinfo
  {journal} {Phys. Rev. A}\ }\textbf {\bibinfo {volume} {97}},\ \bibinfo
  {pages} {043624} (\bibinfo {year} {2018})}\BibitemShut {NoStop}%
\bibitem [{\citenamefont {Biedro\ifmmode~\acute{n}\else \'{n}\fi{}}\ \emph
  {et~al.}(2018)\citenamefont {Biedro\ifmmode~\acute{n}\else \'{n}\fi{}},
  \citenamefont {\L{}\k{a}cki},\ and\ \citenamefont
  {Zakrzewski}}]{Biedron_2018}%
  \BibitemOpen
  \bibfield  {author} {\bibinfo {author} {\bibfnamefont {K.}~\bibnamefont
  {Biedro\ifmmode~\acute{n}\else \'{n}\fi{}}}, \bibinfo {author} {\bibfnamefont
  {M.}~\bibnamefont {\L{}\k{a}cki}}, \ and\ \bibinfo {author} {\bibfnamefont
  {J.}~\bibnamefont {Zakrzewski}},\ }\href {\doibase
  10.1103/PhysRevB.97.245102} {\bibfield  {journal} {\bibinfo  {journal} {Phys.
  Rev. B}\ }\textbf {\bibinfo {volume} {97}},\ \bibinfo {pages} {245102}
  (\bibinfo {year} {2018})}\BibitemShut {NoStop}%
\bibitem [{\citenamefont {Suthar}\ \emph {et~al.}(2020)\citenamefont {Suthar},
  \citenamefont {Sable}, \citenamefont {Bai}, \citenamefont {Bandyopadhyay},
  \citenamefont {Pal},\ and\ \citenamefont {Angom}}]{Suthar_2020}%
  \BibitemOpen
  \bibfield  {author} {\bibinfo {author} {\bibfnamefont {K.}~\bibnamefont
  {Suthar}}, \bibinfo {author} {\bibfnamefont {H.}~\bibnamefont {Sable}},
  \bibinfo {author} {\bibfnamefont {R.}~\bibnamefont {Bai}}, \bibinfo {author}
  {\bibfnamefont {S.}~\bibnamefont {Bandyopadhyay}}, \bibinfo {author}
  {\bibfnamefont {S.}~\bibnamefont {Pal}}, \ and\ \bibinfo {author}
  {\bibfnamefont {D.}~\bibnamefont {Angom}},\ }\href {\doibase
  10.1103/PhysRevA.102.013320} {\bibfield  {journal} {\bibinfo  {journal}
  {Phys. Rev. A}\ }\textbf {\bibinfo {volume} {102}},\ \bibinfo {pages}
  {013320} (\bibinfo {year} {2020})}\BibitemShut {NoStop}%
\bibitem [{\citenamefont {Lewenstein}\ \emph {et~al.}(2007)\citenamefont
  {Lewenstein}, \citenamefont {Sanpera}, \citenamefont {Ahufinger},
  \citenamefont {Damski}, \citenamefont {Sen~De},\ and\ \citenamefont
  {Sen}}]{Lewenstein07}%
  \BibitemOpen
  \bibfield  {author} {\bibinfo {author} {\bibfnamefont {M.}~\bibnamefont
  {Lewenstein}}, \bibinfo {author} {\bibfnamefont {A.}~\bibnamefont {Sanpera}},
  \bibinfo {author} {\bibfnamefont {V.}~\bibnamefont {Ahufinger}}, \bibinfo
  {author} {\bibfnamefont {B.}~\bibnamefont {Damski}}, \bibinfo {author}
  {\bibfnamefont {A.}~\bibnamefont {Sen~De}}, \ and\ \bibinfo {author}
  {\bibfnamefont {U.}~\bibnamefont {Sen}},\ }\href {\doibase
  10.1080/00018730701223200} {\bibfield  {journal} {\bibinfo  {journal} {Adv.
  Phys.}\ }\textbf {\bibinfo {volume} {56}},\ \bibinfo {pages} {243} (\bibinfo
  {year} {2007})}\BibitemShut {NoStop}%
\bibitem [{\citenamefont {Bloch}\ \emph {et~al.}(2008)\citenamefont {Bloch},
  \citenamefont {Dalibard},\ and\ \citenamefont {Zwerger}}]{Bloch_2008}%
  \BibitemOpen
  \bibfield  {author} {\bibinfo {author} {\bibfnamefont {I.}~\bibnamefont
  {Bloch}}, \bibinfo {author} {\bibfnamefont {J.}~\bibnamefont {Dalibard}}, \
  and\ \bibinfo {author} {\bibfnamefont {W.}~\bibnamefont {Zwerger}},\ }\href
  {\doibase 10.1103/RevModPhys.80.885} {\bibfield  {journal} {\bibinfo
  {journal} {Rev. Mod. Phys.}\ }\textbf {\bibinfo {volume} {80}},\ \bibinfo
  {pages} {885} (\bibinfo {year} {2008})}\BibitemShut {NoStop}%
\bibitem [{\citenamefont {Blatt}\ and\ \citenamefont
  {Roos}(2012)}]{Blatt_2012}%
  \BibitemOpen
  \bibfield  {author} {\bibinfo {author} {\bibfnamefont {R.}~\bibnamefont
  {Blatt}}\ and\ \bibinfo {author} {\bibfnamefont {C.~F.}\ \bibnamefont
  {Roos}},\ }\href {\doibase 10.1038/nphys2252} {\bibfield  {journal} {\bibinfo
   {journal} {Nat. Phys.}\ }\textbf {\bibinfo {volume} {8}},\ \bibinfo {pages}
  {277} (\bibinfo {year} {2012})}\BibitemShut {NoStop}%
\bibitem [{\citenamefont {Bloch}\ \emph {et~al.}(2012)\citenamefont {Bloch},
  \citenamefont {Dalibard},\ and\ \citenamefont {Nascimb{\`e}ne}}]{Bloch_2012}%
  \BibitemOpen
  \bibfield  {author} {\bibinfo {author} {\bibfnamefont {I.}~\bibnamefont
  {Bloch}}, \bibinfo {author} {\bibfnamefont {J.}~\bibnamefont {Dalibard}}, \
  and\ \bibinfo {author} {\bibfnamefont {S.}~\bibnamefont {Nascimb{\`e}ne}},\
  }\href {\doibase 10.1038/nphys2259} {\bibfield  {journal} {\bibinfo
  {journal} {Nat. Phys.}\ }\textbf {\bibinfo {volume} {8}},\ \bibinfo {pages}
  {267} (\bibinfo {year} {2012})}\BibitemShut {NoStop}%
\bibitem [{\citenamefont {Schneider}\ \emph {et~al.}(2012)\citenamefont
  {Schneider}, \citenamefont {Porras},\ and\ \citenamefont
  {Schaetz}}]{Schneider_2012}%
  \BibitemOpen
  \bibfield  {author} {\bibinfo {author} {\bibfnamefont {C.}~\bibnamefont
  {Schneider}}, \bibinfo {author} {\bibfnamefont {D.}~\bibnamefont {Porras}}, \
  and\ \bibinfo {author} {\bibfnamefont {T.}~\bibnamefont {Schaetz}},\ }\href
  {\doibase 10.1088/0034-4885/75/2/024401} {\bibfield  {journal} {\bibinfo
  {journal} {Rep. Prog. Phys.}\ }\textbf {\bibinfo {volume} {75}},\ \bibinfo
  {pages} {024401} (\bibinfo {year} {2012})}\BibitemShut {NoStop}%
\bibitem [{\citenamefont {Dutta}\ \emph {et~al.}(2015)\citenamefont {Dutta},
  \citenamefont {Gajda}, \citenamefont {Hauke}, \citenamefont {Lewenstein},
  \citenamefont {Lühmann}, \citenamefont {Malomed}, \citenamefont
  {Sowi{\'{n}}ski},\ and\ \citenamefont {Zakrzewski}}]{Dutta_2015}%
  \BibitemOpen
  \bibfield  {author} {\bibinfo {author} {\bibfnamefont {O.}~\bibnamefont
  {Dutta}}, \bibinfo {author} {\bibfnamefont {M.}~\bibnamefont {Gajda}},
  \bibinfo {author} {\bibfnamefont {P.}~\bibnamefont {Hauke}}, \bibinfo
  {author} {\bibfnamefont {M.}~\bibnamefont {Lewenstein}}, \bibinfo {author}
  {\bibfnamefont {D.-S.}\ \bibnamefont {Lühmann}}, \bibinfo {author}
  {\bibfnamefont {B.~A.}\ \bibnamefont {Malomed}}, \bibinfo {author}
  {\bibfnamefont {T.}~\bibnamefont {Sowi{\'{n}}ski}}, \ and\ \bibinfo {author}
  {\bibfnamefont {J.}~\bibnamefont {Zakrzewski}},\ }\href {\doibase
  10.1088/0034-4885/78/6/066001} {\bibfield  {journal} {\bibinfo  {journal}
  {Rep. Prog. Phys.}\ }\textbf {\bibinfo {volume} {78}},\ \bibinfo {pages}
  {066001} (\bibinfo {year} {2015})}\BibitemShut {NoStop}%
\bibitem [{\citenamefont {Baier}\ \emph {et~al.}(2016)\citenamefont {Baier},
  \citenamefont {Mark}, \citenamefont {Petter}, \citenamefont {Aikawa},
  \citenamefont {Chomaz}, \citenamefont {Cai}, \citenamefont {Baranov},
  \citenamefont {Zoller},\ and\ \citenamefont {Ferlaino}}]{Baier_2016}%
  \BibitemOpen
  \bibfield  {author} {\bibinfo {author} {\bibfnamefont {S.}~\bibnamefont
  {Baier}}, \bibinfo {author} {\bibfnamefont {M.~J.}\ \bibnamefont {Mark}},
  \bibinfo {author} {\bibfnamefont {D.}~\bibnamefont {Petter}}, \bibinfo
  {author} {\bibfnamefont {K.}~\bibnamefont {Aikawa}}, \bibinfo {author}
  {\bibfnamefont {L.}~\bibnamefont {Chomaz}}, \bibinfo {author} {\bibfnamefont
  {Z.}~\bibnamefont {Cai}}, \bibinfo {author} {\bibfnamefont {M.}~\bibnamefont
  {Baranov}}, \bibinfo {author} {\bibfnamefont {P.}~\bibnamefont {Zoller}}, \
  and\ \bibinfo {author} {\bibfnamefont {F.}~\bibnamefont {Ferlaino}},\ }\href
  {\doibase 10.1126/science.aac9812} {\bibfield  {journal} {\bibinfo  {journal}
  {Science}\ }\textbf {\bibinfo {volume} {352}},\ \bibinfo {pages} {201}
  (\bibinfo {year} {2016})}\BibitemShut {NoStop}%
\bibitem [{\citenamefont {Zhang}\ \emph {et~al.}(2017)\citenamefont {Zhang},
  \citenamefont {Pagano}, \citenamefont {Hess}, \citenamefont {Kyprianidis},
  \citenamefont {Becker}, \citenamefont {Kaplan}, \citenamefont {Gorshkov},
  \citenamefont {Gong},\ and\ \citenamefont {Monroe}}]{Zhang_2017}%
  \BibitemOpen
  \bibfield  {author} {\bibinfo {author} {\bibfnamefont {J.}~\bibnamefont
  {Zhang}}, \bibinfo {author} {\bibfnamefont {G.}~\bibnamefont {Pagano}},
  \bibinfo {author} {\bibfnamefont {P.~W.}\ \bibnamefont {Hess}}, \bibinfo
  {author} {\bibfnamefont {A.}~\bibnamefont {Kyprianidis}}, \bibinfo {author}
  {\bibfnamefont {P.}~\bibnamefont {Becker}}, \bibinfo {author} {\bibfnamefont
  {H.}~\bibnamefont {Kaplan}}, \bibinfo {author} {\bibfnamefont {A.~V.}\
  \bibnamefont {Gorshkov}}, \bibinfo {author} {\bibfnamefont {Z.-X.}\
  \bibnamefont {Gong}}, \ and\ \bibinfo {author} {\bibfnamefont
  {C.}~\bibnamefont {Monroe}},\ }\href {\doibase 10.1038/nature24654}
  {\bibfield  {journal} {\bibinfo  {journal} {Nature}\ }\textbf {\bibinfo
  {volume} {551}},\ \bibinfo {pages} {601} (\bibinfo {year}
  {2017})}\BibitemShut {NoStop}%
\bibitem [{\citenamefont {Bernien}\ \emph {et~al.}(2017)\citenamefont
  {Bernien}, \citenamefont {Schwartz}, \citenamefont {Keesling}, \citenamefont
  {Levine}, \citenamefont {Omran}, \citenamefont {Pichler}, \citenamefont
  {Choi}, \citenamefont {Zibrov}, \citenamefont {Endres}, \citenamefont
  {Greiner}, \citenamefont {Vuleti{\'{c}}},\ and\ \citenamefont
  {Lukin}}]{Bernien_2017}%
  \BibitemOpen
  \bibfield  {author} {\bibinfo {author} {\bibfnamefont {H.}~\bibnamefont
  {Bernien}}, \bibinfo {author} {\bibfnamefont {S.}~\bibnamefont {Schwartz}},
  \bibinfo {author} {\bibfnamefont {A.}~\bibnamefont {Keesling}}, \bibinfo
  {author} {\bibfnamefont {H.}~\bibnamefont {Levine}}, \bibinfo {author}
  {\bibfnamefont {A.}~\bibnamefont {Omran}}, \bibinfo {author} {\bibfnamefont
  {H.}~\bibnamefont {Pichler}}, \bibinfo {author} {\bibfnamefont
  {S.}~\bibnamefont {Choi}}, \bibinfo {author} {\bibfnamefont {A.~S.}\
  \bibnamefont {Zibrov}}, \bibinfo {author} {\bibfnamefont {M.}~\bibnamefont
  {Endres}}, \bibinfo {author} {\bibfnamefont {M.}~\bibnamefont {Greiner}},
  \bibinfo {author} {\bibfnamefont {V.}~\bibnamefont {Vuleti{\'{c}}}}, \ and\
  \bibinfo {author} {\bibfnamefont {M.~D.}\ \bibnamefont {Lukin}},\ }\href
  {\doibase 10.1038/nature24622} {\bibfield  {journal} {\bibinfo  {journal}
  {Nature}\ }\textbf {\bibinfo {volume} {551}},\ \bibinfo {pages} {579}
  (\bibinfo {year} {2017})}\BibitemShut {NoStop}%
\bibitem [{\citenamefont {Choi}\ \emph {et~al.}(2016)\citenamefont {Choi},
  \citenamefont {Hild}, \citenamefont {Zeiher}, \citenamefont {Schau{\ss}},
  \citenamefont {Rubio-Abadal}, \citenamefont {Yefsah}, \citenamefont
  {Khemani}, \citenamefont {Huse}, \citenamefont {Bloch},\ and\ \citenamefont
  {Gross}}]{Choi_2016}%
  \BibitemOpen
  \bibfield  {author} {\bibinfo {author} {\bibfnamefont {J.-y.}\ \bibnamefont
  {Choi}}, \bibinfo {author} {\bibfnamefont {S.}~\bibnamefont {Hild}}, \bibinfo
  {author} {\bibfnamefont {J.}~\bibnamefont {Zeiher}}, \bibinfo {author}
  {\bibfnamefont {P.}~\bibnamefont {Schau{\ss}}}, \bibinfo {author}
  {\bibfnamefont {A.}~\bibnamefont {Rubio-Abadal}}, \bibinfo {author}
  {\bibfnamefont {T.}~\bibnamefont {Yefsah}}, \bibinfo {author} {\bibfnamefont
  {V.}~\bibnamefont {Khemani}}, \bibinfo {author} {\bibfnamefont {D.~A.}\
  \bibnamefont {Huse}}, \bibinfo {author} {\bibfnamefont {I.}~\bibnamefont
  {Bloch}}, \ and\ \bibinfo {author} {\bibfnamefont {C.}~\bibnamefont
  {Gross}},\ }\href {\doibase 10.1126/science.aaf8834} {\bibfield  {journal}
  {\bibinfo  {journal} {Science}\ }\textbf {\bibinfo {volume} {352}},\ \bibinfo
  {pages} {1547} (\bibinfo {year} {2016})}\BibitemShut {NoStop}%
\bibitem [{\citenamefont {Fukuhara}\ \emph
  {et~al.}(2013{\natexlab{a}})\citenamefont {Fukuhara}, \citenamefont
  {Schau{\ss}}, \citenamefont {Endres}, \citenamefont {Hild}, \citenamefont
  {Cheneau}, \citenamefont {Bloch},\ and\ \citenamefont
  {Gross}}]{Fukuhara_2013a}%
  \BibitemOpen
  \bibfield  {author} {\bibinfo {author} {\bibfnamefont {T.}~\bibnamefont
  {Fukuhara}}, \bibinfo {author} {\bibfnamefont {P.}~\bibnamefont
  {Schau{\ss}}}, \bibinfo {author} {\bibfnamefont {M.}~\bibnamefont {Endres}},
  \bibinfo {author} {\bibfnamefont {S.}~\bibnamefont {Hild}}, \bibinfo {author}
  {\bibfnamefont {M.}~\bibnamefont {Cheneau}}, \bibinfo {author} {\bibfnamefont
  {I.}~\bibnamefont {Bloch}}, \ and\ \bibinfo {author} {\bibfnamefont
  {C.}~\bibnamefont {Gross}},\ }\href {\doibase 10.1038/nature12541} {\bibfield
   {journal} {\bibinfo  {journal} {Nature}\ }\textbf {\bibinfo {volume}
  {502}},\ \bibinfo {pages} {76} (\bibinfo {year}
  {2013}{\natexlab{a}})}\BibitemShut {NoStop}%
\bibitem [{\citenamefont {Fukuhara}\ \emph
  {et~al.}(2013{\natexlab{b}})\citenamefont {Fukuhara}, \citenamefont
  {Kantian}, \citenamefont {Endres}, \citenamefont {Cheneau}, \citenamefont
  {Schau{\ss}}, \citenamefont {Hild}, \citenamefont {Bellem}, \citenamefont
  {Schollw{\"o}ck}, \citenamefont {Giamarchi}, \citenamefont {Gross},
  \citenamefont {Bloch},\ and\ \citenamefont {Kuhr}}]{Fukuhara_2013b}%
  \BibitemOpen
  \bibfield  {author} {\bibinfo {author} {\bibfnamefont {T.}~\bibnamefont
  {Fukuhara}}, \bibinfo {author} {\bibfnamefont {A.}~\bibnamefont {Kantian}},
  \bibinfo {author} {\bibfnamefont {M.}~\bibnamefont {Endres}}, \bibinfo
  {author} {\bibfnamefont {M.}~\bibnamefont {Cheneau}}, \bibinfo {author}
  {\bibfnamefont {P.}~\bibnamefont {Schau{\ss}}}, \bibinfo {author}
  {\bibfnamefont {S.}~\bibnamefont {Hild}}, \bibinfo {author} {\bibfnamefont
  {D.}~\bibnamefont {Bellem}}, \bibinfo {author} {\bibfnamefont
  {U.}~\bibnamefont {Schollw{\"o}ck}}, \bibinfo {author} {\bibfnamefont
  {T.}~\bibnamefont {Giamarchi}}, \bibinfo {author} {\bibfnamefont
  {C.}~\bibnamefont {Gross}}, \bibinfo {author} {\bibfnamefont
  {I.}~\bibnamefont {Bloch}}, \ and\ \bibinfo {author} {\bibfnamefont
  {S.}~\bibnamefont {Kuhr}},\ }\href {\doibase 10.1038/nphys2561} {\bibfield
  {journal} {\bibinfo  {journal} {Nat. Phys.}\ }\textbf {\bibinfo {volume}
  {9}},\ \bibinfo {pages} {235} (\bibinfo {year}
  {2013}{\natexlab{b}})}\BibitemShut {NoStop}%
\bibitem [{\citenamefont {Greiner}\ \emph {et~al.}(2002)\citenamefont
  {Greiner}, \citenamefont {Mandel}, \citenamefont {Esslinger}, \citenamefont
  {H{\"a}nsch},\ and\ \citenamefont {Bloch}}]{Greiner_2002a}%
  \BibitemOpen
  \bibfield  {author} {\bibinfo {author} {\bibfnamefont {M.}~\bibnamefont
  {Greiner}}, \bibinfo {author} {\bibfnamefont {O.}~\bibnamefont {Mandel}},
  \bibinfo {author} {\bibfnamefont {T.}~\bibnamefont {Esslinger}}, \bibinfo
  {author} {\bibfnamefont {T.~W.}\ \bibnamefont {H{\"a}nsch}}, \ and\ \bibinfo
  {author} {\bibfnamefont {I.}~\bibnamefont {Bloch}},\ }\href {\doibase
  10.1038/415039a} {\bibfield  {journal} {\bibinfo  {journal} {Nature}\
  }\textbf {\bibinfo {volume} {415}},\ \bibinfo {pages} {39} (\bibinfo {year}
  {2002})}\BibitemShut {NoStop}%
\bibitem [{\citenamefont {Mazurenko}\ \emph {et~al.}(2017)\citenamefont
  {Mazurenko}, \citenamefont {Chiu}, \citenamefont {Ji}, \citenamefont
  {Parsons}, \citenamefont {Kan{\'a}sz-Nagy}, \citenamefont {Schmidt},
  \citenamefont {Grusdt}, \citenamefont {Demler}, \citenamefont {Greif},\ and\
  \citenamefont {Greiner}}]{Mazurenko_2017}%
  \BibitemOpen
  \bibfield  {author} {\bibinfo {author} {\bibfnamefont {A.}~\bibnamefont
  {Mazurenko}}, \bibinfo {author} {\bibfnamefont {C.~S.}\ \bibnamefont {Chiu}},
  \bibinfo {author} {\bibfnamefont {G.}~\bibnamefont {Ji}}, \bibinfo {author}
  {\bibfnamefont {M.~F.}\ \bibnamefont {Parsons}}, \bibinfo {author}
  {\bibfnamefont {M.}~\bibnamefont {Kan{\'a}sz-Nagy}}, \bibinfo {author}
  {\bibfnamefont {R.}~\bibnamefont {Schmidt}}, \bibinfo {author} {\bibfnamefont
  {F.}~\bibnamefont {Grusdt}}, \bibinfo {author} {\bibfnamefont
  {E.}~\bibnamefont {Demler}}, \bibinfo {author} {\bibfnamefont
  {D.}~\bibnamefont {Greif}}, \ and\ \bibinfo {author} {\bibfnamefont
  {M.}~\bibnamefont {Greiner}},\ }\href {\doibase 10.1038/nature22362}
  {\bibfield  {journal} {\bibinfo  {journal} {Nature}\ }\textbf {\bibinfo
  {volume} {545}},\ \bibinfo {pages} {462} (\bibinfo {year}
  {2017})}\BibitemShut {NoStop}%
\bibitem [{\citenamefont {Chiu}\ \emph {et~al.}(2019)\citenamefont {Chiu},
  \citenamefont {Ji}, \citenamefont {Bohrdt}, \citenamefont {Xu}, \citenamefont
  {Knap}, \citenamefont {Demler}, \citenamefont {Grusdt}, \citenamefont
  {Greiner},\ and\ \citenamefont {Greif}}]{Chiu_2019}%
  \BibitemOpen
  \bibfield  {author} {\bibinfo {author} {\bibfnamefont {C.~S.}\ \bibnamefont
  {Chiu}}, \bibinfo {author} {\bibfnamefont {G.}~\bibnamefont {Ji}}, \bibinfo
  {author} {\bibfnamefont {A.}~\bibnamefont {Bohrdt}}, \bibinfo {author}
  {\bibfnamefont {M.}~\bibnamefont {Xu}}, \bibinfo {author} {\bibfnamefont
  {M.}~\bibnamefont {Knap}}, \bibinfo {author} {\bibfnamefont {E.}~\bibnamefont
  {Demler}}, \bibinfo {author} {\bibfnamefont {F.}~\bibnamefont {Grusdt}},
  \bibinfo {author} {\bibfnamefont {M.}~\bibnamefont {Greiner}}, \ and\
  \bibinfo {author} {\bibfnamefont {D.}~\bibnamefont {Greif}},\ }\href
  {\doibase 10.1126/science.aav3587} {\bibfield  {journal} {\bibinfo  {journal}
  {Science}\ }\textbf {\bibinfo {volume} {365}},\ \bibinfo {pages} {251}
  (\bibinfo {year} {2019})}\BibitemShut {NoStop}%
\bibitem [{\citenamefont {Ebadi}\ \emph {et~al.}(2021)\citenamefont {Ebadi},
  \citenamefont {Wang}, \citenamefont {Levine}, \citenamefont {Keesling},
  \citenamefont {Semeghini}, \citenamefont {Omran}, \citenamefont {Bluvstein},
  \citenamefont {Samajdar}, \citenamefont {Pichler}, \citenamefont {Ho},
  \citenamefont {Choi}, \citenamefont {Sachdev}, \citenamefont {Greiner},
  \citenamefont {Vuleti{\'{c}}},\ and\ \citenamefont {Lukin}}]{Ebadi_2021}%
  \BibitemOpen
  \bibfield  {author} {\bibinfo {author} {\bibfnamefont {S.}~\bibnamefont
  {Ebadi}}, \bibinfo {author} {\bibfnamefont {T.~T.}\ \bibnamefont {Wang}},
  \bibinfo {author} {\bibfnamefont {H.}~\bibnamefont {Levine}}, \bibinfo
  {author} {\bibfnamefont {A.}~\bibnamefont {Keesling}}, \bibinfo {author}
  {\bibfnamefont {G.}~\bibnamefont {Semeghini}}, \bibinfo {author}
  {\bibfnamefont {A.}~\bibnamefont {Omran}}, \bibinfo {author} {\bibfnamefont
  {D.}~\bibnamefont {Bluvstein}}, \bibinfo {author} {\bibfnamefont
  {R.}~\bibnamefont {Samajdar}}, \bibinfo {author} {\bibfnamefont
  {H.}~\bibnamefont {Pichler}}, \bibinfo {author} {\bibfnamefont {W.~W.}\
  \bibnamefont {Ho}}, \bibinfo {author} {\bibfnamefont {S.}~\bibnamefont
  {Choi}}, \bibinfo {author} {\bibfnamefont {S.}~\bibnamefont {Sachdev}},
  \bibinfo {author} {\bibfnamefont {M.}~\bibnamefont {Greiner}}, \bibinfo
  {author} {\bibfnamefont {V.}~\bibnamefont {Vuleti{\'{c}}}}, \ and\ \bibinfo
  {author} {\bibfnamefont {M.~D.}\ \bibnamefont {Lukin}},\ }\href {\doibase
  10.1038/s41586-021-03582-4} {\bibfield  {journal} {\bibinfo  {journal}
  {Nature}\ }\textbf {\bibinfo {volume} {595}},\ \bibinfo {pages} {227}
  (\bibinfo {year} {2021})}\BibitemShut {NoStop}%
\bibitem [{\citenamefont {Mistakidis}\ \emph {et~al.}(2022)\citenamefont
  {Mistakidis}, \citenamefont {Volosniev}, \citenamefont {Barfknecht},
  \citenamefont {Fogarty}, \citenamefont {Busch}, \citenamefont {Foerster},
  \citenamefont {Schmelcher},\ and\ \citenamefont {Zinner}}]{Mistakidis2022}%
  \BibitemOpen
  \bibfield  {author} {\bibinfo {author} {\bibfnamefont {S.}~\bibnamefont
  {Mistakidis}}, \bibinfo {author} {\bibfnamefont {A.}~\bibnamefont
  {Volosniev}}, \bibinfo {author} {\bibfnamefont {R.}~\bibnamefont
  {Barfknecht}}, \bibinfo {author} {\bibfnamefont {T.}~\bibnamefont {Fogarty}},
  \bibinfo {author} {\bibfnamefont {T.}~\bibnamefont {Busch}}, \bibinfo
  {author} {\bibfnamefont {A.}~\bibnamefont {Foerster}}, \bibinfo {author}
  {\bibfnamefont {P.}~\bibnamefont {Schmelcher}}, \ and\ \bibinfo {author}
  {\bibfnamefont {N.}~\bibnamefont {Zinner}},\ }\href
  {https://doi.org/10.48550/arXiv.2202.11071} {\bibfield  {journal} {\bibinfo
  {journal} {arXiv:2202.11071}\ } (\bibinfo {year} {2022})}\BibitemShut
  {NoStop}%
\bibitem [{\citenamefont {Sowi{\'{n}}ski}\ and\ \citenamefont
  {Garc{\'{\i}}a-March}(2019)}]{Sowinski_2019}%
  \BibitemOpen
  \bibfield  {author} {\bibinfo {author} {\bibfnamefont {T.}~\bibnamefont
  {Sowi{\'{n}}ski}}\ and\ \bibinfo {author} {\bibfnamefont {M.~{\'{A}}.}\
  \bibnamefont {Garc{\'{\i}}a-March}},\ }\href {\doibase
  10.1088/1361-6633/ab3a80} {\bibfield  {journal} {\bibinfo  {journal} {Rep.
  Prog. Phys.}\ }\textbf {\bibinfo {volume} {82}},\ \bibinfo {pages} {104401}
  (\bibinfo {year} {2019})}\BibitemShut {NoStop}%
\bibitem [{\citenamefont {Cazalilla}\ \emph {et~al.}(2011)\citenamefont
  {Cazalilla}, \citenamefont {Citro}, \citenamefont {Giamarchi}, \citenamefont
  {Orignac},\ and\ \citenamefont {Rigol}}]{Cazalilla_2011}%
  \BibitemOpen
  \bibfield  {author} {\bibinfo {author} {\bibfnamefont {M.~A.}\ \bibnamefont
  {Cazalilla}}, \bibinfo {author} {\bibfnamefont {R.}~\bibnamefont {Citro}},
  \bibinfo {author} {\bibfnamefont {T.}~\bibnamefont {Giamarchi}}, \bibinfo
  {author} {\bibfnamefont {E.}~\bibnamefont {Orignac}}, \ and\ \bibinfo
  {author} {\bibfnamefont {M.}~\bibnamefont {Rigol}},\ }\href {\doibase
  10.1103/RevModPhys.83.1405} {\bibfield  {journal} {\bibinfo  {journal} {Rev.
  Mod. Phys.}\ }\textbf {\bibinfo {volume} {83}},\ \bibinfo {pages} {1405}
  (\bibinfo {year} {2011})}\BibitemShut {NoStop}%
\bibitem [{\citenamefont {Busch}\ \emph {et~al.}(1998)\citenamefont {Busch},
  \citenamefont {Englert}, \citenamefont {Rzażewski},\ and\ \citenamefont
  {Wilkens}}]{Busch_1998}%
  \BibitemOpen
  \bibfield  {author} {\bibinfo {author} {\bibfnamefont {T.}~\bibnamefont
  {Busch}}, \bibinfo {author} {\bibfnamefont {B.-G.}\ \bibnamefont {Englert}},
  \bibinfo {author} {\bibfnamefont {K.}~\bibnamefont {Rzażewski}}, \ and\
  \bibinfo {author} {\bibfnamefont {M.}~\bibnamefont {Wilkens}},\ }\href
  {\doibase 10.1023/A:1018705520999} {\bibfield  {journal} {\bibinfo  {journal}
  {Found. Phys.}\ }\textbf {\bibinfo {volume} {28}},\ \bibinfo {pages} {549}
  (\bibinfo {year} {1998})}\BibitemShut {NoStop}%
\bibitem [{\citenamefont {Ko{\'{s}}cik}\ and\ \citenamefont
  {Sowi{\'{n}}ski}(2018)}]{Koscik_2018}%
  \BibitemOpen
  \bibfield  {author} {\bibinfo {author} {\bibfnamefont {P.}~\bibnamefont
  {Ko{\'{s}}cik}}\ and\ \bibinfo {author} {\bibfnamefont {T.}~\bibnamefont
  {Sowi{\'{n}}ski}},\ }\href {\doibase 10.1038/s41598-017-18505-5} {\bibfield
  {journal} {\bibinfo  {journal} {Sci. Rep.}\ }\textbf {\bibinfo {volume}
  {8}},\ \bibinfo {pages} {48} (\bibinfo {year} {2018})}\BibitemShut {NoStop}%
\bibitem [{\citenamefont {Shea}\ \emph {et~al.}(2009)\citenamefont {Shea},
  \citenamefont {van Zyl},\ and\ \citenamefont {Bhaduri}}]{shea_2009}%
  \BibitemOpen
  \bibfield  {author} {\bibinfo {author} {\bibfnamefont {P.}~\bibnamefont
  {Shea}}, \bibinfo {author} {\bibfnamefont {B.~P.}\ \bibnamefont {van Zyl}}, \
  and\ \bibinfo {author} {\bibfnamefont {R.~K.}\ \bibnamefont {Bhaduri}},\
  }\href {\doibase 10.1119/1.3013812} {\bibfield  {journal} {\bibinfo
  {journal} {Am. J. Phys.}\ }\textbf {\bibinfo {volume} {77}},\ \bibinfo
  {pages} {511} (\bibinfo {year} {2009})}\BibitemShut {NoStop}%
\bibitem [{\citenamefont {Sowiński}\ \emph {et~al.}(2010)\citenamefont
  {Sowiński}, \citenamefont {Brewczyk}, \citenamefont {Gajda},\ and\
  \citenamefont {Rzażewski}}]{Sowinski_2010}%
  \BibitemOpen
  \bibfield  {author} {\bibinfo {author} {\bibfnamefont {T.}~\bibnamefont
  {Sowiński}}, \bibinfo {author} {\bibfnamefont {M.}~\bibnamefont {Brewczyk}},
  \bibinfo {author} {\bibfnamefont {M.}~\bibnamefont {Gajda}}, \ and\ \bibinfo
  {author} {\bibfnamefont {K.}~\bibnamefont {Rzażewski}},\ }\href {\doibase
  10.1103/PhysRevA.82.053631} {\bibfield  {journal} {\bibinfo  {journal} {Phys.
  Rev. A}\ }\textbf {\bibinfo {volume} {82}},\ \bibinfo {pages} {053631}
  (\bibinfo {year} {2010})}\BibitemShut {NoStop}%
\bibitem [{\citenamefont {O{\l}dziejewski}\ \emph {et~al.}(2016)\citenamefont
  {O{\l}dziejewski}, \citenamefont {G{\'{o}}recki},\ and\ \citenamefont
  {Rz{\k{a}}{\.{z}}ewski}}]{Oldziejewski_2016}%
  \BibitemOpen
  \bibfield  {author} {\bibinfo {author} {\bibfnamefont {R.}~\bibnamefont
  {O{\l}dziejewski}}, \bibinfo {author} {\bibfnamefont {W.}~\bibnamefont
  {G{\'{o}}recki}}, \ and\ \bibinfo {author} {\bibfnamefont {K.}~\bibnamefont
  {Rz{\k{a}}{\.{z}}ewski}},\ }\href {\doibase 10.1209/0295-5075/114/46003}
  {\bibfield  {journal} {\bibinfo  {journal} {{EPL}}\ }\textbf {\bibinfo
  {volume} {114}},\ \bibinfo {pages} {46003} (\bibinfo {year}
  {2016})}\BibitemShut {NoStop}%
\bibitem [{\citenamefont {G{\'{o}}recki}\ and\ \citenamefont
  {Rz{\k{a}}{\.{z}}ewski}(2016)}]{Gorecki_2016}%
  \BibitemOpen
  \bibfield  {author} {\bibinfo {author} {\bibfnamefont {W.}~\bibnamefont
  {G{\'{o}}recki}}\ and\ \bibinfo {author} {\bibfnamefont {K.}~\bibnamefont
  {Rz{\k{a}}{\.{z}}ewski}},\ }\href {\doibase 10.1209/0295-5075/116/26004}
  {\bibfield  {journal} {\bibinfo  {journal} {{EPL}}\ }\textbf {\bibinfo
  {volume} {116}},\ \bibinfo {pages} {26004} (\bibinfo {year}
  {2016})}\BibitemShut {NoStop}%
\bibitem [{\citenamefont {Budewig}\ \emph {et~al.}(2019)\citenamefont
  {Budewig}, \citenamefont {Mistakidis},\ and\ \citenamefont
  {Schmelcher}}]{Budewig_2019}%
  \BibitemOpen
  \bibfield  {author} {\bibinfo {author} {\bibfnamefont {L.}~\bibnamefont
  {Budewig}}, \bibinfo {author} {\bibfnamefont {S.~I.}\ \bibnamefont
  {Mistakidis}}, \ and\ \bibinfo {author} {\bibfnamefont {P.}~\bibnamefont
  {Schmelcher}},\ }\href {\doibase 10.1080/00268976.2019.1575995} {\bibfield
  {journal} {\bibinfo  {journal} {Mol. Phys.}\ }\textbf {\bibinfo {volume}
  {117}},\ \bibinfo {pages} {2043} (\bibinfo {year} {2019})}\BibitemShut
  {NoStop}%
\bibitem [{\citenamefont {G{\'{o}}recki}\ and\ \citenamefont
  {Rzą{\.{z}}ewski}(2017)}]{Gorecki_2017}%
  \BibitemOpen
  \bibfield  {author} {\bibinfo {author} {\bibfnamefont {W.}~\bibnamefont
  {G{\'{o}}recki}}\ and\ \bibinfo {author} {\bibfnamefont {K.}~\bibnamefont
  {Rzą{\.{z}}ewski}},\ }\href {\doibase 10.1209/0295-5075/118/66002}
  {\bibfield  {journal} {\bibinfo  {journal} {{EPL}}\ }\textbf {\bibinfo
  {volume} {118}},\ \bibinfo {pages} {66002} (\bibinfo {year}
  {2017})}\BibitemShut {NoStop}%
\bibitem [{\citenamefont {Dawid}\ \emph {et~al.}(2018)\citenamefont {Dawid},
  \citenamefont {Lewenstein},\ and\ \citenamefont {Tomza}}]{Dawid_2018}%
  \BibitemOpen
  \bibfield  {author} {\bibinfo {author} {\bibfnamefont {A.}~\bibnamefont
  {Dawid}}, \bibinfo {author} {\bibfnamefont {M.}~\bibnamefont {Lewenstein}}, \
  and\ \bibinfo {author} {\bibfnamefont {M.}~\bibnamefont {Tomza}},\ }\href
  {\doibase 10.1103/PhysRevA.97.063618} {\bibfield  {journal} {\bibinfo
  {journal} {Phys. Rev. A}\ }\textbf {\bibinfo {volume} {97}},\ \bibinfo
  {pages} {063618} (\bibinfo {year} {2018})}\BibitemShut {NoStop}%
\bibitem [{\citenamefont {Dawid}\ and\ \citenamefont
  {Tomza}(2020)}]{Dawid_2020}%
  \BibitemOpen
  \bibfield  {author} {\bibinfo {author} {\bibfnamefont {A.}~\bibnamefont
  {Dawid}}\ and\ \bibinfo {author} {\bibfnamefont {M.}~\bibnamefont {Tomza}},\
  }\href {\doibase 10.1039/D0CP05542E} {\bibfield  {journal} {\bibinfo
  {journal} {Phys. Chem. Chem. Phys.}\ }\textbf {\bibinfo {volume} {22}},\
  \bibinfo {pages} {28140} (\bibinfo {year} {2020})}\BibitemShut {NoStop}%
\bibitem [{\citenamefont {Sroczy{\'{n}}ska}\ \emph {et~al.}(2022)\citenamefont
  {Sroczy{\'{n}}ska}, \citenamefont {Dawid}, \citenamefont {Tomza},
  \citenamefont {Idziaszek}, \citenamefont {Calarco},\ and\ \citenamefont
  {Jachymski}}]{SroczyskaNJP22}%
  \BibitemOpen
  \bibfield  {author} {\bibinfo {author} {\bibfnamefont {M.}~\bibnamefont
  {Sroczy{\'{n}}ska}}, \bibinfo {author} {\bibfnamefont {A.}~\bibnamefont
  {Dawid}}, \bibinfo {author} {\bibfnamefont {M.}~\bibnamefont {Tomza}},
  \bibinfo {author} {\bibfnamefont {Z.}~\bibnamefont {Idziaszek}}, \bibinfo
  {author} {\bibfnamefont {T.}~\bibnamefont {Calarco}}, \ and\ \bibinfo
  {author} {\bibfnamefont {K.}~\bibnamefont {Jachymski}},\ }\href {\doibase
  10.1088/1367-2630/ac434b} {\bibfield  {journal} {\bibinfo  {journal} {New J.
  Phys.}\ }\textbf {\bibinfo {volume} {24}},\ \bibinfo {pages} {015001}
  (\bibinfo {year} {2022})}\BibitemShut {NoStop}%
\bibitem [{\citenamefont {Sowi{\'n}ski}\ \emph {et~al.}(2013)\citenamefont
  {Sowi{\'n}ski}, \citenamefont {Grass}, \citenamefont {Dutta},\ and\
  \citenamefont {Lewenstein}}]{Sowinski_2013}%
  \BibitemOpen
  \bibfield  {author} {\bibinfo {author} {\bibfnamefont {T.}~\bibnamefont
  {Sowi{\'n}ski}}, \bibinfo {author} {\bibfnamefont {T.}~\bibnamefont {Grass}},
  \bibinfo {author} {\bibfnamefont {O.}~\bibnamefont {Dutta}}, \ and\ \bibinfo
  {author} {\bibfnamefont {M.}~\bibnamefont {Lewenstein}},\ }\href {\doibase
  10.1103/PhysRevA.88.033607} {\bibfield  {journal} {\bibinfo  {journal} {Phys.
  Rev. A}\ }\textbf {\bibinfo {volume} {88}},\ \bibinfo {pages} {033607}
  (\bibinfo {year} {2013})}\BibitemShut {NoStop}%
\bibitem [{\citenamefont {Rojo-Francàs}\ \emph {et~al.}(2020)\citenamefont
  {Rojo-Francàs}, \citenamefont {Polls},\ and\ \citenamefont
  {Juliá-Díaz}}]{Rojo-Francas_2020}%
  \BibitemOpen
  \bibfield  {author} {\bibinfo {author} {\bibfnamefont {A.}~\bibnamefont
  {Rojo-Francàs}}, \bibinfo {author} {\bibfnamefont {A.}~\bibnamefont
  {Polls}}, \ and\ \bibinfo {author} {\bibfnamefont {B.}~\bibnamefont
  {Juliá-Díaz}},\ }\href {\doibase 10.3390/math8071196} {\bibfield  {journal}
  {\bibinfo  {journal} {Mathematics}\ }\textbf {\bibinfo {volume} {8}},\
  \bibinfo {pages} {1196} (\bibinfo {year} {2020})}\BibitemShut {NoStop}%
\bibitem [{\citenamefont {Peng}\ and\ \citenamefont {Cui}(2020)}]{Peng_2020}%
  \BibitemOpen
  \bibfield  {author} {\bibinfo {author} {\bibfnamefont {C.}~\bibnamefont
  {Peng}}\ and\ \bibinfo {author} {\bibfnamefont {X.}~\bibnamefont {Cui}},\
  }\href {\doibase 10.1103/PhysRevA.102.033312} {\bibfield  {journal} {\bibinfo
   {journal} {Phys. Rev. A}\ }\textbf {\bibinfo {volume} {102}},\ \bibinfo
  {pages} {033312} (\bibinfo {year} {2020})}\BibitemShut {NoStop}%
\bibitem [{\citenamefont {Laird}\ \emph {et~al.}(2017)\citenamefont {Laird},
  \citenamefont {Shi}, \citenamefont {Parish},\ and\ \citenamefont
  {Levinsen}}]{Laird_2017}%
  \BibitemOpen
  \bibfield  {author} {\bibinfo {author} {\bibfnamefont {E.~K.}\ \bibnamefont
  {Laird}}, \bibinfo {author} {\bibfnamefont {Z.-Y.}\ \bibnamefont {Shi}},
  \bibinfo {author} {\bibfnamefont {M.~M.}\ \bibnamefont {Parish}}, \ and\
  \bibinfo {author} {\bibfnamefont {J.}~\bibnamefont {Levinsen}},\ }\href
  {\doibase 10.1103/PhysRevA.96.032701} {\bibfield  {journal} {\bibinfo
  {journal} {Phys. Rev. A}\ }\textbf {\bibinfo {volume} {96}},\ \bibinfo
  {pages} {032701} (\bibinfo {year} {2017})}\BibitemShut {NoStop}%
\bibitem [{\citenamefont {Deuretzbacher}\ \emph {et~al.}(2014)\citenamefont
  {Deuretzbacher}, \citenamefont {Becker}, \citenamefont {Bjerlin},
  \citenamefont {Reimann},\ and\ \citenamefont {Santos}}]{Deuretzbacher_2014}%
  \BibitemOpen
  \bibfield  {author} {\bibinfo {author} {\bibfnamefont {F.}~\bibnamefont
  {Deuretzbacher}}, \bibinfo {author} {\bibfnamefont {D.}~\bibnamefont
  {Becker}}, \bibinfo {author} {\bibfnamefont {J.}~\bibnamefont {Bjerlin}},
  \bibinfo {author} {\bibfnamefont {S.~M.}\ \bibnamefont {Reimann}}, \ and\
  \bibinfo {author} {\bibfnamefont {L.}~\bibnamefont {Santos}},\ }\href
  {\doibase 10.1103/PhysRevA.90.013611} {\bibfield  {journal} {\bibinfo
  {journal} {Phys. Rev. A}\ }\textbf {\bibinfo {volume} {90}},\ \bibinfo
  {pages} {013611} (\bibinfo {year} {2014})}\BibitemShut {NoStop}%
\bibitem [{\citenamefont {Volosniev}\ \emph {et~al.}(2014)\citenamefont
  {Volosniev}, \citenamefont {Fedorov}, \citenamefont {Jensen}, \citenamefont
  {Valiente},\ and\ \citenamefont {Zinner}}]{Volosniev_2014}%
  \BibitemOpen
  \bibfield  {author} {\bibinfo {author} {\bibfnamefont {A.~G.}\ \bibnamefont
  {Volosniev}}, \bibinfo {author} {\bibfnamefont {D.~V.}\ \bibnamefont
  {Fedorov}}, \bibinfo {author} {\bibfnamefont {A.~S.}\ \bibnamefont {Jensen}},
  \bibinfo {author} {\bibfnamefont {M.}~\bibnamefont {Valiente}}, \ and\
  \bibinfo {author} {\bibfnamefont {N.~T.}\ \bibnamefont {Zinner}},\ }\href
  {\doibase 10.1038/ncomms6300} {\bibfield  {journal} {\bibinfo  {journal}
  {Nat. Commun.}\ }\textbf {\bibinfo {volume} {5}},\ \bibinfo {pages} {5300}
  (\bibinfo {year} {2014})}\BibitemShut {NoStop}%
\bibitem [{\citenamefont {Grining}\ \emph
  {et~al.}(2015{\natexlab{a}})\citenamefont {Grining}, \citenamefont {Tomza},
  \citenamefont {Lesiuk}, \citenamefont {Przybytek}, \citenamefont {Musia{\l}},
  \citenamefont {Massignan}, \citenamefont {Lewenstein},\ and\ \citenamefont
  {Moszynski}}]{Grining_2015}%
  \BibitemOpen
  \bibfield  {author} {\bibinfo {author} {\bibfnamefont {T.}~\bibnamefont
  {Grining}}, \bibinfo {author} {\bibfnamefont {M.}~\bibnamefont {Tomza}},
  \bibinfo {author} {\bibfnamefont {M.}~\bibnamefont {Lesiuk}}, \bibinfo
  {author} {\bibfnamefont {M.}~\bibnamefont {Przybytek}}, \bibinfo {author}
  {\bibfnamefont {M.}~\bibnamefont {Musia{\l}}}, \bibinfo {author}
  {\bibfnamefont {P.}~\bibnamefont {Massignan}}, \bibinfo {author}
  {\bibfnamefont {M.}~\bibnamefont {Lewenstein}}, \ and\ \bibinfo {author}
  {\bibfnamefont {R.}~\bibnamefont {Moszynski}},\ }\href {\doibase
  10.1088/1367-2630/17/11/115001} {\bibfield  {journal} {\bibinfo  {journal}
  {New J. Phys.}\ }\textbf {\bibinfo {volume} {17}},\ \bibinfo {pages} {115001}
  (\bibinfo {year} {2015}{\natexlab{a}})}\BibitemShut {NoStop}%
\bibitem [{\citenamefont {Z\"urn}\ \emph {et~al.}(2012)\citenamefont {Z\"urn},
  \citenamefont {Serwane}, \citenamefont {Lompe}, \citenamefont {Wenz},
  \citenamefont {Ries}, \citenamefont {Bohn},\ and\ \citenamefont
  {Jochim}}]{Zurn_2012}%
  \BibitemOpen
  \bibfield  {author} {\bibinfo {author} {\bibfnamefont {G.}~\bibnamefont
  {Z\"urn}}, \bibinfo {author} {\bibfnamefont {F.}~\bibnamefont {Serwane}},
  \bibinfo {author} {\bibfnamefont {T.}~\bibnamefont {Lompe}}, \bibinfo
  {author} {\bibfnamefont {A.~N.}\ \bibnamefont {Wenz}}, \bibinfo {author}
  {\bibfnamefont {M.~G.}\ \bibnamefont {Ries}}, \bibinfo {author}
  {\bibfnamefont {J.~E.}\ \bibnamefont {Bohn}}, \ and\ \bibinfo {author}
  {\bibfnamefont {S.}~\bibnamefont {Jochim}},\ }\href {\doibase
  10.1103/PhysRevLett.108.075303} {\bibfield  {journal} {\bibinfo  {journal}
  {Phys. Rev. Lett.}\ }\textbf {\bibinfo {volume} {108}},\ \bibinfo {pages}
  {075303} (\bibinfo {year} {2012})}\BibitemShut {NoStop}%
\bibitem [{\citenamefont {Wenz}\ \emph {et~al.}(2013)\citenamefont {Wenz},
  \citenamefont {Z{\"u}rn}, \citenamefont {Murmann}, \citenamefont {Brouzos},
  \citenamefont {Lompe},\ and\ \citenamefont {Jochim}}]{Wenz_2013}%
  \BibitemOpen
  \bibfield  {author} {\bibinfo {author} {\bibfnamefont {A.}~\bibnamefont
  {Wenz}}, \bibinfo {author} {\bibfnamefont {G.}~\bibnamefont {Z{\"u}rn}},
  \bibinfo {author} {\bibfnamefont {S.}~\bibnamefont {Murmann}}, \bibinfo
  {author} {\bibfnamefont {I.}~\bibnamefont {Brouzos}}, \bibinfo {author}
  {\bibfnamefont {T.}~\bibnamefont {Lompe}}, \ and\ \bibinfo {author}
  {\bibfnamefont {S.}~\bibnamefont {Jochim}},\ }\href {\doibase
  10.1126/science.1240516} {\bibfield  {journal} {\bibinfo  {journal}
  {Science}\ }\textbf {\bibinfo {volume} {342}},\ \bibinfo {pages} {457}
  (\bibinfo {year} {2013})}\BibitemShut {NoStop}%
\bibitem [{\citenamefont {Murmann}\ \emph
  {et~al.}(2015{\natexlab{a}})\citenamefont {Murmann}, \citenamefont
  {Bergschneider}, \citenamefont {Klinkhamer}, \citenamefont {Z\"urn},
  \citenamefont {Lompe},\ and\ \citenamefont {Jochim}}]{Murmann_2015a}%
  \BibitemOpen
  \bibfield  {author} {\bibinfo {author} {\bibfnamefont {S.}~\bibnamefont
  {Murmann}}, \bibinfo {author} {\bibfnamefont {A.}~\bibnamefont
  {Bergschneider}}, \bibinfo {author} {\bibfnamefont {V.~M.}\ \bibnamefont
  {Klinkhamer}}, \bibinfo {author} {\bibfnamefont {G.}~\bibnamefont {Z\"urn}},
  \bibinfo {author} {\bibfnamefont {T.}~\bibnamefont {Lompe}}, \ and\ \bibinfo
  {author} {\bibfnamefont {S.}~\bibnamefont {Jochim}},\ }\href {\doibase
  10.1103/PhysRevLett.114.080402} {\bibfield  {journal} {\bibinfo  {journal}
  {Phys. Rev. Lett.}\ }\textbf {\bibinfo {volume} {114}},\ \bibinfo {pages}
  {080402} (\bibinfo {year} {2015}{\natexlab{a}})}\BibitemShut {NoStop}%
\bibitem [{\citenamefont {Z\"urn}\ \emph {et~al.}(2013)\citenamefont {Z\"urn},
  \citenamefont {Wenz}, \citenamefont {Murmann}, \citenamefont {Bergschneider},
  \citenamefont {Lompe},\ and\ \citenamefont {Jochim}}]{Zurn_2013}%
  \BibitemOpen
  \bibfield  {author} {\bibinfo {author} {\bibfnamefont {G.}~\bibnamefont
  {Z\"urn}}, \bibinfo {author} {\bibfnamefont {A.~N.}\ \bibnamefont {Wenz}},
  \bibinfo {author} {\bibfnamefont {S.}~\bibnamefont {Murmann}}, \bibinfo
  {author} {\bibfnamefont {A.}~\bibnamefont {Bergschneider}}, \bibinfo {author}
  {\bibfnamefont {T.}~\bibnamefont {Lompe}}, \ and\ \bibinfo {author}
  {\bibfnamefont {S.}~\bibnamefont {Jochim}},\ }\href {\doibase
  10.1103/PhysRevLett.111.175302} {\bibfield  {journal} {\bibinfo  {journal}
  {Phys. Rev. Lett.}\ }\textbf {\bibinfo {volume} {111}},\ \bibinfo {pages}
  {175302} (\bibinfo {year} {2013})}\BibitemShut {NoStop}%
\bibitem [{\citenamefont {Murmann}\ \emph
  {et~al.}(2015{\natexlab{b}})\citenamefont {Murmann}, \citenamefont
  {Deuretzbacher}, \citenamefont {Z\"urn}, \citenamefont {Bjerlin},
  \citenamefont {Reimann}, \citenamefont {Santos}, \citenamefont {Lompe},\ and\
  \citenamefont {Jochim}}]{Murmann_2015b}%
  \BibitemOpen
  \bibfield  {author} {\bibinfo {author} {\bibfnamefont {S.}~\bibnamefont
  {Murmann}}, \bibinfo {author} {\bibfnamefont {F.}~\bibnamefont
  {Deuretzbacher}}, \bibinfo {author} {\bibfnamefont {G.}~\bibnamefont
  {Z\"urn}}, \bibinfo {author} {\bibfnamefont {J.}~\bibnamefont {Bjerlin}},
  \bibinfo {author} {\bibfnamefont {S.~M.}\ \bibnamefont {Reimann}}, \bibinfo
  {author} {\bibfnamefont {L.}~\bibnamefont {Santos}}, \bibinfo {author}
  {\bibfnamefont {T.}~\bibnamefont {Lompe}}, \ and\ \bibinfo {author}
  {\bibfnamefont {S.}~\bibnamefont {Jochim}},\ }\href {\doibase
  10.1103/PhysRevLett.115.215301} {\bibfield  {journal} {\bibinfo  {journal}
  {Phys. Rev. Lett.}\ }\textbf {\bibinfo {volume} {115}},\ \bibinfo {pages}
  {215301} (\bibinfo {year} {2015}{\natexlab{b}})}\BibitemShut {NoStop}%
\bibitem [{\citenamefont {Ashkin}(1970)}]{Ashkin_1970}%
  \BibitemOpen
  \bibfield  {author} {\bibinfo {author} {\bibfnamefont {A.}~\bibnamefont
  {Ashkin}},\ }\href {\doibase 10.1103/PhysRevLett.24.156} {\bibfield
  {journal} {\bibinfo  {journal} {Phys. Rev. Lett.}\ }\textbf {\bibinfo
  {volume} {24}},\ \bibinfo {pages} {156} (\bibinfo {year} {1970})}\BibitemShut
  {NoStop}%
\bibitem [{\citenamefont {Ashkin}\ \emph {et~al.}(1986)\citenamefont {Ashkin},
  \citenamefont {Dziedzic}, \citenamefont {Bjorkholm},\ and\ \citenamefont
  {Chu}}]{Ashkin_1986}%
  \BibitemOpen
  \bibfield  {author} {\bibinfo {author} {\bibfnamefont {A.}~\bibnamefont
  {Ashkin}}, \bibinfo {author} {\bibfnamefont {J.~M.}\ \bibnamefont
  {Dziedzic}}, \bibinfo {author} {\bibfnamefont {J.~E.}\ \bibnamefont
  {Bjorkholm}}, \ and\ \bibinfo {author} {\bibfnamefont {S.}~\bibnamefont
  {Chu}},\ }\href {\doibase 10.1364/OL.11.000288} {\bibfield  {journal}
  {\bibinfo  {journal} {Opt. Lett.}\ }\textbf {\bibinfo {volume} {11}},\
  \bibinfo {pages} {288} (\bibinfo {year} {1986})}\BibitemShut {NoStop}%
\bibitem [{\citenamefont {Matthews}(2009)}]{Matthews_2009}%
  \BibitemOpen
  \bibfield  {author} {\bibinfo {author} {\bibfnamefont {J.~N.~A.}\
  \bibnamefont {Matthews}},\ }\href {\doibase 10.1063/1.3086092} {\bibfield
  {journal} {\bibinfo  {journal} {Phys. Today}\ }\textbf {\bibinfo {volume}
  {62}},\ \bibinfo {pages} {26} (\bibinfo {year} {2009})}\BibitemShut {NoStop}%
\bibitem [{\citenamefont {Liao}\ \emph {et~al.}(2008)\citenamefont {Liao},
  \citenamefont {Bareil}, \citenamefont {Sheng},\ and\ \citenamefont
  {Chiou}}]{Liao_2008}%
  \BibitemOpen
  \bibfield  {author} {\bibinfo {author} {\bibfnamefont {G.-B.}\ \bibnamefont
  {Liao}}, \bibinfo {author} {\bibfnamefont {P.~B.}\ \bibnamefont {Bareil}},
  \bibinfo {author} {\bibfnamefont {Y.}~\bibnamefont {Sheng}}, \ and\ \bibinfo
  {author} {\bibfnamefont {A.}~\bibnamefont {Chiou}},\ }\href {\doibase
  10.1364/OE.16.001996} {\bibfield  {journal} {\bibinfo  {journal} {Opt.
  Express}\ }\textbf {\bibinfo {volume} {16}},\ \bibinfo {pages} {1996}
  (\bibinfo {year} {2008})}\BibitemShut {NoStop}%
\bibitem [{\citenamefont {Matheson}\ \emph {et~al.}(2021)\citenamefont
  {Matheson}, \citenamefont {Mendonca}, \citenamefont {Gibson}, \citenamefont
  {Dalgarno}, \citenamefont {Wright}, \citenamefont {Paterson},\ and\
  \citenamefont {Tassieri}}]{Matheson_2021}%
  \BibitemOpen
  \bibfield  {author} {\bibinfo {author} {\bibfnamefont {A.}~\bibnamefont
  {Matheson}}, \bibinfo {author} {\bibfnamefont {T.}~\bibnamefont {Mendonca}},
  \bibinfo {author} {\bibfnamefont {G.}~\bibnamefont {Gibson}}, \bibinfo
  {author} {\bibfnamefont {P.}~\bibnamefont {Dalgarno}}, \bibinfo {author}
  {\bibfnamefont {A.}~\bibnamefont {Wright}}, \bibinfo {author} {\bibfnamefont
  {L.}~\bibnamefont {Paterson}}, \ and\ \bibinfo {author} {\bibfnamefont
  {M.}~\bibnamefont {Tassieri}},\ }\href {\doibase 10.3389/fphy.2021.621512}
  {\bibfield  {journal} {\bibinfo  {journal} {Front. Phys.}\ }\textbf {\bibinfo
  {volume} {9}},\ \bibinfo {pages} {621512} (\bibinfo {year}
  {2021})}\BibitemShut {NoStop}%
\bibitem [{\citenamefont {Kaufman}\ \emph {et~al.}(2014)\citenamefont
  {Kaufman}, \citenamefont {Lester}, \citenamefont {Reynolds}, \citenamefont
  {Wall}, \citenamefont {Foss-Feig}, \citenamefont {Hazzard}, \citenamefont
  {Rey},\ and\ \citenamefont {Regal}}]{Kaufman_2014}%
  \BibitemOpen
  \bibfield  {author} {\bibinfo {author} {\bibfnamefont {A.~M.}\ \bibnamefont
  {Kaufman}}, \bibinfo {author} {\bibfnamefont {B.~J.}\ \bibnamefont {Lester}},
  \bibinfo {author} {\bibfnamefont {C.~M.}\ \bibnamefont {Reynolds}}, \bibinfo
  {author} {\bibfnamefont {M.~L.}\ \bibnamefont {Wall}}, \bibinfo {author}
  {\bibfnamefont {M.}~\bibnamefont {Foss-Feig}}, \bibinfo {author}
  {\bibfnamefont {K.~R.~A.}\ \bibnamefont {Hazzard}}, \bibinfo {author}
  {\bibfnamefont {A.~M.}\ \bibnamefont {Rey}}, \ and\ \bibinfo {author}
  {\bibfnamefont {C.~A.}\ \bibnamefont {Regal}},\ }\href {\doibase
  10.1126/science.1250057} {\bibfield  {journal} {\bibinfo  {journal}
  {Science}\ }\textbf {\bibinfo {volume} {345}},\ \bibinfo {pages} {306}
  (\bibinfo {year} {2014})}\BibitemShut {NoStop}%
\bibitem [{\citenamefont {Kaufman}\ \emph {et~al.}(2015)\citenamefont
  {Kaufman}, \citenamefont {Lester}, \citenamefont {Foss-Feig}, \citenamefont
  {Wall}, \citenamefont {Rey},\ and\ \citenamefont {Regal}}]{Kaufman_2016}%
  \BibitemOpen
  \bibfield  {author} {\bibinfo {author} {\bibfnamefont {A.~M.}\ \bibnamefont
  {Kaufman}}, \bibinfo {author} {\bibfnamefont {B.~J.}\ \bibnamefont {Lester}},
  \bibinfo {author} {\bibfnamefont {M.}~\bibnamefont {Foss-Feig}}, \bibinfo
  {author} {\bibfnamefont {M.~L.}\ \bibnamefont {Wall}}, \bibinfo {author}
  {\bibfnamefont {A.~M.}\ \bibnamefont {Rey}}, \ and\ \bibinfo {author}
  {\bibfnamefont {C.~A.}\ \bibnamefont {Regal}},\ }\href {\doibase
  10.1038/nature16073} {\bibfield  {journal} {\bibinfo  {journal} {Nature}\
  }\textbf {\bibinfo {volume} {527}},\ \bibinfo {pages} {208} (\bibinfo {year}
  {2015})}\BibitemShut {NoStop}%
\bibitem [{\citenamefont {Browaeys}\ and\ \citenamefont
  {Lahaye}(2020)}]{Browaeys_2020}%
  \BibitemOpen
  \bibfield  {author} {\bibinfo {author} {\bibfnamefont {A.}~\bibnamefont
  {Browaeys}}\ and\ \bibinfo {author} {\bibfnamefont {T.}~\bibnamefont
  {Lahaye}},\ }\href {\doibase 10.1038/s41567-019-0733-z} {\bibfield  {journal}
  {\bibinfo  {journal} {Nat. Phys.}\ }\textbf {\bibinfo {volume} {16}},\
  \bibinfo {pages} {1745} (\bibinfo {year} {2020})}\BibitemShut {NoStop}%
\bibitem [{\citenamefont {Barbier}\ \emph {et~al.}(2021)\citenamefont
  {Barbier}, \citenamefont {Hollerith},\ and\ \citenamefont
  {Hofstetter}}]{Barbier_2021}%
  \BibitemOpen
  \bibfield  {author} {\bibinfo {author} {\bibfnamefont {M.}~\bibnamefont
  {Barbier}}, \bibinfo {author} {\bibfnamefont {S.}~\bibnamefont {Hollerith}},
  \ and\ \bibinfo {author} {\bibfnamefont {W.}~\bibnamefont {Hofstetter}},\
  }\href {\doibase 10.1103/PhysRevA.104.053304} {\bibfield  {journal} {\bibinfo
   {journal} {Phys. Rev. A}\ }\textbf {\bibinfo {volume} {104}},\ \bibinfo
  {pages} {053304} (\bibinfo {year} {2021})}\BibitemShut {NoStop}%
\bibitem [{\citenamefont {Ortner}\ \emph {et~al.}(2009)\citenamefont {Ortner},
  \citenamefont {Micheli}, \citenamefont {Pupillo},\ and\ \citenamefont
  {Zoller}}]{Ortner_2009}%
  \BibitemOpen
  \bibfield  {author} {\bibinfo {author} {\bibfnamefont {M.}~\bibnamefont
  {Ortner}}, \bibinfo {author} {\bibfnamefont {A.}~\bibnamefont {Micheli}},
  \bibinfo {author} {\bibfnamefont {G.}~\bibnamefont {Pupillo}}, \ and\
  \bibinfo {author} {\bibfnamefont {P.}~\bibnamefont {Zoller}},\ }\href
  {\doibase 10.1088/1367-2630/11/5/055045} {\bibfield  {journal} {\bibinfo
  {journal} {New J. Phys.}\ }\textbf {\bibinfo {volume} {11}},\ \bibinfo
  {pages} {055045} (\bibinfo {year} {2009})}\BibitemShut {NoStop}%
\bibitem [{\citenamefont {Blackmore}\ \emph {et~al.}(2019)\citenamefont
  {Blackmore}, \citenamefont {Caldwell}, \citenamefont {Gregory}, \citenamefont
  {Bridge}, \citenamefont {Sawant}, \citenamefont {Aldegunde}, \citenamefont
  {Mur-Petit}, \citenamefont {Jaksch}, \citenamefont {Hutson}, \citenamefont
  {Sauer}, \citenamefont {Tarbutt},\ and\ \citenamefont
  {Cornish}}]{Blackmore_2019}%
  \BibitemOpen
  \bibfield  {author} {\bibinfo {author} {\bibfnamefont {J.~A.}\ \bibnamefont
  {Blackmore}}, \bibinfo {author} {\bibfnamefont {L.}~\bibnamefont {Caldwell}},
  \bibinfo {author} {\bibfnamefont {P.~D.}\ \bibnamefont {Gregory}}, \bibinfo
  {author} {\bibfnamefont {E.~M.}\ \bibnamefont {Bridge}}, \bibinfo {author}
  {\bibfnamefont {R.}~\bibnamefont {Sawant}}, \bibinfo {author} {\bibfnamefont
  {J.}~\bibnamefont {Aldegunde}}, \bibinfo {author} {\bibfnamefont
  {J.}~\bibnamefont {Mur-Petit}}, \bibinfo {author} {\bibfnamefont
  {D.}~\bibnamefont {Jaksch}}, \bibinfo {author} {\bibfnamefont {J.~M.}\
  \bibnamefont {Hutson}}, \bibinfo {author} {\bibfnamefont {B.~E.}\
  \bibnamefont {Sauer}}, \bibinfo {author} {\bibfnamefont {M.~R.}\ \bibnamefont
  {Tarbutt}}, \ and\ \bibinfo {author} {\bibfnamefont {S.~L.}\ \bibnamefont
  {Cornish}},\ }\href {\doibase 10.1088/2058-9565/aaee35} {\bibfield  {journal}
  {\bibinfo  {journal} {Quantum Sci. Technol.}\ }\textbf {\bibinfo {volume}
  {4}},\ \bibinfo {pages} {014010} (\bibinfo {year} {2019})}\BibitemShut
  {NoStop}%
\bibitem [{\citenamefont {Griesmaier}\ \emph {et~al.}(2005)\citenamefont
  {Griesmaier}, \citenamefont {Werner}, \citenamefont {Hensler}, \citenamefont
  {Stuhler},\ and\ \citenamefont {Pfau}}]{Griesmaier_2005}%
  \BibitemOpen
  \bibfield  {author} {\bibinfo {author} {\bibfnamefont {A.}~\bibnamefont
  {Griesmaier}}, \bibinfo {author} {\bibfnamefont {J.}~\bibnamefont {Werner}},
  \bibinfo {author} {\bibfnamefont {S.}~\bibnamefont {Hensler}}, \bibinfo
  {author} {\bibfnamefont {J.}~\bibnamefont {Stuhler}}, \ and\ \bibinfo
  {author} {\bibfnamefont {T.}~\bibnamefont {Pfau}},\ }\href {\doibase
  10.1103/PhysRevLett.94.160401} {\bibfield  {journal} {\bibinfo  {journal}
  {Phys. Rev. Lett.}\ }\textbf {\bibinfo {volume} {94}},\ \bibinfo {pages}
  {160401} (\bibinfo {year} {2005})}\BibitemShut {NoStop}%
\bibitem [{\citenamefont {Lu}\ \emph {et~al.}(2011)\citenamefont {Lu},
  \citenamefont {Burdick}, \citenamefont {Youn},\ and\ \citenamefont
  {Lev}}]{Lu_2011}%
  \BibitemOpen
  \bibfield  {author} {\bibinfo {author} {\bibfnamefont {M.}~\bibnamefont
  {Lu}}, \bibinfo {author} {\bibfnamefont {N.~Q.}\ \bibnamefont {Burdick}},
  \bibinfo {author} {\bibfnamefont {S.~H.}\ \bibnamefont {Youn}}, \ and\
  \bibinfo {author} {\bibfnamefont {B.~L.}\ \bibnamefont {Lev}},\ }\href
  {\doibase 10.1103/PhysRevLett.107.190401} {\bibfield  {journal} {\bibinfo
  {journal} {Phys. Rev. Lett.}\ }\textbf {\bibinfo {volume} {107}},\ \bibinfo
  {pages} {190401} (\bibinfo {year} {2011})}\BibitemShut {NoStop}%
\bibitem [{\citenamefont {Aikawa}\ \emph {et~al.}(2012)\citenamefont {Aikawa},
  \citenamefont {Frisch}, \citenamefont {Mark}, \citenamefont {Baier},
  \citenamefont {Rietzler}, \citenamefont {Grimm},\ and\ \citenamefont
  {Ferlaino}}]{Aikawa_2012}%
  \BibitemOpen
  \bibfield  {author} {\bibinfo {author} {\bibfnamefont {K.}~\bibnamefont
  {Aikawa}}, \bibinfo {author} {\bibfnamefont {A.}~\bibnamefont {Frisch}},
  \bibinfo {author} {\bibfnamefont {M.}~\bibnamefont {Mark}}, \bibinfo {author}
  {\bibfnamefont {S.}~\bibnamefont {Baier}}, \bibinfo {author} {\bibfnamefont
  {A.}~\bibnamefont {Rietzler}}, \bibinfo {author} {\bibfnamefont
  {R.}~\bibnamefont {Grimm}}, \ and\ \bibinfo {author} {\bibfnamefont
  {F.}~\bibnamefont {Ferlaino}},\ }\href {\doibase
  10.1103/PhysRevLett.108.210401} {\bibfield  {journal} {\bibinfo  {journal}
  {Phys. Rev. Lett.}\ }\textbf {\bibinfo {volume} {108}},\ \bibinfo {pages}
  {210401} (\bibinfo {year} {2012})}\BibitemShut {NoStop}%
\bibitem [{\citenamefont {Miyazawa}\ \emph {et~al.}(2022)\citenamefont
  {Miyazawa}, \citenamefont {Inoue}, \citenamefont {Matsui}, \citenamefont
  {Nomura},\ and\ \citenamefont {Kozuma}}]{MiyazawaARXIV2022}%
  \BibitemOpen
  \bibfield  {author} {\bibinfo {author} {\bibfnamefont {Y.}~\bibnamefont
  {Miyazawa}}, \bibinfo {author} {\bibfnamefont {R.}~\bibnamefont {Inoue}},
  \bibinfo {author} {\bibfnamefont {H.}~\bibnamefont {Matsui}}, \bibinfo
  {author} {\bibfnamefont {G.}~\bibnamefont {Nomura}}, \ and\ \bibinfo {author}
  {\bibfnamefont {M.}~\bibnamefont {Kozuma}},\ }\href
  {https://doi.org/10.48550/arXiv.2207.11692} {\bibfield  {journal} {\bibinfo
  {journal} {arXiv:2207.11692}\ } (\bibinfo {year} {2022})}\BibitemShut
  {NoStop}%
\bibitem [{\citenamefont {Norcia}\ and\ \citenamefont
  {Ferlaino}(2021)}]{Norcia_2021}%
  \BibitemOpen
  \bibfield  {author} {\bibinfo {author} {\bibfnamefont {M.~A.}\ \bibnamefont
  {Norcia}}\ and\ \bibinfo {author} {\bibfnamefont {F.}~\bibnamefont
  {Ferlaino}},\ }\href {\doibase 10.1038/s41567-021-01398-7} {\bibfield
  {journal} {\bibinfo  {journal} {Nature Physics}\ }\textbf {\bibinfo {volume}
  {17}},\ \bibinfo {pages} {1349} (\bibinfo {year} {2021})}\BibitemShut
  {NoStop}%
\bibitem [{\citenamefont {Chomaz}\ \emph {et~al.}(2022)\citenamefont {Chomaz},
  \citenamefont {Ferrier-Barbut}, \citenamefont {Ferlaino}, \citenamefont
  {Laburthe-Tolra}, \citenamefont {Lev},\ and\ \citenamefont
  {Pfau}}]{Chomaz_2022}%
  \BibitemOpen
  \bibfield  {author} {\bibinfo {author} {\bibfnamefont {L.}~\bibnamefont
  {Chomaz}}, \bibinfo {author} {\bibfnamefont {I.}~\bibnamefont
  {Ferrier-Barbut}}, \bibinfo {author} {\bibfnamefont {F.}~\bibnamefont
  {Ferlaino}}, \bibinfo {author} {\bibfnamefont {B.}~\bibnamefont
  {Laburthe-Tolra}}, \bibinfo {author} {\bibfnamefont {B.~L.}\ \bibnamefont
  {Lev}}, \ and\ \bibinfo {author} {\bibfnamefont {T.}~\bibnamefont {Pfau}},\
  }\href {https://doi.org/10.48550/arXiv.2201.02672} {\bibfield  {journal}
  {\bibinfo  {journal} {arXiv:2201.02672}\ } (\bibinfo {year}
  {2022})}\BibitemShut {NoStop}%
\bibitem [{\citenamefont {Frisch}\ \emph {et~al.}(2014)\citenamefont {Frisch},
  \citenamefont {Mark}, \citenamefont {Aikawa}, \citenamefont {Ferlaino},
  \citenamefont {Bohn}, \citenamefont {Makrides}, \citenamefont {Petrov},\ and\
  \citenamefont {Kotochigova}}]{Frisch_2014}%
  \BibitemOpen
  \bibfield  {author} {\bibinfo {author} {\bibfnamefont {A.}~\bibnamefont
  {Frisch}}, \bibinfo {author} {\bibfnamefont {M.}~\bibnamefont {Mark}},
  \bibinfo {author} {\bibfnamefont {K.}~\bibnamefont {Aikawa}}, \bibinfo
  {author} {\bibfnamefont {F.}~\bibnamefont {Ferlaino}}, \bibinfo {author}
  {\bibfnamefont {J.~L.}\ \bibnamefont {Bohn}}, \bibinfo {author}
  {\bibfnamefont {C.}~\bibnamefont {Makrides}}, \bibinfo {author}
  {\bibfnamefont {A.}~\bibnamefont {Petrov}}, \ and\ \bibinfo {author}
  {\bibfnamefont {S.}~\bibnamefont {Kotochigova}},\ }\href {\doibase
  10.1038/nature13137} {\bibfield  {journal} {\bibinfo  {journal} {Nature}\
  }\textbf {\bibinfo {volume} {507}},\ \bibinfo {pages} {475} (\bibinfo {year}
  {2014})}\BibitemShut {NoStop}%
\bibitem [{\citenamefont {Chomaz}\ \emph {et~al.}(2018)\citenamefont {Chomaz},
  \citenamefont {van Bijnen}, \citenamefont {Petter}, \citenamefont {Faraoni},
  \citenamefont {Baier}, \citenamefont {Becher}, \citenamefont {Mark},
  \citenamefont {Waechtler}, \citenamefont {Santos},\ and\ \citenamefont
  {Ferlaino}}]{Chomaz_2018}%
  \BibitemOpen
  \bibfield  {author} {\bibinfo {author} {\bibfnamefont {L.}~\bibnamefont
  {Chomaz}}, \bibinfo {author} {\bibfnamefont {R.~M.~W.}\ \bibnamefont {van
  Bijnen}}, \bibinfo {author} {\bibfnamefont {D.}~\bibnamefont {Petter}},
  \bibinfo {author} {\bibfnamefont {G.}~\bibnamefont {Faraoni}}, \bibinfo
  {author} {\bibfnamefont {S.}~\bibnamefont {Baier}}, \bibinfo {author}
  {\bibfnamefont {J.~H.}\ \bibnamefont {Becher}}, \bibinfo {author}
  {\bibfnamefont {M.~J.}\ \bibnamefont {Mark}}, \bibinfo {author}
  {\bibfnamefont {F.}~\bibnamefont {Waechtler}}, \bibinfo {author}
  {\bibfnamefont {L.}~\bibnamefont {Santos}}, \ and\ \bibinfo {author}
  {\bibfnamefont {F.}~\bibnamefont {Ferlaino}},\ }\href {\doibase
  10.1038/s41567-018-0054-7} {\bibfield  {journal} {\bibinfo  {journal} {Nat.
  Phys.}\ }\textbf {\bibinfo {volume} {14}},\ \bibinfo {pages} {442} (\bibinfo
  {year} {2018})}\BibitemShut {NoStop}%
\bibitem [{\citenamefont {Kadau}\ \emph {et~al.}(2016)\citenamefont {Kadau},
  \citenamefont {Schmitt}, \citenamefont {Wenzel}, \citenamefont {Wink},
  \citenamefont {Maier}, \citenamefont {Ferrier-Barbut},\ and\ \citenamefont
  {Pfau}}]{Kadau_2016}%
  \BibitemOpen
  \bibfield  {author} {\bibinfo {author} {\bibfnamefont {H.}~\bibnamefont
  {Kadau}}, \bibinfo {author} {\bibfnamefont {M.}~\bibnamefont {Schmitt}},
  \bibinfo {author} {\bibfnamefont {M.}~\bibnamefont {Wenzel}}, \bibinfo
  {author} {\bibfnamefont {C.}~\bibnamefont {Wink}}, \bibinfo {author}
  {\bibfnamefont {T.}~\bibnamefont {Maier}}, \bibinfo {author} {\bibfnamefont
  {I.}~\bibnamefont {Ferrier-Barbut}}, \ and\ \bibinfo {author} {\bibfnamefont
  {T.}~\bibnamefont {Pfau}},\ }\href {\doibase 10.1038/nature16485} {\bibfield
  {journal} {\bibinfo  {journal} {Nature}\ }\textbf {\bibinfo {volume} {530}},\
  \bibinfo {pages} {194} (\bibinfo {year} {2016})}\BibitemShut {NoStop}%
\bibitem [{\citenamefont {Schmitt}\ \emph {et~al.}(2016)\citenamefont
  {Schmitt}, \citenamefont {Wenzel}, \citenamefont {Böttcher}, \citenamefont
  {Ferrier-Barbut},\ and\ \citenamefont {Pfau}}]{Schmitt_2016}%
  \BibitemOpen
  \bibfield  {author} {\bibinfo {author} {\bibfnamefont {M.}~\bibnamefont
  {Schmitt}}, \bibinfo {author} {\bibfnamefont {M.}~\bibnamefont {Wenzel}},
  \bibinfo {author} {\bibfnamefont {F.}~\bibnamefont {Böttcher}}, \bibinfo
  {author} {\bibfnamefont {I.}~\bibnamefont {Ferrier-Barbut}}, \ and\ \bibinfo
  {author} {\bibfnamefont {T.}~\bibnamefont {Pfau}},\ }\href {\doibase
  10.1038/nature20126} {\bibfield  {journal} {\bibinfo  {journal} {Nature}\
  }\textbf {\bibinfo {volume} {539}},\ \bibinfo {pages} {259} (\bibinfo {year}
  {2016})}\BibitemShut {NoStop}%
\bibitem [{\citenamefont {Koch}\ \emph {et~al.}(2008)\citenamefont {Koch},
  \citenamefont {Lahaye}, \citenamefont {Metz}, \citenamefont {Fröhlich},
  \citenamefont {Griesmaier},\ and\ \citenamefont {Pfau}}]{Koch_2008}%
  \BibitemOpen
  \bibfield  {author} {\bibinfo {author} {\bibfnamefont {T.}~\bibnamefont
  {Koch}}, \bibinfo {author} {\bibfnamefont {T.}~\bibnamefont {Lahaye}},
  \bibinfo {author} {\bibfnamefont {J.}~\bibnamefont {Metz}}, \bibinfo {author}
  {\bibfnamefont {B.}~\bibnamefont {Fröhlich}}, \bibinfo {author}
  {\bibfnamefont {A.}~\bibnamefont {Griesmaier}}, \ and\ \bibinfo {author}
  {\bibfnamefont {T.}~\bibnamefont {Pfau}},\ }\href {\doibase 10.1038/nphys887}
  {\bibfield  {journal} {\bibinfo  {journal} {Nat. Phys.}\ }\textbf {\bibinfo
  {volume} {4}},\ \bibinfo {pages} {218} (\bibinfo {year} {2008})}\BibitemShut
  {NoStop}%
\bibitem [{\citenamefont {de~Paz}\ \emph {et~al.}(2013)\citenamefont {de~Paz},
  \citenamefont {Sharma}, \citenamefont {Chotia}, \citenamefont {Mar\'echal},
  \citenamefont {Huckans}, \citenamefont {Pedri}, \citenamefont {Santos},
  \citenamefont {Gorceix}, \citenamefont {Vernac},\ and\ \citenamefont
  {Laburthe-Tolra}}]{dePazPRL13}%
  \BibitemOpen
  \bibfield  {author} {\bibinfo {author} {\bibfnamefont {A.}~\bibnamefont
  {de~Paz}}, \bibinfo {author} {\bibfnamefont {A.}~\bibnamefont {Sharma}},
  \bibinfo {author} {\bibfnamefont {A.}~\bibnamefont {Chotia}}, \bibinfo
  {author} {\bibfnamefont {E.}~\bibnamefont {Mar\'echal}}, \bibinfo {author}
  {\bibfnamefont {J.~H.}\ \bibnamefont {Huckans}}, \bibinfo {author}
  {\bibfnamefont {P.}~\bibnamefont {Pedri}}, \bibinfo {author} {\bibfnamefont
  {L.}~\bibnamefont {Santos}}, \bibinfo {author} {\bibfnamefont
  {O.}~\bibnamefont {Gorceix}}, \bibinfo {author} {\bibfnamefont
  {L.}~\bibnamefont {Vernac}}, \ and\ \bibinfo {author} {\bibfnamefont
  {B.}~\bibnamefont {Laburthe-Tolra}},\ }\href {\doibase
  10.1103/PhysRevLett.111.185305} {\bibfield  {journal} {\bibinfo  {journal}
  {Phys. Rev. Lett.}\ }\textbf {\bibinfo {volume} {111}},\ \bibinfo {pages}
  {185305} (\bibinfo {year} {2013})}\BibitemShut {NoStop}%
\bibitem [{\citenamefont {Lepoutre}\ \emph {et~al.}(2018)\citenamefont
  {Lepoutre}, \citenamefont {Kechadi}, \citenamefont {Naylor}, \citenamefont
  {Zhu}, \citenamefont {Gabardos}, \citenamefont {Isaev}, \citenamefont
  {Pedri}, \citenamefont {Rey}, \citenamefont {Vernac},\ and\ \citenamefont
  {Laburthe-Tolra}}]{Lepoutre_2018}%
  \BibitemOpen
  \bibfield  {author} {\bibinfo {author} {\bibfnamefont {S.}~\bibnamefont
  {Lepoutre}}, \bibinfo {author} {\bibfnamefont {K.}~\bibnamefont {Kechadi}},
  \bibinfo {author} {\bibfnamefont {B.}~\bibnamefont {Naylor}}, \bibinfo
  {author} {\bibfnamefont {B.}~\bibnamefont {Zhu}}, \bibinfo {author}
  {\bibfnamefont {L.}~\bibnamefont {Gabardos}}, \bibinfo {author}
  {\bibfnamefont {L.}~\bibnamefont {Isaev}}, \bibinfo {author} {\bibfnamefont
  {P.}~\bibnamefont {Pedri}}, \bibinfo {author} {\bibfnamefont {A.~M.}\
  \bibnamefont {Rey}}, \bibinfo {author} {\bibfnamefont {L.}~\bibnamefont
  {Vernac}}, \ and\ \bibinfo {author} {\bibfnamefont {B.}~\bibnamefont
  {Laburthe-Tolra}},\ }\href {\doibase 10.1103/PhysRevA.97.023610} {\bibfield
  {journal} {\bibinfo  {journal} {Phys. Rev. A}\ }\textbf {\bibinfo {volume}
  {97}},\ \bibinfo {pages} {023610} (\bibinfo {year} {2018})}\BibitemShut
  {NoStop}%
\bibitem [{\citenamefont {Lepoutre}\ \emph {et~al.}(2019)\citenamefont
  {Lepoutre}, \citenamefont {Schachenmayer}, \citenamefont {Gabardos},
  \citenamefont {Zhu}, \citenamefont {Naylor}, \citenamefont {Maréchal},
  \citenamefont {Gorceix}, \citenamefont {Rey}, \citenamefont {Vernac},\ and\
  \citenamefont {Laburthe-Tolra}}]{Lepoutre_2019}%
  \BibitemOpen
  \bibfield  {author} {\bibinfo {author} {\bibfnamefont {S.}~\bibnamefont
  {Lepoutre}}, \bibinfo {author} {\bibfnamefont {J.}~\bibnamefont
  {Schachenmayer}}, \bibinfo {author} {\bibfnamefont {L.}~\bibnamefont
  {Gabardos}}, \bibinfo {author} {\bibfnamefont {B.}~\bibnamefont {Zhu}},
  \bibinfo {author} {\bibfnamefont {B.}~\bibnamefont {Naylor}}, \bibinfo
  {author} {\bibfnamefont {E.}~\bibnamefont {Maréchal}}, \bibinfo {author}
  {\bibfnamefont {O.}~\bibnamefont {Gorceix}}, \bibinfo {author} {\bibfnamefont
  {A.~M.}\ \bibnamefont {Rey}}, \bibinfo {author} {\bibfnamefont
  {L.}~\bibnamefont {Vernac}}, \ and\ \bibinfo {author} {\bibfnamefont
  {B.}~\bibnamefont {Laburthe-Tolra}},\ }\href {\doibase
  10.1038/s41467-019-09699-5} {\bibfield  {journal} {\bibinfo  {journal} {Nat.
  Commun.}\ }\textbf {\bibinfo {volume} {10}},\ \bibinfo {pages} {1714}
  (\bibinfo {year} {2019})}\BibitemShut {NoStop}%
\bibitem [{\citenamefont {Patscheider}\ \emph {et~al.}(2020)\citenamefont
  {Patscheider}, \citenamefont {Zhu}, \citenamefont {Chomaz}, \citenamefont
  {Petter}, \citenamefont {Baier}, \citenamefont {Rey}, \citenamefont
  {Ferlaino},\ and\ \citenamefont {Mark}}]{Patscheider_2020}%
  \BibitemOpen
  \bibfield  {author} {\bibinfo {author} {\bibfnamefont {A.}~\bibnamefont
  {Patscheider}}, \bibinfo {author} {\bibfnamefont {B.}~\bibnamefont {Zhu}},
  \bibinfo {author} {\bibfnamefont {L.}~\bibnamefont {Chomaz}}, \bibinfo
  {author} {\bibfnamefont {D.}~\bibnamefont {Petter}}, \bibinfo {author}
  {\bibfnamefont {S.}~\bibnamefont {Baier}}, \bibinfo {author} {\bibfnamefont
  {A.-M.}\ \bibnamefont {Rey}}, \bibinfo {author} {\bibfnamefont
  {F.}~\bibnamefont {Ferlaino}}, \ and\ \bibinfo {author} {\bibfnamefont
  {M.~J.}\ \bibnamefont {Mark}},\ }\href {\doibase
  10.1103/PhysRevResearch.2.023050} {\bibfield  {journal} {\bibinfo  {journal}
  {Phys. Rev. Res.}\ }\textbf {\bibinfo {volume} {2}},\ \bibinfo {pages}
  {023050} (\bibinfo {year} {2020})}\BibitemShut {NoStop}%
\bibitem [{\citenamefont {Gabardos}\ \emph {et~al.}(2020)\citenamefont
  {Gabardos}, \citenamefont {Zhu}, \citenamefont {Lepoutre}, \citenamefont
  {Rey}, \citenamefont {Laburthe-Tolra},\ and\ \citenamefont
  {Vernac}}]{GabardosPRL20}%
  \BibitemOpen
  \bibfield  {author} {\bibinfo {author} {\bibfnamefont {L.}~\bibnamefont
  {Gabardos}}, \bibinfo {author} {\bibfnamefont {B.}~\bibnamefont {Zhu}},
  \bibinfo {author} {\bibfnamefont {S.}~\bibnamefont {Lepoutre}}, \bibinfo
  {author} {\bibfnamefont {A.~M.}\ \bibnamefont {Rey}}, \bibinfo {author}
  {\bibfnamefont {B.}~\bibnamefont {Laburthe-Tolra}}, \ and\ \bibinfo {author}
  {\bibfnamefont {L.}~\bibnamefont {Vernac}},\ }\href {\doibase
  10.1103/PhysRevLett.125.143401} {\bibfield  {journal} {\bibinfo  {journal}
  {Phys. Rev. Lett.}\ }\textbf {\bibinfo {volume} {125}},\ \bibinfo {pages}
  {143401} (\bibinfo {year} {2020})}\BibitemShut {NoStop}%
\bibitem [{\citenamefont {Li}\ \emph {et~al.}(2018)\citenamefont {Li},
  \citenamefont {Tiesinga},\ and\ \citenamefont {Kotochigova}}]{Li_2018}%
  \BibitemOpen
  \bibfield  {author} {\bibinfo {author} {\bibfnamefont {M.}~\bibnamefont
  {Li}}, \bibinfo {author} {\bibfnamefont {E.}~\bibnamefont {Tiesinga}}, \ and\
  \bibinfo {author} {\bibfnamefont {S.}~\bibnamefont {Kotochigova}},\ }\href
  {\doibase 10.1103/PhysRevA.97.053627} {\bibfield  {journal} {\bibinfo
  {journal} {Phys. Rev. A}\ }\textbf {\bibinfo {volume} {97}},\ \bibinfo
  {pages} {053627} (\bibinfo {year} {2018})}\BibitemShut {NoStop}%
\bibitem [{\citenamefont {Tiesinga}\ \emph {et~al.}(2021)\citenamefont
  {Tiesinga}, \citenamefont {Kłos}, \citenamefont {Li}, \citenamefont
  {Petrov},\ and\ \citenamefont {Kotochigova}}]{Tiesinga_2021}%
  \BibitemOpen
  \bibfield  {author} {\bibinfo {author} {\bibfnamefont {E.}~\bibnamefont
  {Tiesinga}}, \bibinfo {author} {\bibfnamefont {J.}~\bibnamefont {Kłos}},
  \bibinfo {author} {\bibfnamefont {M.}~\bibnamefont {Li}}, \bibinfo {author}
  {\bibfnamefont {A.}~\bibnamefont {Petrov}}, \ and\ \bibinfo {author}
  {\bibfnamefont {S.}~\bibnamefont {Kotochigova}},\ }\href {\doibase
  10.1088/1367-2630/ac1a9a} {\bibfield  {journal} {\bibinfo  {journal} {New J.
  Phys.}\ }\textbf {\bibinfo {volume} {23}},\ \bibinfo {pages} {085007}
  (\bibinfo {year} {2021})}\BibitemShut {NoStop}%
\bibitem [{\citenamefont {Zhu}\ \emph {et~al.}(2019)\citenamefont {Zhu},
  \citenamefont {Rey},\ and\ \citenamefont {Schachenmayer}}]{Zhu_2019}%
  \BibitemOpen
  \bibfield  {author} {\bibinfo {author} {\bibfnamefont {B.}~\bibnamefont
  {Zhu}}, \bibinfo {author} {\bibfnamefont {A.~M.}\ \bibnamefont {Rey}}, \ and\
  \bibinfo {author} {\bibfnamefont {J.}~\bibnamefont {Schachenmayer}},\ }\href
  {\doibase 10.1088/1367-2630/ab354d} {\bibfield  {journal} {\bibinfo
  {journal} {New J. Phys.}\ }\textbf {\bibinfo {volume} {21}},\ \bibinfo
  {pages} {082001} (\bibinfo {year} {2019})}\BibitemShut {NoStop}%
\bibitem [{\citenamefont {Fersterer}\ \emph {et~al.}(2019)\citenamefont
  {Fersterer}, \citenamefont {Safavi-Naini}, \citenamefont {Zhu}, \citenamefont
  {Gabardos}, \citenamefont {Lepoutre}, \citenamefont {Vernac}, \citenamefont
  {Laburthe-Tolra}, \citenamefont {Blakie},\ and\ \citenamefont
  {Rey}}]{Fersterer_2019}%
  \BibitemOpen
  \bibfield  {author} {\bibinfo {author} {\bibfnamefont {P.}~\bibnamefont
  {Fersterer}}, \bibinfo {author} {\bibfnamefont {A.}~\bibnamefont
  {Safavi-Naini}}, \bibinfo {author} {\bibfnamefont {B.}~\bibnamefont {Zhu}},
  \bibinfo {author} {\bibfnamefont {L.}~\bibnamefont {Gabardos}}, \bibinfo
  {author} {\bibfnamefont {S.}~\bibnamefont {Lepoutre}}, \bibinfo {author}
  {\bibfnamefont {L.}~\bibnamefont {Vernac}}, \bibinfo {author} {\bibfnamefont
  {B.}~\bibnamefont {Laburthe-Tolra}}, \bibinfo {author} {\bibfnamefont
  {P.~B.}\ \bibnamefont {Blakie}}, \ and\ \bibinfo {author} {\bibfnamefont
  {A.~M.}\ \bibnamefont {Rey}},\ }\href {\doibase 10.1103/PhysRevA.100.033609}
  {\bibfield  {journal} {\bibinfo  {journal} {Phys. Rev. A}\ }\textbf {\bibinfo
  {volume} {100}},\ \bibinfo {pages} {033609} (\bibinfo {year}
  {2019})}\BibitemShut {NoStop}%
\bibitem [{\citenamefont {Suchorowski}\ \emph {et~al.}(2021)\citenamefont
  {Suchorowski}, \citenamefont {Dawid},\ and\ \citenamefont {Tomza}}]{OurRepo}%
  \BibitemOpen
  \bibfield  {author} {\bibinfo {author} {\bibfnamefont {M.}~\bibnamefont
  {Suchorowski}}, \bibinfo {author} {\bibfnamefont {A.}~\bibnamefont {Dawid}},
  \ and\ \bibinfo {author} {\bibfnamefont {M.}~\bibnamefont {Tomza}},\ }\href
  {\doibase 0.5281/zenodo.7182351} {\enquote {\bibinfo {title} {”gitlab
  repository: Two ultracold highly magnetic atoms in 1d harmonic trap},}\
  }\bibinfo {howpublished}
  {\url{https://gitlab.com/msuchorowski/two-cold-atoms-in-harmonic-trap}}
  (\bibinfo {year} {2021})\BibitemShut {NoStop}%
\bibitem [{\citenamefont {Gadway}\ and\ \citenamefont
  {Yan}(2016)}]{Gadway_2016}%
  \BibitemOpen
  \bibfield  {author} {\bibinfo {author} {\bibfnamefont {B.}~\bibnamefont
  {Gadway}}\ and\ \bibinfo {author} {\bibfnamefont {B.}~\bibnamefont {Yan}},\
  }\href {\doibase 10.1088/0953-4075/49/15/152002} {\bibfield  {journal}
  {\bibinfo  {journal} {J. Phys. B At. Mol. Opt. Phys.}\ }\textbf {\bibinfo
  {volume} {49}},\ \bibinfo {pages} {152002} (\bibinfo {year}
  {2016})}\BibitemShut {NoStop}%
\bibitem [{\citenamefont {Deuretzbacher}\ \emph {et~al.}(2010)\citenamefont
  {Deuretzbacher}, \citenamefont {Cremon},\ and\ \citenamefont
  {Reimann}}]{Deuretzbacher_2010}%
  \BibitemOpen
  \bibfield  {author} {\bibinfo {author} {\bibfnamefont {F.}~\bibnamefont
  {Deuretzbacher}}, \bibinfo {author} {\bibfnamefont {J.~C.}\ \bibnamefont
  {Cremon}}, \ and\ \bibinfo {author} {\bibfnamefont {S.~M.}\ \bibnamefont
  {Reimann}},\ }\href {\doibase 10.1103/PhysRevA.81.063616} {\bibfield
  {journal} {\bibinfo  {journal} {Phys. Rev. A}\ }\textbf {\bibinfo {volume}
  {81}},\ \bibinfo {pages} {063616} (\bibinfo {year} {2010})}\BibitemShut
  {NoStop}%
\bibitem [{\citenamefont {Sinha}\ and\ \citenamefont
  {Santos}(2007)}]{Sinha_2007}%
  \BibitemOpen
  \bibfield  {author} {\bibinfo {author} {\bibfnamefont {S.}~\bibnamefont
  {Sinha}}\ and\ \bibinfo {author} {\bibfnamefont {L.}~\bibnamefont {Santos}},\
  }\href {\doibase 10.1103/PhysRevLett.99.140406} {\bibfield  {journal}
  {\bibinfo  {journal} {Phys. Rev. Lett.}\ }\textbf {\bibinfo {volume} {99}},\
  \bibinfo {pages} {140406} (\bibinfo {year} {2007})}\BibitemShut {NoStop}%
\bibitem [{\citenamefont {Virtanen}\ \emph {et~al.}(2020)\citenamefont
  {Virtanen}, \citenamefont {Gommers}, \citenamefont {Oliphant}, \citenamefont
  {Haberland}, \citenamefont {Reddy}, \citenamefont {Cournapeau}, \citenamefont
  {Burovski}, \citenamefont {Peterson}, \citenamefont {Weckesser},
  \citenamefont {Bright}, \citenamefont {van~der Walt}, \citenamefont {Brett},
  \citenamefont {Wilson}, \citenamefont {Millman}, \citenamefont {Mayorov},
  \citenamefont {Nelson}, \citenamefont {Jones}, \citenamefont {Kern},
  \citenamefont {Larson}, \citenamefont {Carey}, \citenamefont {Polat},
  \citenamefont {Feng}, \citenamefont {Moore}, \citenamefont {VanderPlas},
  \citenamefont {Laxalde}, \citenamefont {Perktold}, \citenamefont {Cimrman},
  \citenamefont {Henriksen}, \citenamefont {Quintero}, \citenamefont {Harris},
  \citenamefont {Archibald}, \citenamefont {Ribeiro}, \citenamefont
  {Pedregosa},\ and\ \citenamefont {van Mulbregt}}]{scipy}%
  \BibitemOpen
  \bibfield  {author} {\bibinfo {author} {\bibfnamefont {P.}~\bibnamefont
  {Virtanen}}, \bibinfo {author} {\bibfnamefont {R.}~\bibnamefont {Gommers}},
  \bibinfo {author} {\bibfnamefont {T.~E.}\ \bibnamefont {Oliphant}}, \bibinfo
  {author} {\bibfnamefont {M.}~\bibnamefont {Haberland}}, \bibinfo {author}
  {\bibfnamefont {T.}~\bibnamefont {Reddy}}, \bibinfo {author} {\bibfnamefont
  {D.}~\bibnamefont {Cournapeau}}, \bibinfo {author} {\bibfnamefont
  {E.}~\bibnamefont {Burovski}}, \bibinfo {author} {\bibfnamefont
  {P.}~\bibnamefont {Peterson}}, \bibinfo {author} {\bibfnamefont
  {W.}~\bibnamefont {Weckesser}}, \bibinfo {author} {\bibfnamefont
  {J.}~\bibnamefont {Bright}}, \bibinfo {author} {\bibfnamefont {S.~J.}\
  \bibnamefont {van~der Walt}}, \bibinfo {author} {\bibfnamefont
  {M.}~\bibnamefont {Brett}}, \bibinfo {author} {\bibfnamefont
  {J.}~\bibnamefont {Wilson}}, \bibinfo {author} {\bibfnamefont {K.~J.}\
  \bibnamefont {Millman}}, \bibinfo {author} {\bibfnamefont {N.}~\bibnamefont
  {Mayorov}}, \bibinfo {author} {\bibfnamefont {A.~R.~J.}\ \bibnamefont
  {Nelson}}, \bibinfo {author} {\bibfnamefont {E.}~\bibnamefont {Jones}},
  \bibinfo {author} {\bibfnamefont {R.}~\bibnamefont {Kern}}, \bibinfo {author}
  {\bibfnamefont {E.}~\bibnamefont {Larson}}, \bibinfo {author} {\bibfnamefont
  {C.~J.}\ \bibnamefont {Carey}}, \bibinfo {author} {\bibfnamefont
  {{\.{I}}.}~\bibnamefont {Polat}}, \bibinfo {author} {\bibfnamefont
  {Y.}~\bibnamefont {Feng}}, \bibinfo {author} {\bibfnamefont {E.~W.}\
  \bibnamefont {Moore}}, \bibinfo {author} {\bibfnamefont {J.}~\bibnamefont
  {VanderPlas}}, \bibinfo {author} {\bibfnamefont {D.}~\bibnamefont {Laxalde}},
  \bibinfo {author} {\bibfnamefont {J.}~\bibnamefont {Perktold}}, \bibinfo
  {author} {\bibfnamefont {R.}~\bibnamefont {Cimrman}}, \bibinfo {author}
  {\bibfnamefont {I.}~\bibnamefont {Henriksen}}, \bibinfo {author}
  {\bibfnamefont {E.~A.}\ \bibnamefont {Quintero}}, \bibinfo {author}
  {\bibfnamefont {C.~R.}\ \bibnamefont {Harris}}, \bibinfo {author}
  {\bibfnamefont {A.~M.}\ \bibnamefont {Archibald}}, \bibinfo {author}
  {\bibfnamefont {A.~H.}\ \bibnamefont {Ribeiro}}, \bibinfo {author}
  {\bibfnamefont {F.}~\bibnamefont {Pedregosa}}, \ and\ \bibinfo {author}
  {\bibfnamefont {P.}~\bibnamefont {van Mulbregt}},\ }\href {\doibase
  10.1038/s41592-019-0686-2} {\bibfield  {journal} {\bibinfo  {journal} {Nat.
  Methods}\ }\textbf {\bibinfo {volume} {17}},\ \bibinfo {pages} {261}
  (\bibinfo {year} {2020})}\BibitemShut {NoStop}%
\bibitem [{\citenamefont {Grining}\ \emph
  {et~al.}(2015{\natexlab{b}})\citenamefont {Grining}, \citenamefont {Tomza},
  \citenamefont {Lesiuk}, \citenamefont {Przybytek}, \citenamefont {Musia\l{}},
  \citenamefont {Moszynski}, \citenamefont {Lewenstein},\ and\ \citenamefont
  {Massignan}}]{GriningPRA2015}%
  \BibitemOpen
  \bibfield  {author} {\bibinfo {author} {\bibfnamefont {T.}~\bibnamefont
  {Grining}}, \bibinfo {author} {\bibfnamefont {M.}~\bibnamefont {Tomza}},
  \bibinfo {author} {\bibfnamefont {M.}~\bibnamefont {Lesiuk}}, \bibinfo
  {author} {\bibfnamefont {M.}~\bibnamefont {Przybytek}}, \bibinfo {author}
  {\bibfnamefont {M.}~\bibnamefont {Musia\l{}}}, \bibinfo {author}
  {\bibfnamefont {R.}~\bibnamefont {Moszynski}}, \bibinfo {author}
  {\bibfnamefont {M.}~\bibnamefont {Lewenstein}}, \ and\ \bibinfo {author}
  {\bibfnamefont {P.}~\bibnamefont {Massignan}},\ }\href {\doibase
  10.1103/PhysRevA.92.061601} {\bibfield  {journal} {\bibinfo  {journal} {Phys.
  Rev. A}\ }\textbf {\bibinfo {volume} {92}},\ \bibinfo {pages} {061601}
  (\bibinfo {year} {2015}{\natexlab{b}})}\BibitemShut {NoStop}%
\bibitem [{\citenamefont {Bolda}\ \emph {et~al.}(2002)\citenamefont {Bolda},
  \citenamefont {Tiesinga},\ and\ \citenamefont {Julienne}}]{BoldaPRA2002}%
  \BibitemOpen
  \bibfield  {author} {\bibinfo {author} {\bibfnamefont {E.~L.}\ \bibnamefont
  {Bolda}}, \bibinfo {author} {\bibfnamefont {E.}~\bibnamefont {Tiesinga}}, \
  and\ \bibinfo {author} {\bibfnamefont {P.~S.}\ \bibnamefont {Julienne}},\
  }\href {\doibase 10.1103/PhysRevA.66.013403} {\bibfield  {journal} {\bibinfo
  {journal} {Phys. Rev. A}\ }\textbf {\bibinfo {volume} {66}},\ \bibinfo
  {pages} {013403} (\bibinfo {year} {2002})}\BibitemShut {NoStop}%
\bibitem [{\citenamefont {Kinoshita}\ \emph {et~al.}(2004)\citenamefont
  {Kinoshita}, \citenamefont {Wenger},\ and\ \citenamefont
  {Weiss}}]{Kinoshita_2004}%
  \BibitemOpen
  \bibfield  {author} {\bibinfo {author} {\bibfnamefont {T.}~\bibnamefont
  {Kinoshita}}, \bibinfo {author} {\bibfnamefont {T.}~\bibnamefont {Wenger}}, \
  and\ \bibinfo {author} {\bibfnamefont {D.~S.}\ \bibnamefont {Weiss}},\ }\href
  {\doibase 10.1126/science.1100700} {\bibfield  {journal} {\bibinfo  {journal}
  {Science}\ }\textbf {\bibinfo {volume} {305}},\ \bibinfo {pages} {1125}
  (\bibinfo {year} {2004})}\BibitemShut {NoStop}%
\bibitem [{\citenamefont {Hamilton}\ and\ \citenamefont
  {Yang}(2010)}]{hamilton_2010}%
  \BibitemOpen
  \bibfield  {author} {\bibinfo {author} {\bibfnamefont {J.~H.}\ \bibnamefont
  {Hamilton}}\ and\ \bibinfo {author} {\bibfnamefont {F.}~\bibnamefont
  {Yang}},\ }\href@noop {} {\emph {\bibinfo {title} {Modern Atomic and Nuclear
  Physics}}}\ (\bibinfo  {publisher} {World Scientific Publishing Company},\
  \bibinfo {year} {2010})\BibitemShut {NoStop}%
\bibitem [{\citenamefont {Haake}(2010)}]{haake_qc}%
  \BibitemOpen
  \bibfield  {author} {\bibinfo {author} {\bibfnamefont {F.}~\bibnamefont
  {Haake}},\ }\href@noop {} {\emph {\bibinfo {title} {Quantum Signatures of
  Chaos}}},\ \bibinfo {edition} {3rd}\ ed.\ (\bibinfo  {publisher} {Springer},\
  \bibinfo {address} {Berlin, Heidelberg},\ \bibinfo {year} {2010})\BibitemShut
  {NoStop}%
\end{thebibliography}%

\end{document}